\documentclass[final,3p,times]{elsarticle}
\usepackage{subcaption}
\usepackage{cprotect}
\usepackage[linesnumbered,ruled]{algorithm2e}

\usepackage{amssymb}
\usepackage{amsmath}

\journal{Journal of Computational Physics}

\begin{document}
	
\begin{frontmatter}
		
				
\title{Accurate and efficient surface reconstruction from volume fraction data on general meshes}
		

\author[a]{Henning Scheufler \corref{author}}
\author[b]{Johan Roenby}

\cortext[author] {Corresponding author.\\\textit{E-mail address:} Henning.Scheufler@dlr.de}
\address[a]{DLR German Aerospace Center, Institute of Space Systems, 28359 Bremen, Germany}
\address[b]{Department of Mathematical Sciences, Aalborg University, Frederikskaj 10A, 2450 K\o benhavn SV, Denmark}

\begin{abstract}
Simulations involving free surfaces and fluid interfaces are important in many areas of engineering. There is, however, still a need for improved simulation methods. Recently, a new efficient geometric VOF method called isoAdvector for general polyhedral meshes was published. We investigate the interface reconstruction step of isoAdvector, and demonstrate that especially for unstructured meshes the applied isosurface based approach can lead to noisy interface orientations. We then introduce a novel computational interface reconstruction scheme based on calculation of a reconstructed distance function (RDF). By iterating over the RDF calculation and interface reconstruction, we obtain second order convergence of both the interface normal and position within cells even with a strict $L_{\infty}$ error norm. In 2D this is verified with reconstruction of a circle on Cartesian meshes and on unstructured triangular and polygonal prism meshes. In 3D the second order convergence is verified with reconstruction of a sphere on Cartesian meshes and on unstructured tetrahedral and polyhedral meshes. The new scheme is combined with the interface advection step of the isoAdvector algorithm. Significantly reduced absolute advection errors are obtained, and for CFL number 0.2 and below we demonstrate second order convergence on all the mentioned mesh types in 2D and 3D. The implementation of the proposed interface reconstruction schemes is straightforward and the computational cost is significantly reduced compared to contemporary methods. The schemes are implemented as an extension to the Computational Fluid Dynamics (CFD) Open Source software package, OpenFOAM\textregistered{}. The extension module and all test cases presented in this paper are released as open source.
\end{abstract}
		

\end{frontmatter}

\section{Introduction}\label{Introduction}
Free surface simulations can provide highly valuable quantitative and qualitative insights in a large variety of engineering applications. The first numerical models were developed in the 1970's and the difficulties of this subject are reflected in the continued high level of activity within the research field. Several different approaches are used to model the free surface \cite{Worner.2012}: moving meshes, marker points, level set, phase fields and volume of fluid. The first two methods are not optimal for large deformations or topological changes of the interface. The volume of fluid and level set approaches do not suffer from these problems and are the most commonly used interface methods for practical engineering applications. A prominent weakness of the level set method is its lack of strict mass conservation which is caused by the fundamental problem that the signed distance function is not a passive tracer field. However, in practice it allows for good predictions if the discretization is fine enough so that the mass loss can be neglected. The volume of fluid method is strictly mass conservative but accurate models are harder to implement. The volume of fluid method based on the approach of Hirt \cite{Hirt.1981} can be divided into two categories: One with a finite interface thickness and the other with infinitesimal interface thickness. Models using finite interface thickness are easier to implement and are still used today e.g. in OpenFOAM and Fluent\textregistered, but are often not sufficiently accurate. Models assuming an infinitesimal interface thickness comprise of two steps: Interface reconstruction and interface advection. The most common reconstruction method is the piecewise linear interface reconstruction method (PLIC), where the interface within a cell is represented by a plane cutting the cell into two subcells. The resulting surface lacks $C^{0}$ continuity. Several approaches for the computation of the normal vector of the plane can be found in literature. In the widely used method by Youngs \cite{Youngs.1982}, the normal vector is calculated from the gradient of the volume fraction field. This method is easy to implement, also on three dimensional unstructured meshes, but lacks accuracy. In the least squares volume-of-fluid interface reconstruction algorithm (LVIRA) {\cite{Pilliod.1997}} an interface plane is projected into the neighbouring cells. The plane is used to calculate a volume fraction. The error of the calculated volume fraction in the neighbouring cells is minimized by changing the orientation of the plane. This delivers accurate results on all mesh types in three dimensions.  However, the method needs a two dimensional minimizer in three dimensions which complicates the implementation and makes the method computationally demanding. Currently, the most accurate approach for the estimation of the normal vector and the curvature is the Height Function method in which the volume fractions in columns of cells are integrated by direction (x,y,z). The derivative is calculated by finite difference operators. Height Functions are second order accurate in curvature and interface normal calculation but only work with hexahedral structured meshes in 3D. In 2D the method has been extended to unstructured triangular prism meshes by Ito et al. \cite{Ito.2014}. In 3D a so-called Embedded Height Function (EHF) approach was invented in Ivey and Moin \cite{Ivey.2015}, where the VOF field is geometrically mapped onto an overlapping hexahedral mesh on which the Height Function and its derivatives can then be calculated.  A less complex alternative are the variants of the Swartz algorithm {\cite{Swartz.1989}}. This method iteratively increases accuracy of the interface normal based on the interface centroids in the surrounding cells. The variant by Dyadechko and Shashkov {\cite{Dyadechko.2005}} showed second order convergence for structured and unstructured meshes in two dimensions. Liovic et al. {\cite{Liovic.2006}} presented a variant for three dimensional structured meshes with the same level of accuracy. The Patterned Interface Reconstruction by Mosso et al. {\cite{Mosso.2009}} demonstrated second order convergence on tetrahedral and hexahedral meshes. Recently, Maric et al. {\cite{Maric.2018}} presented a less expensive variant for unstructured hexahedral meshes but without showing results for the convergence of the reconstruction scheme (i.e. without the advection step). Lopez et al. {\cite{Lopez.2008} proposed among others the conservative level-contour interface reconstruction method (CLCIR). In this method the local interface normal is calculated based on the orientation of a triangulated surface which is initially an isosurface. The surface is adjusted to more accurately match the volume fraction where more than two iterations are needed to fulfill the criterion for the planar preservation defined by Liovic et al. {\cite{Liovic.2006}}. The method is significantly faster than ELVIRA and achieved second order convergence in pure reconstruction and advection benchmarks on structured meshes.} Dyadechko and Shashkov {\cite{Dyadechko.2005} introduced their variant of the Swartz method in the context of the moment of fluid method (MoF) in which also the centroid of the fluid is advected. In the MoF approach the reconstructed interface is chosen as the one that exactly matches the volume fraction, while giving the best fit for the volume centroid.} The method showed accurate results on unstructured three dimensional meshes {\cite{Ahn.2007}}, but it requires a minimization, which increases implementation complexity and computational cost. 

Another approach for determining the interface is to calculate a signed distance function or a level set function. Two different strategies for obtaining the distance function can be found in literature: Either solving the defining PDE or reconstructing the distance function geometrically. The first strategy was originally proposed by Sussman and Puckett {\cite{Sussman.2000}} in connection with their coupled level set/volume-of-fluid (CLSVOF) method. Several variants of the CLSVOF approach can be found in literature, e.g. Le Chenadec and Pitsch {\cite{LeChenadec.2013}}, Menard et al. {\cite{Menard.2007}} and Wang et al. {\cite{Wang.2009}}. They obtain improved mass conservation compared to the standard level set method. The solution of both a VOF and a level set equation, however, complicates implementation and increasing the computational cost. The second approach, where the so-called reconstructed distance function (RDF) is calculated was taken by Cummins et al. {\cite{Cummins.2005}} , who used an interface reconstructed with the Youngs method as the basis for their RDF calculation on a two dimensional Cartesian mesh. More recently, Sun and Tao {\cite{Sun.2010}} proposed an iterative variant of this RDF method, which was later extended to three dimensional structured meshes by Ling et al. {\cite{Ling.2015}}. Reconstruction convergence results were not reported, but second order convergence of the produced advection method was achieved on a three dimensional deformation test case from {\cite{Leveque.1996}}. Cao et al. {\cite{Cao.2018}} tested this approach on unstructured triangular prism meshes in two dimensions. They demonstrated greatly reduced spurious currents due to surface tension and curvature errors when comparing to the classical approach by Brackbill et al. {\cite{Brackbill.1992}}.

The methods proposed in this work are also iterative variants of the RDF method {\cite{Cummins.2005}}, but with focus on the interface reconstruction on unstructured three dimensional meshes consisting of general polyhedral cells. In contrast to, Sun and Tao {\cite{Sun.2010}}, Ling et al. {\cite{Ling.2015}} and Cao et al. {\cite{Cao.2018}} our reconstruction of  the signed distance function in an interface cell only requires information from its point neighbours (i.e. cells with which it shares a vertex). This allows for easy and efficient implementation and parallelization. Additionally, the iterative procedure is based on a residual stopping criterion rather than a predefined number of iterations. Another novelty in our interface advection method is the usage of interface normal information from the previous time step in the initial interface normal estimation for the current time step.  As will be demonstrated, this leads to second order convergence with mesh refinement, while still retaining the computational expenses at a moderate level.

The paper is organized as follows: In Section~\ref{sec:VOFEqn}, we clarify the role and importance of accurate interface reconstruction in the advection step of interfacial flow simulations. In Section~\ref{sec:recon}, we present the three variants of interface reconstruction procedures under investigation. In Section~\ref{Sec:Results}, we demonstrate the accuracy and convergence properties of these reconstruction schemes and compare with results from literature. In Section~\ref{sec:advectionResults}, we investigate their convergence properties in a number of advection test cases. Finally, in Section~\ref{Sec:Conclusion} we summarize our findings.
	
\section{The fundamental VOF equation}\label{sec:VOFEqn}

We consider a system of two immiscible, incompressible fluids occupying the space of a computational domain. The state of such a system is described by a velocity field, $\mathbf u(\mathbf x,t)$, a pressure field, $p(\mathbf x,t)$, and finally the instantaneous position of the dynamically evolving sharp fluid interface. The time evolution of the velocity and pressure field are determined by the Navier--Stokes equations together with the incompressibility condition and appropriate boundary conditions on the domain boundaries and on the fluid interface. The fluid interface may be represented by the fluid density field, $\rho(\mathbf x,t)$, taking a constant value in each of the two fluids, and thus jumping from one to the other value at the fluid interface. With this representation, the interface evolution in time is determined by the continuity equation. This is the basis of the Volume--of--Fluid (VOF) method, where, in the spirit of the Finite Volume Method (FVM), the equation is integrated over a control volume in order to account for the change of mass within the volume. With the control volume being the $i$'th cell of an FVM mesh, the volume integrated continuity equation can be written,
\begin{equation}\label{eq:volIntContEqn}
	\frac{d}{dt}\int_{V_i}\rho(\mathbf x,t)dV + \sum_f \int_f\rho(\mathbf x,t)\mathbf u(\mathbf x,t) \cdot d\mathbf S = 0,
\end{equation}
where $V_i$ is the cell volume and the sum is over the polygonal faces comprising the cell boundary. Let us call the reference fluid A and the other fluid B and denote their densities $\rho_A$ and $\rho_B$, respectively. Then we can write the density field in terms of the 3--dimensional Heaviside function, which is $1$ in fluid A and $0$ in fluid B,
\begin{equation}\label{eq:rhoOfH}
	\rho(\mathbf x,t) = \rho_A H(\mathbf x,t) + \rho_B (1 - H(\mathbf x,t)).
\end{equation}
Inserting Eqn.~\eqref{eq:rhoOfH} into Eqn.~\eqref{eq:volIntContEqn} and formally integrating in time, we get after some rearrangement
\begin{equation}\label{eq:alphaEvol}
	\alpha_i(t+\Delta t) = \alpha_i(t) - \frac1{V_i}\sum_f \int_t^{t+\Delta t}\int_f H(\mathbf x,\tau)\mathbf u(\mathbf x,\tau)\cdot d\mathbf S d\tau,
\end{equation}
where we have defined the volume fraction (of fluid A) in cell $i$ as 
\begin{equation}\label{eq:alphaDef}
	\alpha_i(t) = \frac{1}{V_i}\int_{V_i}H(\mathbf x,t) dV.
\end{equation}
In the VOF method the cell volume fractions, $\alpha_i$, are used to implicitly represent the fluid interface, and so Eqn.~\eqref{eq:alphaEvol} is the exact evolution equation which is to be discretized and solved approximately as part of our interfacial flow solution algorithm.

\begin{figure}[tb!]
		\includegraphics[width=\textwidth]{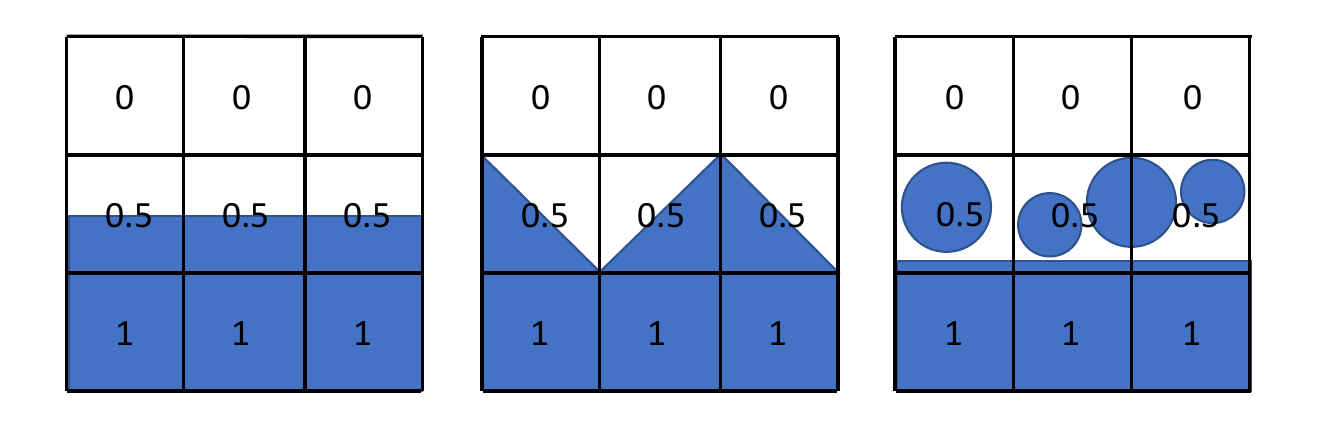}
		\caption{Three examples of fluid distributions giving rise to the same volume fractions on a 2D Cartesian mesh.}
		\label{fig:FluidDists}
\end{figure}

Having only the volume fraction information available and not the actual interface, we have no information about the configuration of fluid A and fluid B inside the cells. Fig.~\ref{fig:FluidDists} shows three different interfaces, and resulting intracellular fluid configurations, that all give rise to the same volume fractions on a Cartesian mesh. In particular, we have no information about the instantaneous distribution of the two fluids at the cell faces. This is required to calculate the integral on the right hand side of Eqn.~\eqref{eq:alphaEvol}. We must therefore produce a subgrid scale model of this intracellular configuration -- or equivalently try to reconstruct the interface inside each cell with $0 < \alpha_i < 1$. The only data available for such a reconstruction are the volume fraction values in the surrounding cells. There is in general an infinite number of interfaces that can produce a given volume fraction field. But if we assume that the interface is smooth, then at each point on the interface there is a characteristic length scale given by the local radius of curvature, $R$, such that if the local cell size, $\Delta x$, is much smaller than $R$, then the piece of interface inside a cell can be approximated well by a plane (in the sense that the angular variations of the interface normal around its average orientation inside the cell are $\ll$ 1 radian). The rightmost panel in Fig.~\ref{fig:FluidDists} shows an example of a mesh that is so coarse ($\Delta x \sim R$) that a planar approximation of the interface inside a cell is not adequate. The cell sizes we will consider in the following will thus be bounded from above by the requirement that $\Delta x < R$ and from below by a practical limit set by our available computer resources.

\section{Interface reconstruction schemes}\label{sec:recon}

If the interface inside a cell is approximately planar, then the role of the reconstruction step in a geometric VOF algorithm is to provide the interface centre positions, $\mathbf x_S$, and unit normals, $\hat{\mathbf n}_S$, inside each interface cell at the beginning of a time step. In this work we regard cell $i$ as an interface cell if
\begin{equation}
	\epsilon < \alpha_i < 1 - \epsilon,
\end{equation}
where $\epsilon$ is a user specified tolerance, which we typically set to $10^{-8}$. By convention the unit normal, $\hat{\mathbf n}_{S,i}$ in cell $i$ points away from fluid A and into fluid B. In the following three sections, we give an algorithmic overview of the three proposed procedures for obtaining $\mathbf x_S$ and $\hat{\mathbf n}_S$.

\subsection{Outline of the \texttt{iso-Alpha} method}\label{sec:iso-Alpha}

The first procedure is the $\alpha$-isosurface based method described in \cite{Roenby.2016}, where it forms the reconstruction step in the isoAdvector VOF algorithm.  In this approach, the cell volume fraction data is interpolated to the cell vertices which allows calculation of internal approximately planar isosurfaces in the interface cells. The numerically calculated isosurface inside a cell cuts it into two subcells which we will assume to coincide with the fluid interface so that one subcells is completely immersed in fluid A and the other subcell is completely immersed in fluid B. If the same isovalue is chosen for all cells the so constructed interface will be a continuous surface. However, with a global isovalue the cut cells will not in general be cut into subvolumes in correspondence to the cell’s volume fraction value. Since this is crucial for the interface advection step, we seek a distinct isovalue in each interface cell such that the resulting intracellular interface generates subcells of volumes in accordance with the cell’s volume fraction. We note that with distinct isovalues in two adjacent cells, the interface segments in such adjacent cells no longer meet up exactly at the face shared by the two cells, and so the global interface representation is not continuous. The method is outlined in Algorithm \ref{alg:isoAlpha} and the details are given in the following subsections.

{\SetAlgoNoLine%
	\begin{algorithm}
	\DontPrintSemicolon
	Calculate interfaceCells: A list of all the interface cell indices, i.e. $i\in$ interfaceCells if $\epsilon < \alpha_i < 1-\epsilon$.\\
	\For {$i \in$ interfaceCells}
	{
		For each of the $N_v$ vertices of interface cell $i$ interpolate the surrounding cell volume fractions to the vertex,
		\begin{equation}\label{eq:interpolation}
		      f_v = \frac{\sum_{k\in \mathcal C_v} w_k \alpha_k}{\sum_{k\in \mathcal C_v} w_k},
		\end{equation}
		where $\mathcal C_v$ is the set of cells to which the vertex belongs and the interpolation weights, $w_k$, are the linear inverse of the cell--centre--to--vertex distances.\label{isoAlphaStp:interpol}\\
		From the range of sorted vertex values, $f_1,...,f_{N_v}$, choose a middle value, $f_M$, where $M = round(N_v/2)$. \label{isoAlphaStp:chooseInitAv}\\
		Calculate the $f_M$-isosurface cutting the cell into two subcells (see Section \ref{ssec:isoFromVertData} for details).\label{isoAlphaStp:isosurface}\\
		Geometrically calculate the volume fraction as functions of the isovalue, $A(f)$, for the particular isovalue $f_M$ (See Section \ref{ssec:subVolume} for details). \label{isoAlphaStp:volRecon}\\
		Use $f_M$ as the initial guess in a root finding procedure to find the (cell specific) isovalue, $f^*$, such that the $f^*$-isosurface in cell $i$ results in the correct volume fraction, $A(f^*) = \alpha_i$ (see Section \ref{ssec:rootFinding} for details).\\
		Calculate the face centre, $\mathbf x_{S,i}$, and face unit normal, $\hat{\mathbf n}_{S,i}$, of this  $f^*$-isosurface in cell $i$ (see Section \ref{ssec:xSAndnS} for details).\label{isoAlphaStp:calcxSandnS}
	}
	\caption{The iso-Alpha interface reconstruction algorithm}
	\label{alg:isoAlpha}
\end{algorithm}} 

\subsubsection{Isosurface reconstruction from vertex data}\label{ssec:isoFromVertData}

Here we describe the procedure used to obtain the $f_0$-isosurface inside a polyhedral cell of a function $f$ whose values are specified at the cell vertices. This is used in Step \ref{isoAlphaStp:isosurface} in Algorithm \ref{alg:isoAlpha}, where the vertex values are the volume fractions, $\alpha$, interpolated to the vertices. It is also used in Step \ref{isoRDFStp:calcxSandnS} of Algorithm \ref{alg:isoRDF}, for the vertex values of the reconstructed distance function, $\Psi$, and in Algorithm {\ref{alg:plicRDF}} with the distance to the interface plane, $d^*$, as vertex values.

Consider the edge between the two vertices, $\mathbf x_j$ and $\mathbf x_k$, with function values $f_j$ and $f_k$, respectively. If $\min(f_j,f_k) \leq f_0 < \max(f_j,f_k)$, then the edge is cut by the $f_0$-isosurface at the point
\begin{equation}
		\mathbf x_{\textrm{cut}} = \mathbf x_j + \lambda_{jk}(\mathbf x_k - \mathbf x_j),\quad \textrm{   where   } \quad \lambda_{jk} = \frac{f_0 - f_j}{f_k - f_j}.
\end{equation}
After finding all such edge cut points for the cell, the cut points can be connected to form the periphery of an internal ``isoface'' as illustrated in Fig.~\ref{fig:slicedCellwithFace}(left). 

This reconstruction method is similar to step 1 in the CLCIR method used in Lopez et al. {\cite{Lopez.2008}} to obtain the $\alpha = 0.5$ isosurface in hexahedral cells for their initial interface normal estimate. We note that the so constructed $\alpha = 0.5$ isosurface generally does not pass through all cells with $0 < \alpha < 1$. Hence there will be surface cells, where the $\alpha = 0.5$ isosurface does not provide the required model for the interface position and orientation in the cell. This is why we have chosen to operate with a general $f_0$ isosurface. 

\subsubsection{Calculating the volume fraction from the isosurface}\label{ssec:subVolume}

\begin{figure}
	\centering
	\includegraphics[width=1.0\textwidth]{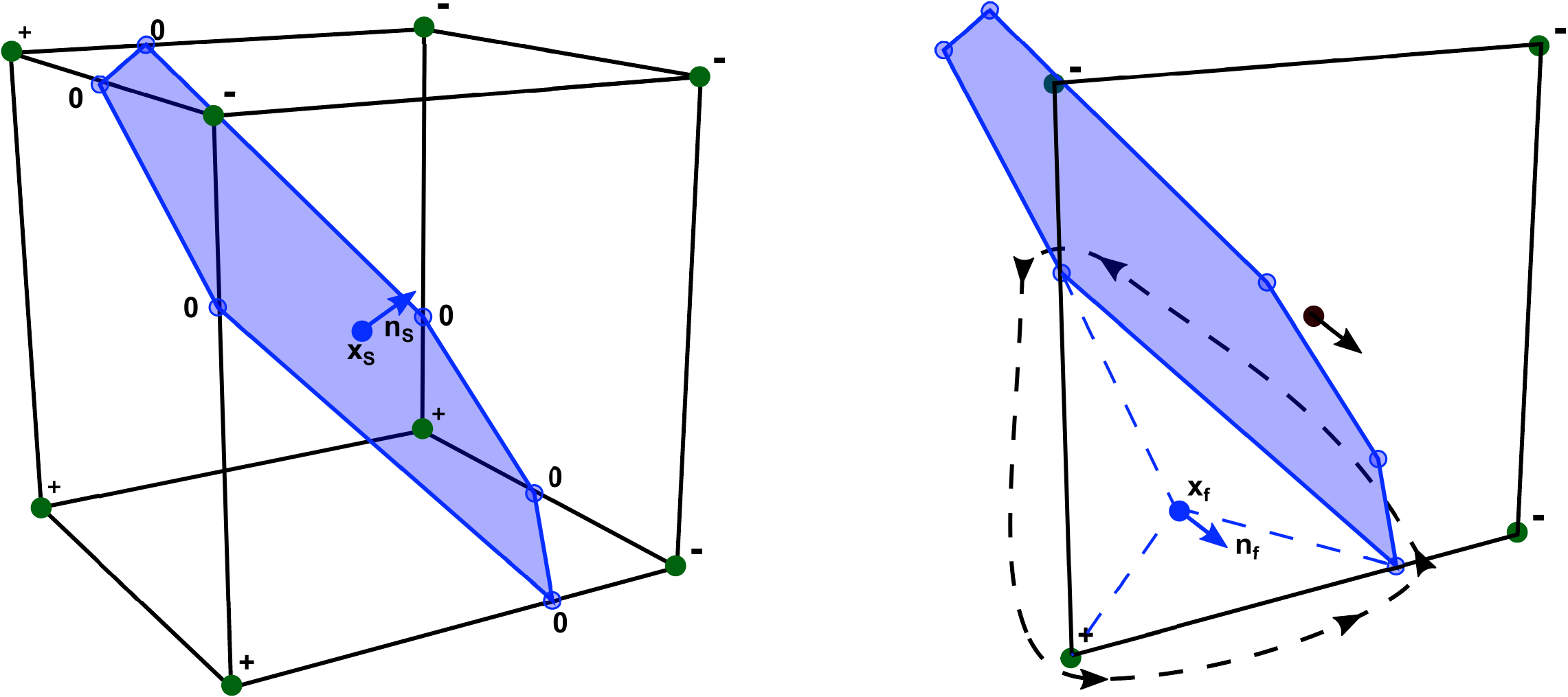}
	\caption{Isosurface inside a hexahedral cell. Left: ``isoface'' constructed from connecting its cut points along the cell edges. Right: Construction of a subface as part of the boundary of the subcell submerged in fluid A.}
	\label{fig:slicedCellwithFace}
\end{figure}

We now describe how we calculate the volume of the subcell of cell $i$ which is submerged in fluid A. This subcell is defined by cutting the cell with the $f_0$-isosurface passing through it. 
In general the volume of a polyhedral cell defined by its $N_f$ polygonal faces can be calculated using a pyramid decomposition based on an arbitrarily chosen point inside the cell. In this work, we calculate the volume as
\begin{equation}
	V = \sum_f \frac13|\mathbf n_f\cdot (\mathbf x_f - \bar{\mathbf x})|, \quad \textrm{ where } \quad \bar{\mathbf x} = \frac1{N_f}\sum_f \mathbf x_f,
\end{equation}
the sums are over all the cell's faces, and $\mathbf x_f$ and $\mathbf n_f$ are the face centres and face area vectors (pointing out of the cell), respectively. In particular, for a subcell defined by the cutting of a cell by an isosurface, the face list in the sum consists of
\begin{itemize}
\item The isoface inside the cell (Fig.~\ref{fig:slicedCellwithFace}(left)).
\item The ``submerged'' part of all the cell faces that are cut by the isosurface (Fig.~\ref{fig:slicedCellwithFace}(right)).
\item All the cell faces that are fully submerged in fluid A.
\end{itemize}
From the resulting subvolume, $V_A$, of cell $i$ submerged in fluid A, the corresponding volume fraction is then simply $V_A/V_i$, where $V_i$ is the volume of cell $i$.

\subsubsection{Calculating the centre and unit normal of a polygonal face}\label{ssec:xSAndnS}

We now describe the calculation used to find the centre, $\mathbf x_f$, and unit normal, $\hat{\mathbf n}_f$, for a polygonal face with $N_v$ vertices. In particular, this is needed for the isoface, which is in general not exactly planar. First, the face area vector is defined as the area weighted average of its triangular decomposition based on an average point,
\begin{equation}
	\mathbf n_f = \sum_{k=1}^{N_v} \mathbf n_{f,k}, \ \textrm{where} \ \mathbf n_{f,k} = \frac12 (\mathbf x_{k+1} - \mathbf x_k)\times(\bar{\mathbf x} - \mathbf x_k) \ \textrm{and} \
	\bar{\mathbf x} = \frac1{N_v} \sum_{k = 1}^{N_v} \mathbf x_k.
\end{equation}
Here we use $\mathbf x_{N_v+1} = \mathbf x_{1}$, and the vertex points are assumed to be ordered such that each $\mathbf n_{f,k}$ points out of fluid A. The unit normal vector and face centre point are then calculated as
\begin{equation}
	\hat{\mathbf n}_f = \frac{\mathbf n_f}{|\mathbf n_f|} \quad \textrm{and} \quad
	\mathbf x_f = \sum_{k = 1}^{N_v} \frac{|\mathbf n_{f,k}|}{|\mathbf n_f|}\frac{\mathbf x_k + \mathbf x_{k+1} + \bar{\mathbf x}}{3}.
\end{equation}

We note that for planar polygons these definitions are independent of the choice of the internal decomposition point, $\bar{\mathbf x}$, as long as this lies in the interface plane. For non-planar polygons, this is not the case, and our choice of $\bar{\mathbf x}$ above constitutes a convention. This convention is directly adopted from OpenFOAM.

\subsubsection{Root finding procedure}\label{ssec:rootFinding}

The procedure outlined in Section \ref{ssec:subVolume} above gives the volume, $V_A$, of the subcell submerged in fluid A as a function of the isovalue, $f$, chosen for the isosurface. A numerical root finding procedure can then be used to find the isovalue, $f^*$, such that $V_A(f^*)/V_i = \alpha_i$, where $V_i$ is the volume of cell $i$. This will involve repeated geometric subvolume calculation which becomes computationally costly. As described in \cite{Roenby.2016} the number of necessary geometric subvolume calculations can be limited by exploiting that $V_A$ is a piecewise cubic polynomial function of the isovalue, $f$, with polynomial coefficients being constant on the $f$-intervals between the vertex values. Thus, once we have done the expensive $V_A$ calculation a few times to find the correct $f$-subinterval, we only need two further evaluations (four in all on the subinterval) to obtain the polynomial coefficients of the function $V_A(f)$. Then we can use an inexpensive polynomial root finding algorithm to quickly find the isovalue, $f^*$, such that $|V_A(f^*)/V_i - \alpha_i| < \epsilon_{iso}$, where $\epsilon_{iso}$ is a user specified tolerance (with default value $10^{-8}$).

\subsection{Outline of the \texttt{iso-RDF} method}\label{iso-RDF}

As will be demonstrated in Section \ref{sec:NumTests}, the local interface orientation error produced by \verb|iso-Alpha| converges only slowly on structured meshes and not at all on unstructured meshes. We therefore seek to improve it by extending it with an iterative process. The new reconstruction method, \verb|iso-RDF|, uses the $\mathbf x_S$ and $\hat{\mathbf n}_S$ obtained with \verb|iso-Alpha| as the initial guess. The reconstructed distance function is then calculated and used to obtain improved estimates of $\mathbf x_S$ and $\hat{\mathbf n}_S$. This is repeated until the change in interface orientation between successive iterations goes below a specified tolerance. The method is described in Algorithm \ref{alg:isoRDF}.

{\SetAlgoNoLine%
	\begin{algorithm}[h!]
	\DontPrintSemicolon
	Find interfaceCells, and calculate initial $\mathbf x_S$ and $\hat{\mathbf n}_S$ with an algorithm which is identical to Algorithm \ref{alg:isoAlpha} except that least squares weights are used in Eqn.~\eqref{eq:interpolation} instead of inverse distance weights.\\
	Calculate the reconstructed distance function, $\Psi$, in the centres of all interfaceCells and in all their point neighbours based on $\mathbf x_S$ and $\hat{\mathbf n}_S$ (see Section \ref{ssec:reconDistFunc} for details).\\
	\For {$iter < maxIter$}
	{
		\For {$i \in$ interfaceCells}
		{
			For each of the $N_v$ vertices of interface cell $i$ interpolate the surrounding cell volume fractions to the vertex,
			\begin{equation}
		   		f_v = \frac{\sum_{k\in \mathcal C_v} w_k \Psi_k}{\sum_{k\in \mathcal C_v} w_k},
			\end{equation}
			where $\mathcal C_v$ is the set of cells to which the vertex belongs and the interpolation weights, $w_k$, are obtained using the least squares method.\label{isoRDFStp:interpol}\\
			Calculate the new face centre, $\mathbf x_{S,i}^{\textrm{new}}$, and face unit normal, $\hat{\mathbf n}_{S,i}^{\textrm{new}}$ of the $\Psi^*$-isosurface with volume fraction $A(\Psi^*)= \alpha_i$ (follow Step \ref{isoAlphaStp:chooseInitAv}-\ref{isoAlphaStp:calcxSandnS} of Algorithm \ref{alg:isoAlpha} with $f_v = \Psi_v$ as vertex values).\label{isoRDFStp:calcxSandnS}\\
		}
		Calculate the residual $res = \frac1N\sum_i^N |1-\hat{\mathbf n}_{S,i} \cdot \hat{\mathbf n}_{S,i}^{\textrm{new}}|$. (see Section \ref{ssec:residual}).\\
		Set $\mathbf x_S = \mathbf x_S^{\textrm{new}}$ and $\hat{\mathbf n}_S = \hat{\mathbf n}_S^{\textrm{new}}$.\\
		\uIf {{$res > tol$}}
		{
			Use new $\mathbf x_S$ and $\hat{\mathbf n}_S$ to recalculate the reconstructed distance function, $\Psi$, in all interfaceCells and in all their point neighbours (see Section \ref{ssec:reconDistFunc} for details).
		}
		\Else
		{
			\textbf{break}
		}
	}
	\caption{The iso-RDF interface reconstruction algorithm}
	\label{alg:isoRDF}
\end{algorithm}}

\subsubsection{Reconstructing the signed distance function}\label{ssec:reconDistFunc}

A good VOF based interface advection algorithm ensures that the jump from $\alpha = 0$ to $\alpha = 1$ happens over one cell everywhere along the interface. This sharpness of the volume fraction field comprises a significant challenge when derivatives are required e.g. for curvature calculations. As is well known from the Level Set literature, the signed distance function is much better suited for this kind of calculations. This is the motivation for attempting to numerically reconstruct the signed distance function from the volume fraction data. 

\begin{figure}[!tbp]
	\centering
	\begin{minipage}{0.45\linewidth}
		\includegraphics[width=\textwidth]{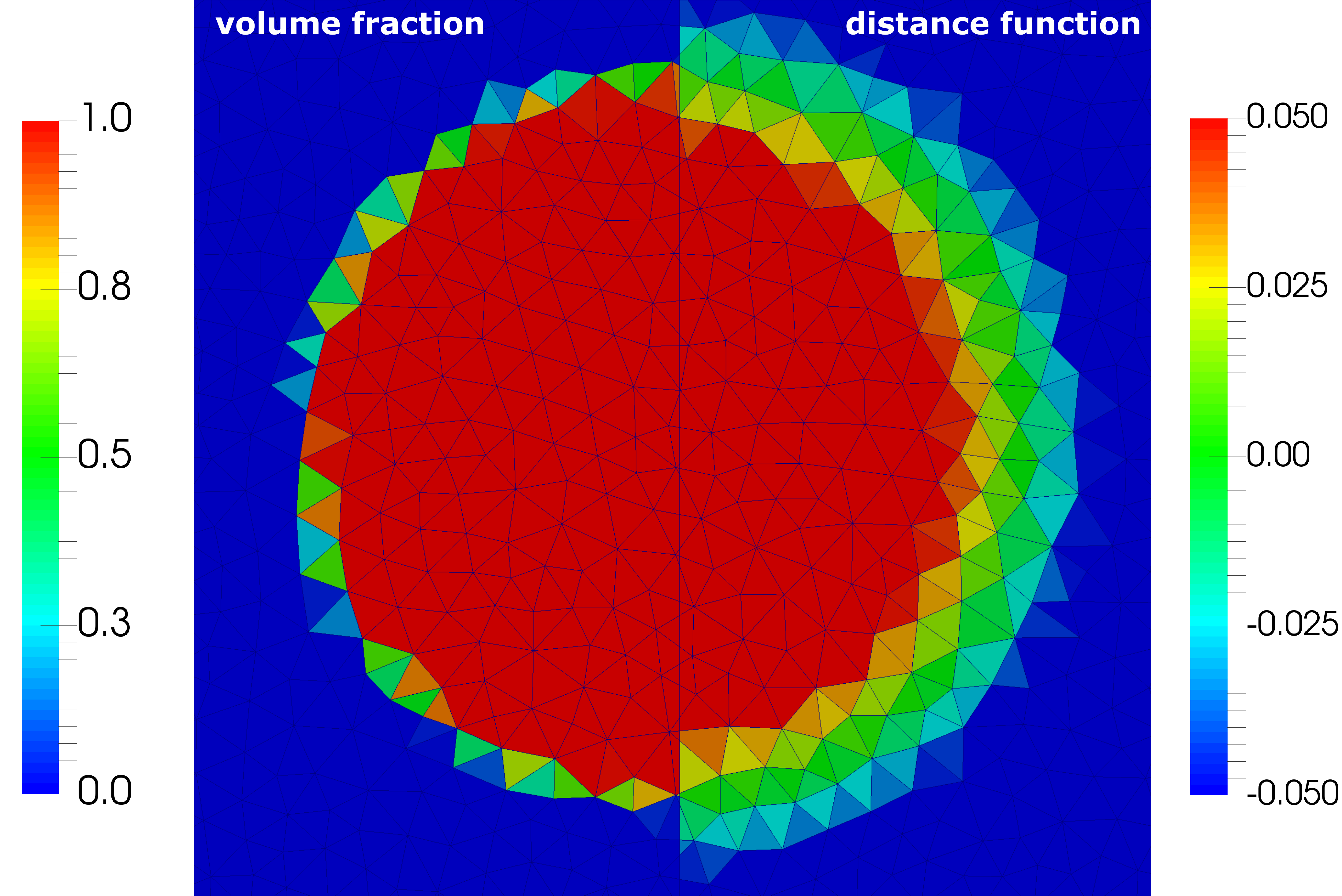}
		\caption{Representation of a circle with the volume fraction field and the distance function field.}
		\label{fig:CompareAlphaHeight}
	\end{minipage}
	\hfill
	\begin{minipage}{0.45\linewidth}
		\includegraphics[width=\textwidth]{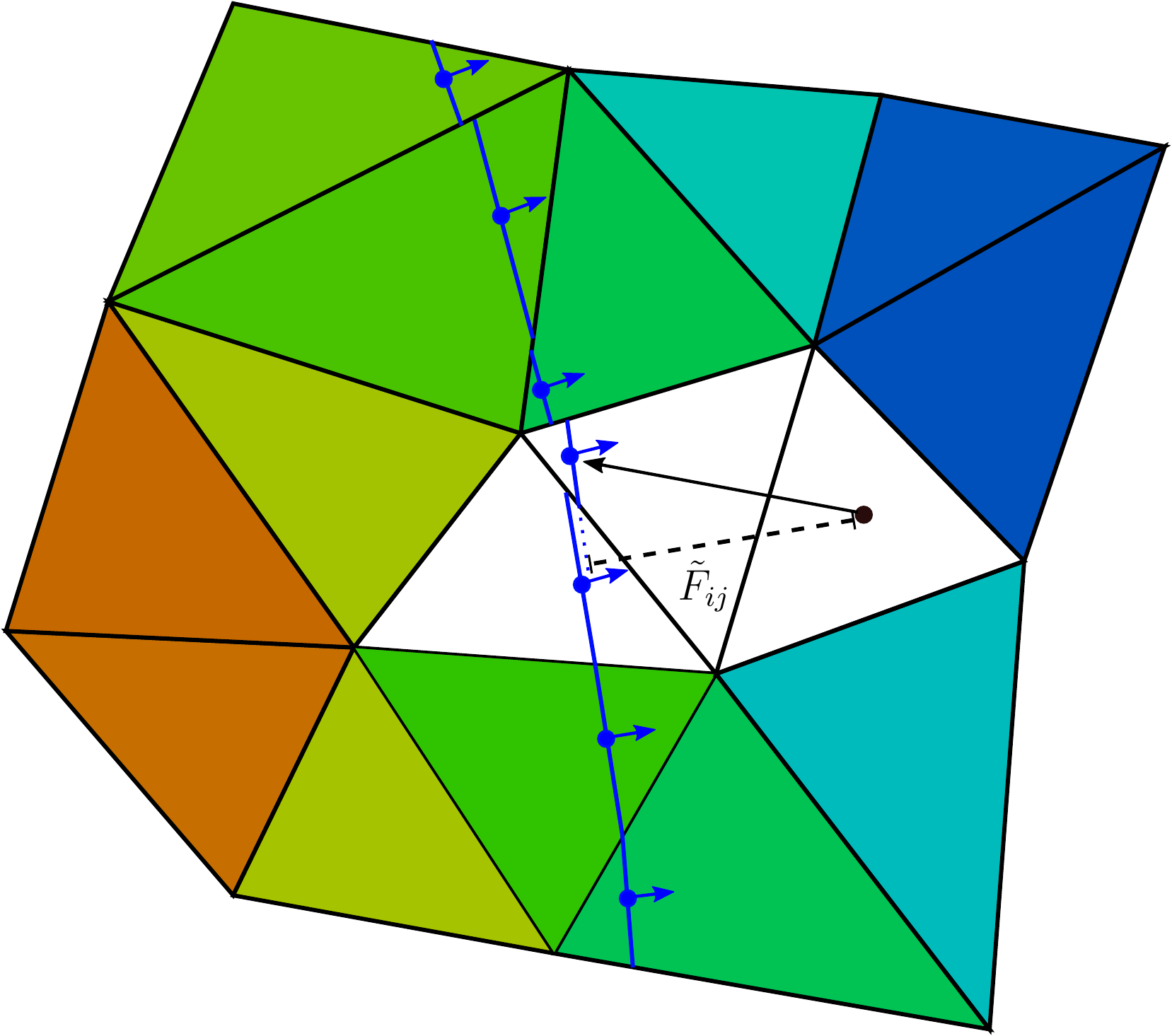}
		\caption{Reconstruction of the signed distance function.}
		\label{fig:SignedDistConstruct}
	\end{minipage}
\end{figure}

Fig.~\ref{fig:CompareAlphaHeight} illustrates the difference in smoothness of the raw VOF data and an RDF for a circle on a triangular prism mesh. An RDF is similar to a signed distance function, but where the length of the gradient of the signed distance function is equal to 1, this is not necessary exactly fulfilled for an RDF.

When constructing the RDF in a cell we calculate the shortest distance from the cell centre to the plane. However, a cell may in general have several neighbour cells holding an interface as illustrated in Fig.~\ref{fig:SignedDistConstruct}. The question is then which interface to use to calculate the RDF in the cell centre? In the current implementation the RDF in a cell is obtained as a weighted average of the distances to the interfaces in the cell itself and its point neighbours. So for cell $i$, the distance to the interface in cell $j$ is calculated as
\begin{equation}
	\tilde \Psi_{ij} =  \hat{\mathbf n}_{S,j} \cdot (\mathbf x_i -  \mathbf x_{S,j}),
\end{equation}
where $\mathbf x_{S,j}$ and $\hat{\mathbf n}_{S,j}$ denote interface centre and unit normal in cell $j$, and $\mathbf x_i$ is the cell centre of cell $i$. From these distances, the RDF in the centre of cell $i$ is calculated as
\begin{equation}\label{eq:cellCentrePsi}
	\Psi_i =  \frac{\sum_j w_{ij} \tilde \Psi_{ij}}{\sum_j w_{ij}},
r\end{equation}
where the sum is over all point neighbours of cell $i$ that are interface cells, and the weighting factor is chosen to be
\begin{equation}
	w_{ij} =  \frac{|\hat{\mathbf n}_{S,j} \cdot (\mathbf x_i -  \mathbf x_{S,j})|^A}{|\mathbf x_i -  \mathbf x_{S,j}|^A},\quad \textrm{with} \quad A = 2.
\end{equation}
This weighting function is similar to the one in Cummins et al. \cite{Cummins.2005}, but they found that their RDF approach worked best with a higher exponent, $A=25$. In our proposed iterative algorithm, we have found the exponent to have no influence on the accuracy for well--resolved interfaces. For under--resolved regions we have found that a higher exponent value can cause the method to diverge, and so in this work we use $A = 2$.

To obtain the RDF value in the vertices of cell $i$, we first use Eqn.~\eqref{eq:cellCentrePsi} to calculate $\Psi$ in cell $i$ and in all its point neighbours (i.e. cells with which it shares a vertex). Then, for each vertex of cell $i$, $\Psi$ is obtained using the least squares method with all the surrounding cell values.

\subsection{Outline of the \texttt{plic-RDF} method}\label{plic-RDF}

Our final reconstruction variant is also iterative. Here, we replace the isosurface calculations with the more traditional gradient calculations in the estimations of the interface normal. Furthermore, in all but the first time step in an interface advection simulation, there is an option of initialising $\hat{\mathbf n}_S$ by interpolating it from the previous time step. As will be seen, this leads to an improved initial $\hat{\mathbf n}_S$ guess and therefore reduces the number of iterations needed to reach convergence. Also a more advanced residual calculation is used which takes into account an estimate of the local radius of curvature relative to cell size. In regions where the interface is highly under--resolved, there is no point in spending a lot of iterations on getting the interface normal to converge to a fine tolerance. Therefore, interface cells where the average normal angle to the surrounding interface cell normals ($\beta_i$ defined in Eqn.~\eqref{eq:beta} below) is greater than 30 degrees are not reiterated and are excluded from the residual calculations. An algorithmic overview of the method is given in Algorithm ~\ref{alg:plicRDF}. Details are given in the following subsections.

{\SetAlgoNoLine%
	\begin{algorithm}[h!]
	\DontPrintSemicolon
	Boolean user input prevTimeNormal: Should $\hat{\mathbf n}_S$ be initialised by interpolation from previous time step?\\
	Calculate interfaceCells: A list of all the interface cell indices, i.e. $i\in$ interfaceCells if $\epsilon < \alpha_i < 1-\epsilon$.\\
	\For {$i \in$ interfaceCells}
	{
		\uIf(\tcp*[f]{See Eqn.~\eqref{eq:beta} for $\beta_i$}){prevTimeNormal and $\beta_i$ less than 10 degrees}
		{
			Interpolate initial $\hat{\mathbf n}_{S,i}$ guess from previous time step (see details in Section \ref{ssec:initnS}).
		}
		\Else
		{
			Use initial guess $\hat{\mathbf n}_{S,i} = \nabla \alpha_i/|\nabla\alpha_i|$.
		}
	}
	\For {($iter < maxIter$)}
	{
		\For {$i\in$ interfaceCells}
		{
			\If(\tcp*[f]{See Eqn.~\eqref{eq:beta} for $\beta_i$}){$\beta_i$ less than 30 degrees}	{
				Calculate the vertex position, $d_v = \mathbf x_v \cdot \hat{\mathbf n}_{S,i}$, along $\hat{\mathbf n}_{S,i}$ for each vertex, $\mathbf x_v$, of cell $i$.\\
				Calculate the face centre, $\mathbf x_{S,i}$ of the (planar) $d^*$-isosurface with volume fraction $A(d^*) = \alpha_i$ (follow Step \ref{isoAlphaStp:chooseInitAv}-\ref{isoAlphaStp:calcxSandnS} of Algorithm \ref{alg:isoAlpha} with $f_v = d_v$ as vertex values).	
			}
		}
		Calculate the reconstructed distance function, $\Psi$, in the centres of all interfaceCells and in all their point neighbours based on $\mathbf x_S$ and $\hat{\mathbf n}_S$ (see Section \ref{ssec:reconDistFunc} for details).\\
		Calculate the gradient of the reconstructed distance function, $\nabla \Psi$, in interfaceCells using the least-squares method (and the $\Psi$-values in the point neighbour stencils.)\\
		Calculate the new estimated interface normal, $\hat{\mathbf n}_{S}^{\textrm{new}} = \nabla \Psi/|\nabla \Psi|$.\\
		Calculate the residual, $res$, and curvature related residual, $res_{\textrm{curv}}$ from the difference between $\hat{\mathbf n}_S$ and $\hat{\mathbf n}_S^{\textrm{new}}$ (see Section \ref{ssec:residual} for details).\\
		\uIf(\tcp*[f]{See Section~\ref{ssec:residual} for $tol_{\textrm{curv}}$}){($res < tol$ or $res_{\textrm{curv}} < tol_{\textrm{curv}}$)}
		{
			\textbf{break}
		}
		\Else
		{
			Set $\hat{\mathbf n}_S = \hat{\mathbf n}_S^{\textrm{new}}$.
		}
	}
	\caption{The PLIC-RDF interface reconstruction algorithm}
	\label{alg:plicRDF}
\end{algorithm}}

\subsubsection{Initial estimate of $\hat{\mathbf n}_S$ in \texttt{plic-RDF}}\label{ssec:initnS}
In the first step of \verb|plic-RDF| the initial guess of the interface orientation inside the interface cells must be specified. If we use the gradient of the volume fraction data the method is similar to the approach by Cummins et al. \cite{Cummins.2005} and Sun et al. \cite{Sun.2010}. However, utilizing the gradient from the volume fraction data is known to provide an erroneous initial guess on fine meshes. Hence, the gradient calculation will only be applied in the first time step of interface advection simulations. For subsequent time steps we have found that an orientation estimate based on the previous time step is more accurate and thus requires fewer iterations. In the current implementation the first guess of $\hat{\mathbf n}_{S,i}$ is obtained from the previous time step normals, $\hat{\mathbf n}_S^{\textrm{old}}$, as
\begin{equation}
 \hat{\mathbf n}_{S,i}  =  \frac{\sum_j w_{ij} \hat{\mathbf n}_{S,j}^{\textrm{old}}}{\sum_j w_{ij}},
\end{equation}
where the sum is over all point neighbours of cell $i$ that are interface cells and the weighting factor is chosen to be
\begin{equation}
w_{ij} =  \left|\mathbf n_{S,j}^{\textrm{old}}\times \left[(\mathbf x_i - \mathbf u_i \Delta t)   -  \mathbf x_{S,j}\right]\right|,
\end{equation}
where $\mathbf u_i$ is the velocity in cell $i$. Thus, the weight is proportional to the area, $\mathbf n_{S,j}^{\textrm{old}}$, and to the distance between $\mathbf x_{S,j}$ and the backwards advected cell $i$ centre along $\hat{\mathbf n}_{S,j}^{\textrm{old}}$. If the interface is highly under--resolved then such interpolation becomes too noisy. Hence, if the average angular difference in interface normal in the surrounding interface cells is more than 10 degrees then we fall back to using the gradient of volume fraction as initial guess. This choice is justified by the fact that volume fraction gradients typically give reasonable results on coarse meshes.

\subsubsection{Estimation of the residuals}\label{ssec:residual}

In the following, the two residuals used in \verb|plic-RDF| are presented.  The first residual is the average difference of the normal vector between successive iterations:
\begin{equation}
 res = \frac1N \sum_i^N | 1 - \hat{\mathbf n}_{S,i}  \cdot \hat{\mathbf n}_{S,i}^{\textrm{new}}|.
\label{eq:absRes}
\end{equation}
The default tolerance for this residual is $tol = 10^{-6}$. We have found that it sometimes requires an excessive number of iterations to reach this tolerance in cases where the interface is poorly resolved. In under-resolved regions an increased number of iterations cannot make up for the lack of information about the interface orientation in the coarse volume fraction data. Therefore, to prevent wasting computational resources, our second residual is based on a local ``under-resolvedness'' parameter,
\begin{equation}\label{eq:beta}
	\beta_i =  \frac{\sum_j \arccos ( \hat{\mathbf n}_{S,i} \cdot \hat{\mathbf n}_{S,j}) |\mathbf n_{S,j}|}{\sum_j |\mathbf n_{S,j}|}.
\end{equation}
Here the sum is over all cell $i$'s point-neighbour cells that are interface cells. The $j$'th term in Eqn.~\eqref{eq:beta} is the angle between the interface normals in cell $i$ and $j$ weighted by the area of the interface in cell $j$. Thus, $\beta_i$ is a measure of the angle between interface normals in neighbouring cells and will be large if the local interface curvature is under-resolved.

For a second order scheme, the discretisation error associated with an under-resolved radius of curvature will be proportional to $\beta^2$. We therefore define a modelled discretisation error in each cell,
\begin{equation}
	E^{\textrm{curv}}_i = C \beta_i^2,
\end{equation}
where $C$ is a properly chosen constant. Our second residual will then be normalized with $E_i^{\textrm{curv}}$
\begin{equation}
res_{\textrm{curv}} = \frac1N \sum_i^N \frac{| 1 - \hat{\mathbf n}_{S,i}  \cdot \hat{\mathbf n}_{S,i}^{\textrm{new}} |}{\max(E^{\textrm{curv}}_{i},tol)},
\label{eq:normRes}
\end{equation}
where $E_i^{\textrm{curv}}$ is replaced by $tol$ in the denominator whenever $E_i^{\textrm{curv}} < tol$ in order to avoid large contributions to the residual from regions where the interface is almost planar. Empirically, we have found that a proper value for the constant is $C = 0.01$. The default tolerance used in the stopping criterion in Algorithm {\ref{alg:plicRDF}} is chosen to be $tol_{\textrm{curv}} = 0.1$.

\section{Numerical tests}\label{sec:NumTests}

Our code implementation is based in OpenFOAM--v1706 and is compiled with g++ (gcc) version 4.8.4  with the default OpenFOAM optimization flag: -O3. The performance data are gathered on a dual Intel Xeon 2687W v2 with 64 GB of DDR3-1600 MHz. All reconstruction and two dimensional advection test are performed on one core whereas the three dimensional advection test cases are performed on 4 cores.

The triangle prism meshes are generated with the open source mesh generator gmsh v3.06. The polygonal prism meshes are generated as the polygonal dual meshes of the triangle prism meshes. This is done using OpenFOAM's  \texttt{polyDualMesh} tool. All two-dimensional meshes have a thickness of 1. The tetrahedral meshes are also generated with gmsh v3.06 with the flag, optimize\_netgen. The polyhedral meshes are generated with cfMesh. The number of cells in the Tables {\ref{tab:rotSphere}} to {\ref{tab:deformation} for the unstructured meshes is similar to the number of cells in the Cartesian meshes used in Jofre et al. {\cite{Jofre.2014}}.

\subsection{Numerical results for interface reconstruction}\label{Sec:Results}

The accuracy of the three proposed methods is now investigated by comparison to simple analytical solutions. First, we reconstruct a circle on 2D meshes and a sphere on 3D meshes of varying type and resolution, and compare with results from literature. 

For our convergence tests we use the $L_{\infty}$ error norm which picks out the maximum error in the domain and therefore is the most difficult norm to get to converge. For the interface position, $\mathbf x_S$, and normal, $\hat{\mathbf n}_S$, inside a cell we define the two norms,
\begin{equation}\label{eq:ErrRecon}
	L_{\infty}^x= \max_i(|\mathbf x_{S,i} - \mathbf x_{S,i}^{\textrm{exact}}|) \quad \textrm{and} \quad 
	L_{\infty}^n= \max_i(1 - \hat{\mathbf n}_{S,i} \cdot \hat{\mathbf n}_{S,i}^{\textrm{exact}}),
\end{equation}
where the maximum is taken over all interface cells, $\mathbf x_{S,i}^{\textrm{exact}}$ is the exact interface position, and $\hat{\mathbf n}_{S,i}^{\textrm{exact}}$ is the exact interface unit normal. 

\subsubsection{Reconstruction of a circle on 2D meshes}

\begin{figure}[tb!]
		\includegraphics[width=\textwidth]{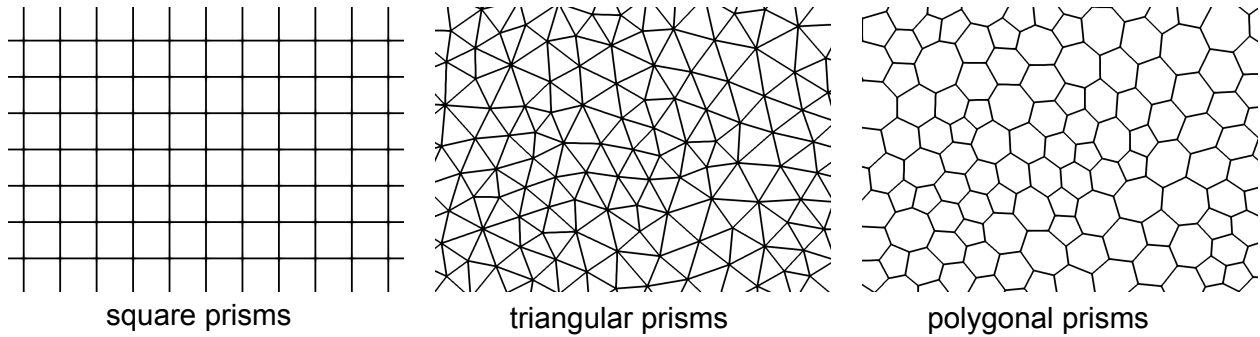}
		\caption{Zoomed view of examples of 2D meshes used for reconstruction test.}
		\label{fig:Meshes2D}
\end{figure}

In the 2D reconstruction test the VOF field is calculated and the surface is then reconstructed for a circular disc of radius 0.5 in a square domain $[0, 2]\times [0, 2]$. The $L_\infty^x$ and $L_\infty^n$ error norms are calculated, and the exercise is repeated on increasingly fine meshes. All this is done for three mesh types consisting of square, triangular, and polygonal prisms, respectively, with the two latter mesh types being unstructured.  Examples of the three types of meshes are shown in Fig.~\ref{fig:Meshes2D}. To avoid any bias in the error estimates due to the choice of mesh-interface combination, we repeat each reconstruction 100 times on the same mesh with randomly chosen centre position for the circular interface. The $L_\infty^x$ and $L_\infty^n$ errors shown below are then the largest values from these 100 repetitions. 

\begin{figure}[tb!]
	\subcaptionbox{$L_{\infty}^n$ for triangular prisms
	\label{fig:Recon-Circle-tri-iter}}
	[.5\linewidth]{\includegraphics[width=0.5\textwidth]{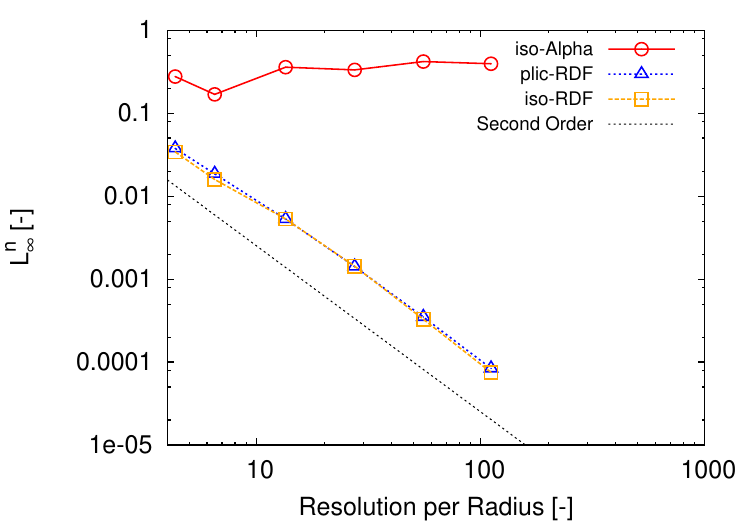}}
	\subcaptionbox{$L_{\infty}^x$ for triangular prisms\label{fig:Recon-Circle-tri-pos}} 
	[.5\linewidth]{\includegraphics[width=0.5\textwidth]{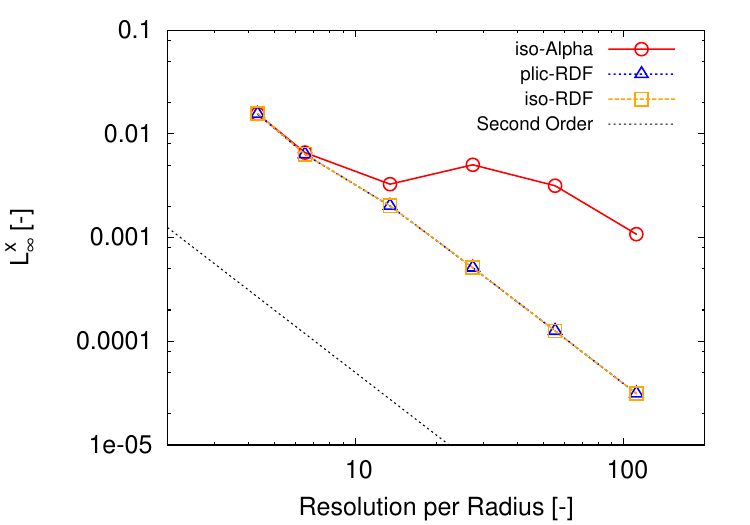}}
	\subcaptionbox{$L_{\infty}^n$ for square prisms\label{fig:Recon-Circle-hex}}
	[.5\linewidth]{\includegraphics[width=0.5\textwidth]{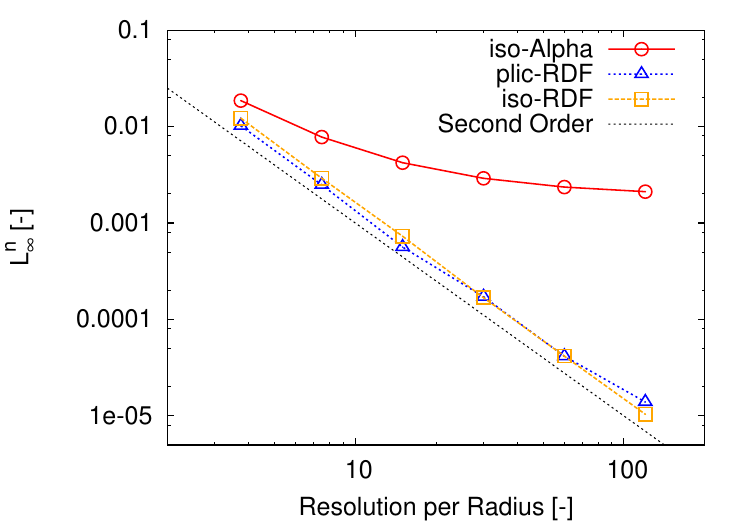}}
	\subcaptionbox{$L_{\infty}^x$ for square prisms\label{fig:Recon-Circle-hex-pos}} 
	[.5\linewidth]{\includegraphics[width=0.5\textwidth]{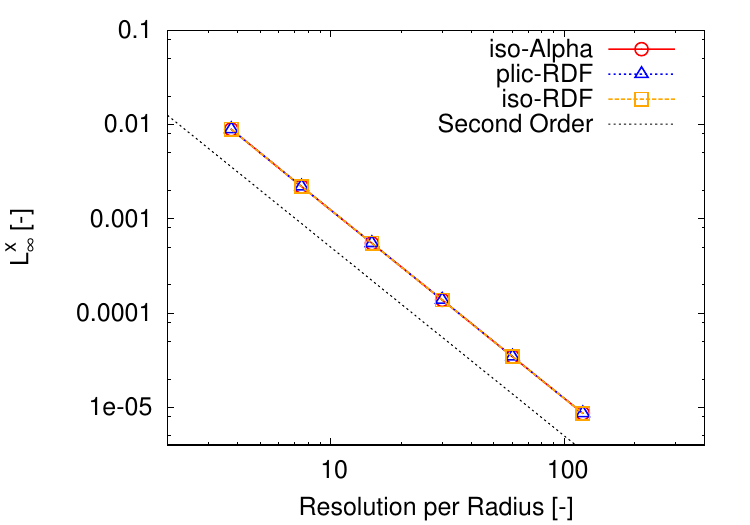}}
	\subcaptionbox{$L_{\infty}^n$ for polygonal prisms\label{fig:Recon-Circle-poly}} 
	[.5\linewidth]{\includegraphics[width=0.5\textwidth]{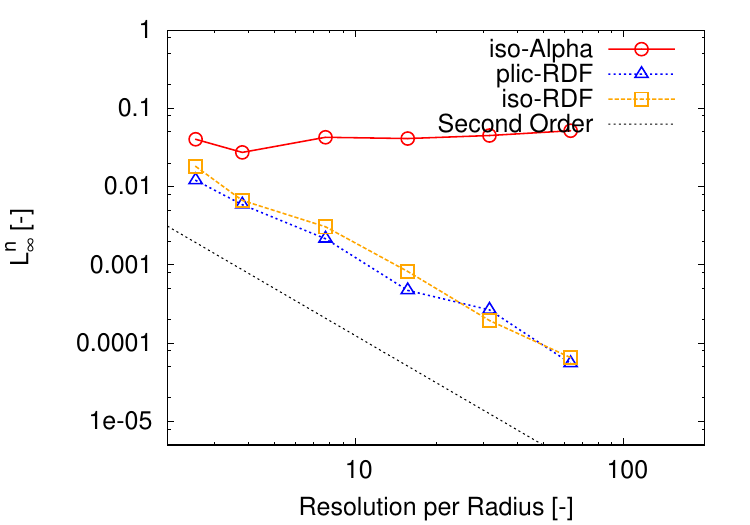}}
	\subcaptionbox{$L_{\infty}^x$ for polygonal prisms\label{fig:Recon-Circle-poly-pos}} 
	[.5\linewidth]{\includegraphics[width=0.5\textwidth]{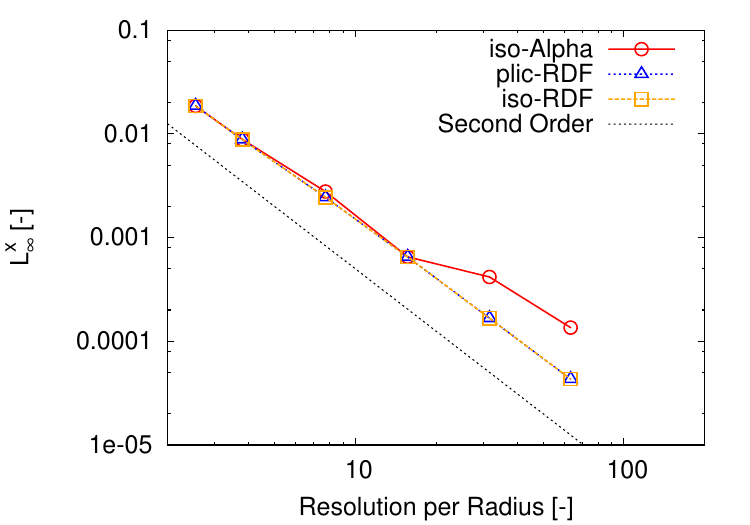}}
	\caption{$L_\infty$ for circle reconstructed on various mesh types.}
	\label{fig:Recon2D} 
\end{figure}
	
Results are shown in Fig.~\ref{fig:Recon2D} with the $L_\infty^n$ in the left panels (a, c and e) and $L_\infty^x$ in the right panels (b, d and f). The first panel, Fig.~\ref{fig:Recon-Circle-tri-iter}, shows the orientation error on the unstructured triangular prism meshes for all three methods. We observe that \verb|iso-Alpha| does not converge in this norm, whereas both \verb|iso-RDF| and \verb|plic-RDF| show second order convergence rate on the tested resolution range. In Fig.~\ref{fig:Recon-Circle-tri-pos} for $L_\infty^x$ we see poor convergence of \verb|iso-Alpha|, and second order convergence for \verb|iso-RDF| and \verb|plic-RDF| on the triangular prism meshes.

For the orientation error on square prism meshes in Fig.~\ref{fig:Recon-Circle-hex}, \verb|iso-Alpha| performs slightly better than on the triangular prism meshes, now converging albeit at a slow rate that is decreasing with mesh refinement. \verb|plic-RDF| and \verb|iso-RDF| both exhibit second order convergence on the full resolution range. As for the position error in Fig.~\ref{fig:Recon-Circle-hex-pos} all methods are second order on the full resolution range for square prism meshes.

The errors on unstructured polygonal prism meshes are shown in Figs.~\ref{fig:Recon-Circle-poly} and \ref{fig:Recon-Circle-poly-pos}. As for the triangular prism meshes, the $L_\infty^n$ error does not converge with \verb|iso-Alpha|, and is second order convergent with \verb|iso-RDF| and \verb|plic-RDF|. The $L_\infty^x$ error behaves better for \verb|iso-Alpha| with near second order convergence. Again, both \verb|iso-RDF| and \verb|plic-RDF| exhibit second order convergence in $L_\infty^x$ on all tested polygonal mesh resolutions. 

In summary, the iterative RDF based methods show second order convergence of both interface orientation and position on all mesh types in 2D, whereas the interface orientations obtained with \verb|iso-Alpha| does not converge on the unstructured meshes and has only poor convergence on the structured meshes.

\begin{figure}[htb!]
	\subcaptionbox{$L_\infty^n$ for tetrahedral cells\label{fig:Recon3D-Circle-tri}} 
	[.5\linewidth]{\includegraphics[width=0.5\textwidth]{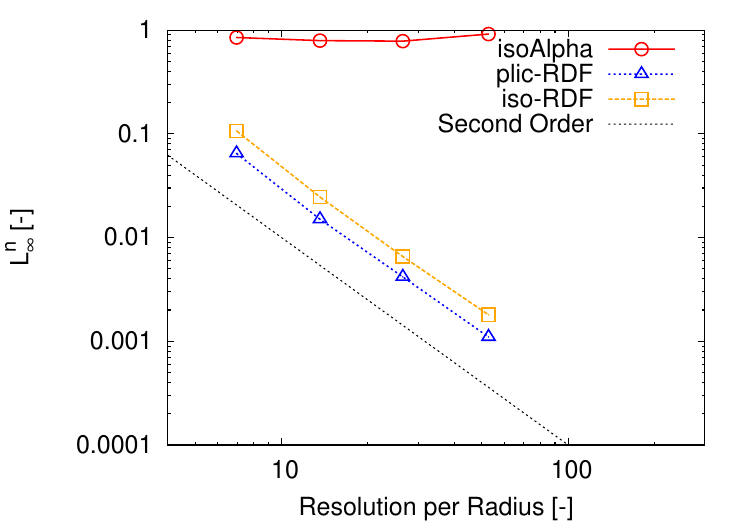}}
	\subcaptionbox{$L_\infty^x$ for tetrahedral cells\label{fig:Recon3D-Circle-tri-pos}} 
	[.5\linewidth]{\includegraphics[width=0.5\textwidth]{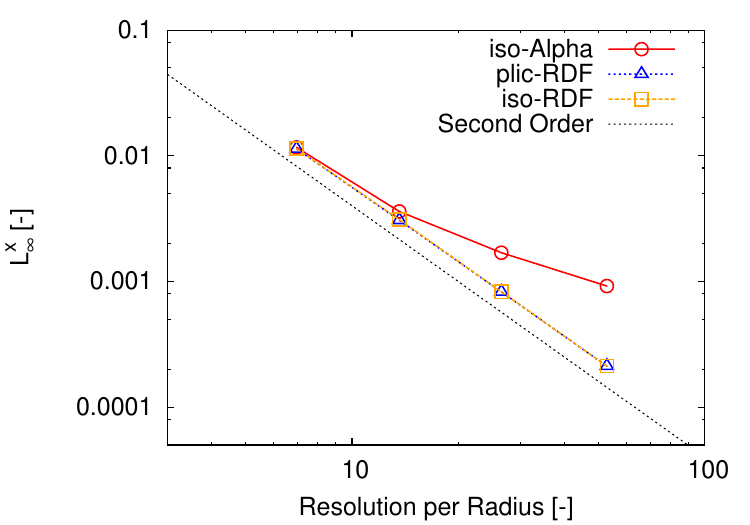}}
	\subcaptionbox{$L_\infty^n$ for hexahedral cells\label{fig:Recon3D-Circle-hex}} 
	[.5\linewidth]{\includegraphics[width=0.5\textwidth]{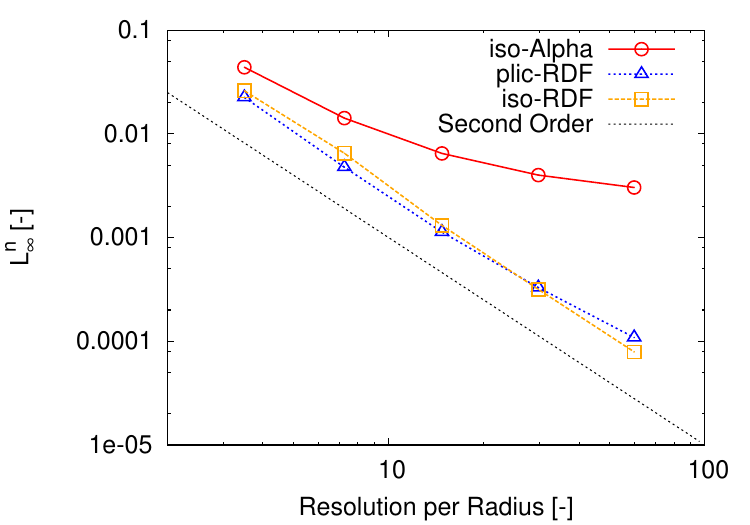}} 
	\subcaptionbox{$L_\infty^x$ for hexahedral cells\label{fig:Recon3D-Circle-hex-pos}} 
	[.5\linewidth]{\includegraphics[width=0.5\textwidth]{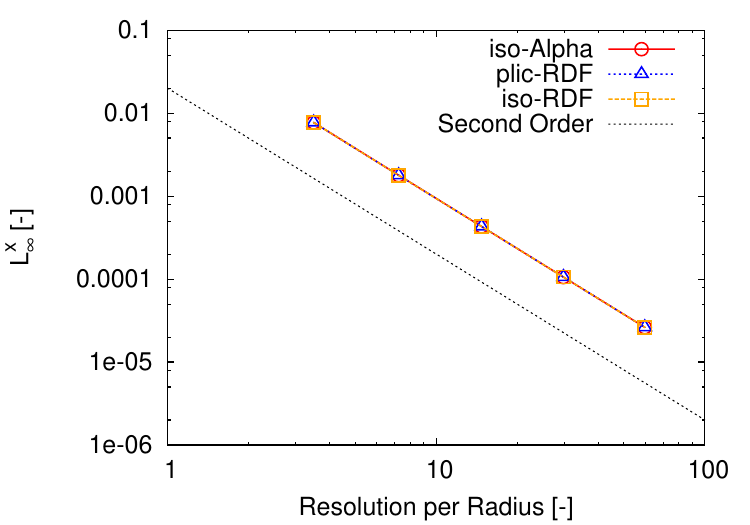}}
    \subcaptionbox{$L_\infty^n$ for polyhedral cells \label{fig:Recon3D-Circle-poly}} 
	[.5\linewidth]{\includegraphics[width=0.5\textwidth]{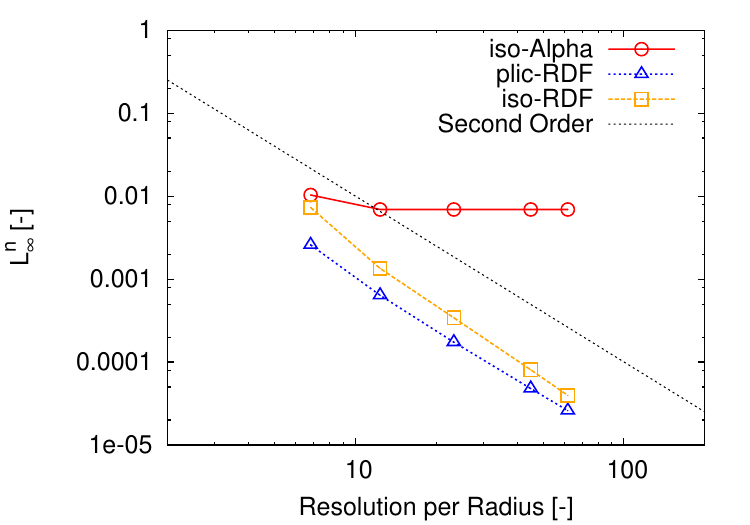}}
	\subcaptionbox{$L_\infty^x$ for polyhedral cells\label{fig:Recon3D-Circle-poly-pos}} 
	[.5\linewidth]{\includegraphics[width=0.5\textwidth]{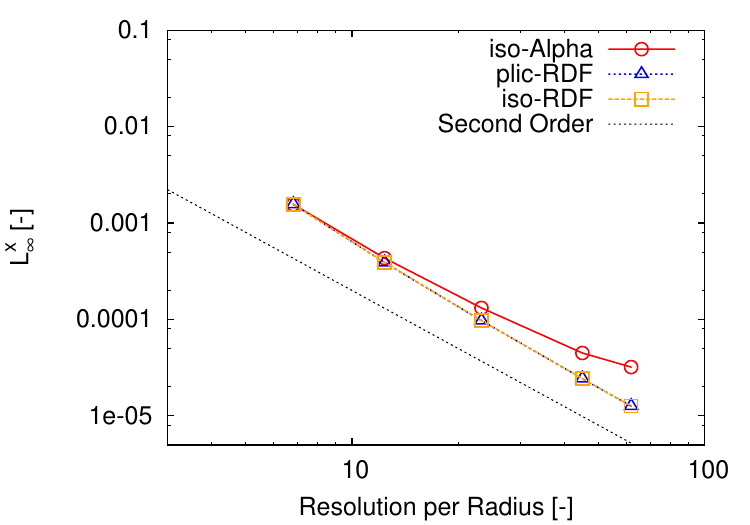}}
	\caption{$L_\infty$ for reconstructed sphere on various mesh types.}\label{fig:Recontriandpoly} 
\end{figure}

\subsubsection{Reconstruction of a sphere on 3D meshes}

To test the accuracy of \verb|iso-Alpha|, \verb|plic-RDF| and \verb|iso-RDF| in 3D a sphere of radius 0.25 is reconstructed in a unit cube domain tessellated by either cubic hexahedra, random tetrahedra or polyhedra. Again the interface is reconstructed on a range of mesh resolutions and for each mesh the reconstruction is repeated 100 times with random centre position of the spherical interface.  The largest error norms from these 100 runs are shown in Fig.~\ref{fig:Recontriandpoly} for unstructured tetrahedral meshes in the two top panels, structured hexahedral meshes in the two middle panels, and unstructured polyhedral meshes in the two bottom panels. The performance of all three schemes is similar to the 2D reconstruction case: The orientation error, $L_\infty^n$, of \verb|iso-Alpha| only converges on the square prism mesh and only with a rate that is falling off for finer meshes. The $L_\infty^x$ norm of \verb|iso-Alpha| reduces with second order on all coarse meshes but convergence rate falls off for all but the hexahedral mesh when resolution is increased. Both \verb|iso-RDF| and \verb|plic-RDF| exhibit second order convergence in both interface orientation and position for all three mesh types on the full resolution range.
	
\begin{figure}[htb!]
	\subcaptionbox{Circle on triangular prisms: Data from Ito et al. \cite{Ito.2014}\label{fig:Recon-Comp-tri}} 
	[.5\linewidth]{\includegraphics[width=0.5\textwidth]{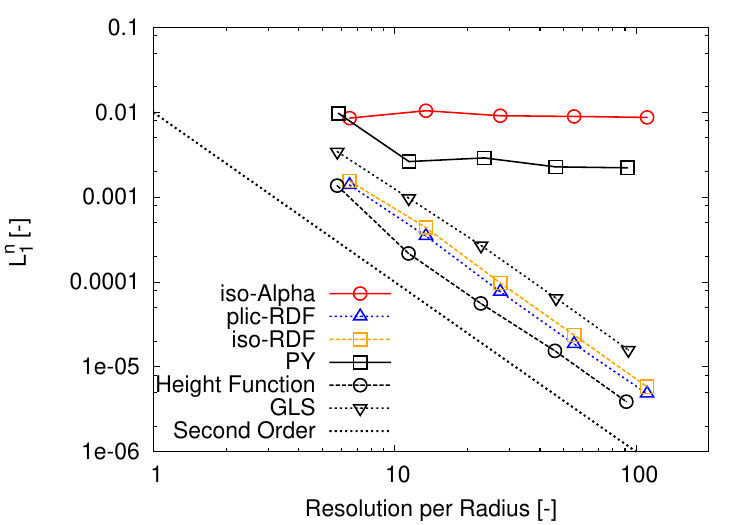}}
	\subcaptionbox{Sphere on hexahedral cells: Data from Ivey and Moin \cite{Ivey.2015} \label{fig:Recon3D-Comp-hex}} 
	[.5\linewidth]{\includegraphics[width=0.5\textwidth]{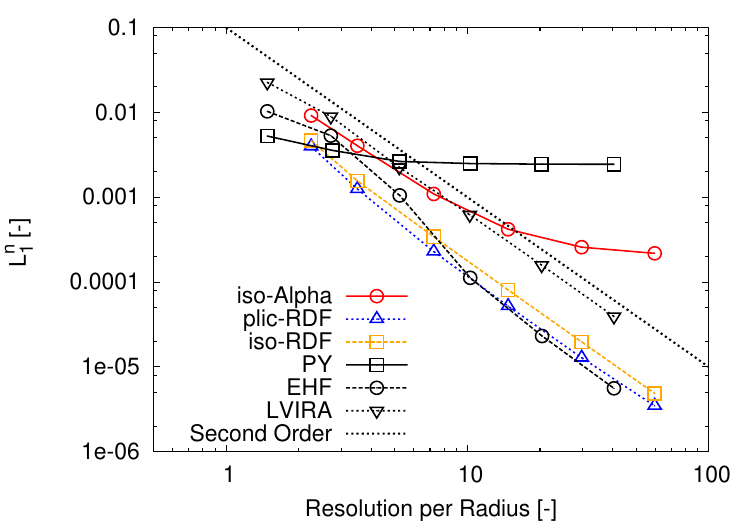}}
	\caption{$L_1^n$ reconstruction errors compared with literature.}	\label{fig:Recon3D-Comp} 
\end{figure}

\subsubsection{Comparison with other methods}

We now compare the accuracy of the proposed reconstruction methods with the data presented in Ito et al. \cite{Ito.2014} and Ivey and Moin \cite{Ivey.2015}.  In these papers the less strict 1-norm is used for the orientation error:
\begin{equation}\label{eq:ErrReconL1}
	L_1^n= \frac1N \sum_i | 1 - \hat{\mathbf n}_{S,i}  \cdot \hat{\mathbf n}_{S,i}^\textrm{exact}|,
\end{equation}
where $\hat{\mathbf n}_{S,i}$ and $\hat{\mathbf n}_{S,i}^\textrm{exact}$ are, respectively, the calculated and exact interface unit normal in interface cell $i$, and the sum is over all the $N$ interface cells. In Ivey and Moin \cite{Ivey.2015} the terms in Eqn.~\eqref{eq:ErrReconL1} are weighted by the cell volumes. For the cubic meshes used below this leads to the same definition.

In the 2D case a circle of radius 0.5 is reconstructed in the domain $[0, 2]\times [0, 2]$, which is covered by an unstructured mesh of triangular prisms. This is repeated 100 times with random location of the sphere centre. Fig.~\ref{fig:Recon-Comp-tri} depicts the largest of the 100 $L_1^n$ errors for various mesh resolutions. Also with this error norm \verb|iso-Alpha| shows lack of convergence in the orientation as does The Parker--Youngs method (PY). Both \verb|iso-RDF| and \verb|plic-RDF| achieve second order convergence as do the Height Function method and the geometric least square (GLS) method \cite{Shahbazi.2003}. The absolute errors of \verb|iso-RDF| and \verb|plic-RDF| are between those of GSL and the Height Function method. 

In the 3D case a sphere of radius 2 initialized in a cubic box of side length 8. This is repeated 100 times with random location of the sphere centre. Fig.~\ref{fig:Recon3D-Comp-hex} shows the largest of the 100 $L_1^n$ errors at various mesh resolutions. In contrast to PY, the \verb|iso-Alpha| method converges but at a slow rate. Both \verb|iso-RDF| and \verb|plic-RDF| achieve second order convergence as do the EHF method and LVIRA. It has to be noted that the EHF method is identical with the Height Function method for this particular case. The absolute errors of \verb|iso-RDF| and \verb|plic-RDF| are significantly lower than LVIRA's and similar to those of the Height Function method.

For a reconstruction method to be useful in practical engineering applications, it is vital that it is sufficiently efficient, preferably with a computational cost comprising only a minor fraction of the total simulation time. Comparing CPU times for different algorithms directly can give an indication of the relative cost, but will also depend on the hardware, compiler and compiler settings used in the particular implementation. Lopez et al. \cite{Lopez.2008} stated computational costs relative to the well-established Youngs algorithm \cite{Youngs.1982}. This normalizes out computer system specific variations. In Table~\ref{tab:perfRecon3D}, we show the results from Lopez et al. \cite{Lopez.2008} (for a hollow sphere case) together with the slowest and fastest sphere reconstruction times of our three methods normalized by our implementation of the Youngs algorithm. \verb|iso-Alpha| is seen to be the fastest method except on polyhedral meshes, where the Youngs algorithm is fastest. \verb|iso-RDF| is 6.7-11.4 times slower, and \verb|plic-RDF| is 2.4-7.3 times slower than the Youngs algorithm. The performance improvement from \verb|iso-RDF| to \verb|plic-RDF| was mainly achieved by introducing the normalized residual, $res_{curv}$, and the requirement of sufficient local interface resolution in order for a cell to be iterated.
Data for comparison is only available for hexahedral meshes, where \verb|iso-Alpha| is significantly faster and \verb|plic-RDF| is 25-50\% slower than the fastest method, CLCIR \cite{Lopez.2008}. Improved ELVIRA \cite{Lopez.2008} is roughly 800 times slower than the Youngs method. It is an efficiency improved version of the LVIRA method, which was used by Jofre et al. \cite{Jofre.2014} on unstructured meshes. Ivey and Moin \cite{Ivey.2015} compared the performance of their EHF method to LVIRA, and found that EHF was 20-100 times faster.

\begin{table}
\centering
	\caption{CPU times relative to the Youngs algorithm}
	\begin{tabular}{lrrrrrr}
		Mesh type & \verb|iso-Alpha| & \verb|iso-RDF| & \verb|plic-RDF| & Improved ELVIRA \cite{Lopez.2008} & CVNTA \cite{Liovic.2006} & CLCIR \cite{Lopez.2008}\\
		\hline
		\hline
		hexahedral & 0.4 - 1.0 & 6.7 - 7.5 & 3.0 - 5.1 & 797 - 899 & 43.7 - 65.7 & 2.6 - 4.0 \\
		tetrahedral & 0.17 - 0.64 & 8.9 - 11.4 & 5.4 - 7.3 & - & - & -\\
		polyhedral & 1.2 - 1.8 & 8.4 - 9.8 & 2.4 - 4.2 & - & - & - \\
		\label{tab:perfRecon3D} 
	\end{tabular}
\end{table}

\subsection{Numerical results for interface advection}\label{sec:advectionResults}

We now investigate the effect of replacing \verb|iso-Alpha| with \verb|plic-RDF| in the isoAdvector algorithm described in \cite{Roenby.2016}. As mentioned in Section \ref{ssec:initnS}, we have found that the initial residual in \verb|plic-RDF|, and thus the computational time, can be reduced by improving the initial guess for $\hat{\mathbf n}_S$. In the following, we therefore use \verb|plic-RDF| with interpolation of $\hat{\mathbf n}_S$ from the previous time step as initial guess instead of the $\nabla\alpha$ values. We define the shape error norm as
\begin{equation}\label{eq:E1AdvectError}
	E_{1}(t) = \sum_{i}V_{i}\left|\alpha_{i}(t)-\alpha_{i}^{exact}(t)\right|,
\end{equation}
where the sums are over all cells, and $\alpha_{i}^{exact}(t)$ is the volume fraction in cell $i$ corresponding to the known exact solution. The volume conservation $E_{Vol}$ is estimated with:

\begin{equation}\label{eq:Evol}
	E_{Vol}(t) = \left|\sum_{i}V_{i}\left[\alpha_{i}(t) - \alpha_{i}^{exact}(t)\right]\right|,
\end{equation}Another important metric is the boundedness of the solution, $E_{bound}(t)$ , measuring the degree of over-- and under--filling of cells \cite{Owkes.2014}:
\begin{equation}\label{eq:Ebound}
	E_{bound}(t) = \max(- \min(V_{i}\alpha_{i}), \max(V_{i}\alpha_{i}))
\end{equation}
The order of convergence is defined as \cite{Tryggvason.2011}:
\begin{equation}\label{eq:convRate}
\mathcal O(E_1) = \frac{\ln[E_1(\Delta x)/E_1(\Delta x/2)]}{\ln 2},
\end{equation}
Furthermore, we adopt the measures for reconstruction time, $T_r$, and execution time, $T_e$, used in \cite{Maric.2018}, where $T_r$ is the average time spent on the reconstruction step in a time step, and $T_e$ is the average total execution time (i.e. reconstruction + advection) per time step.

For time step control, one can either use a fixed time step throughout a simulation, or we can adjust the time step dynamically to ensure that the CFL number is kept at all times below a user speficied value. Further, since pure advection tests only require calculations near the interface, we can employ a CFL limiting time step criterion that only considers cells near the interface (the relevant OpenFOAM input parameter specifying the maximum allowed CFL number in the interface region is called \verb|maxAlphaCo| ). Here we will refer to such a limit as CFL$_{surf}$.

\subsubsection{Circle in constant flow}
\label{sec:circleConstFlow}

\begin{figure}[tb!]
	\subcaptionbox{reciprocal CFL numbers 1 to 250}
	[.5\linewidth]{\includegraphics[width=0.5\textwidth]{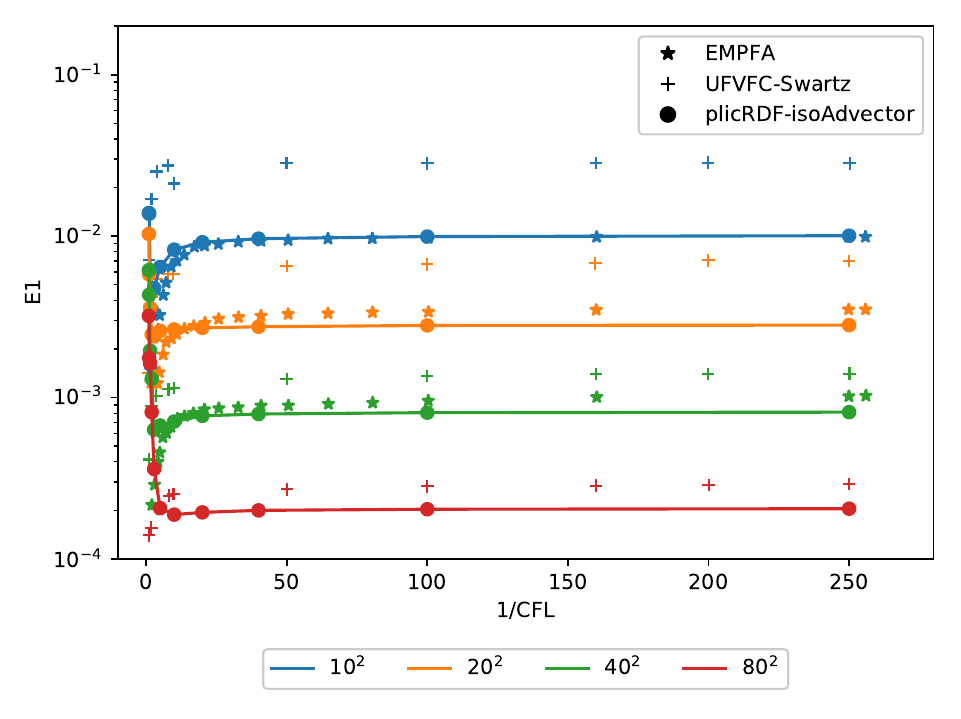}}
	\subcaptionbox{reciprocal CFL numbers 1 to 15}
	[.5\linewidth]{\includegraphics[width=0.5\textwidth]{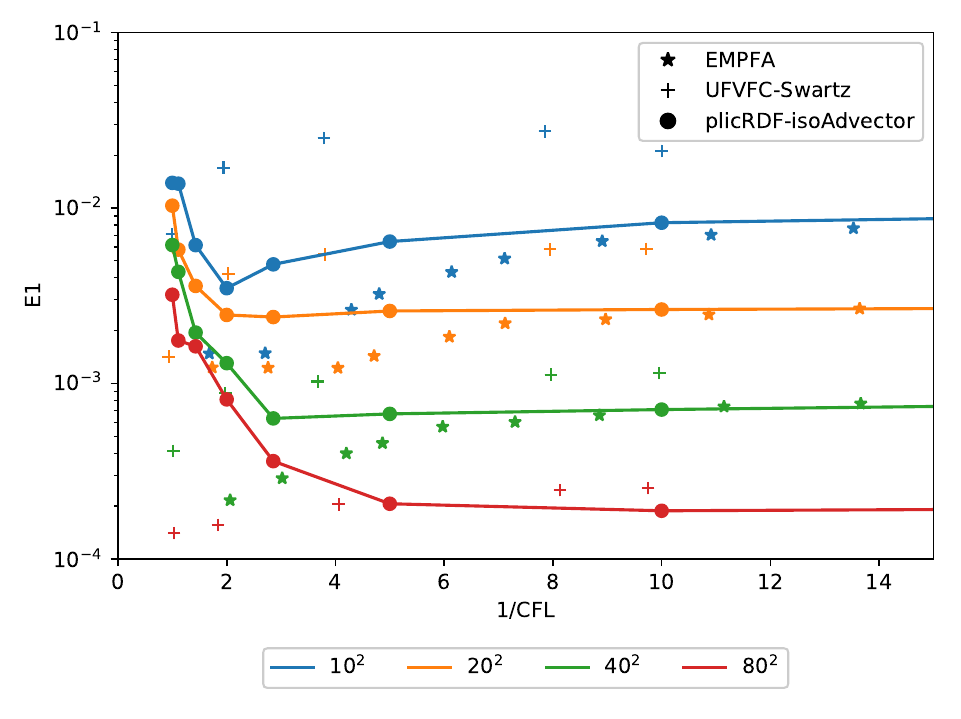}}
	\caption{$E_1$ error as a function of 1/CFL for circle advected in constant flow}
	\label{fig:E1overCFL} 
\end{figure}

\begin{figure}[ht]
	\subcaptionbox{Triangular prism mesh\label{fig:Recon-Circle-triIsoVsRDF}} 
	{\includegraphics[width=0.5\linewidth]{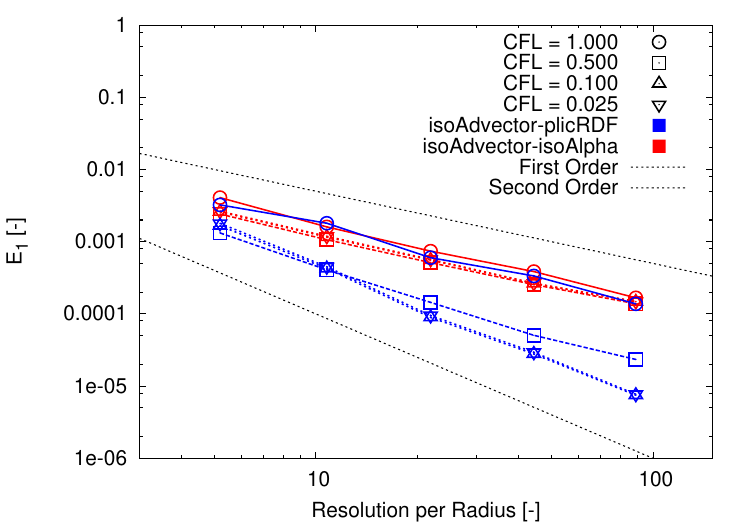}}
	\subcaptionbox{Zoom of interface at $T = 4 s$\label{fig:interfaceAdvectCricleTri3}} 
	{\includegraphics[width=0.5\linewidth]{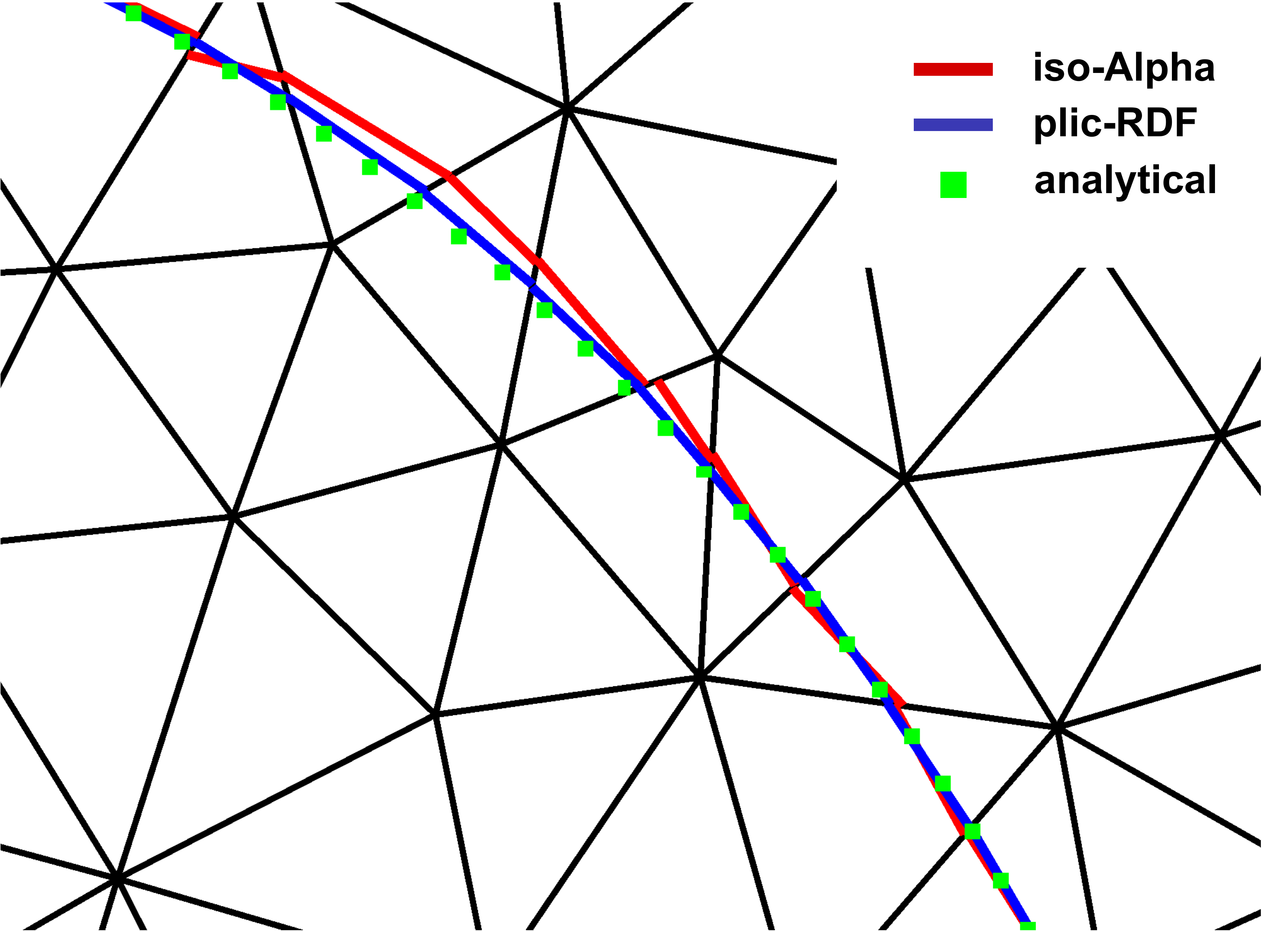}} 
	\subcaptionbox{Square prism mesh\label{fig:Recon-Circle-hexIsoVsRDF}} 
	{\includegraphics[width=0.5\linewidth]{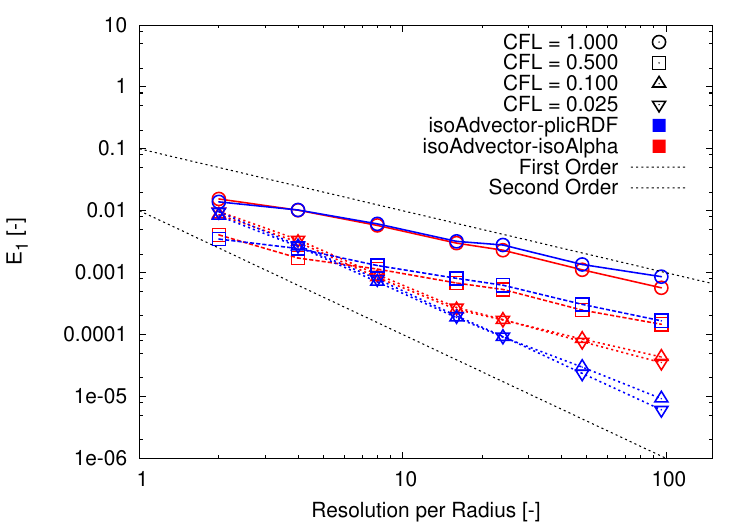}}
	\subcaptionbox{Polygonal prism mesh\label{fig:Recon-Circle-polyIsoVsRDF}} 
	{\includegraphics[width=0.5\linewidth]{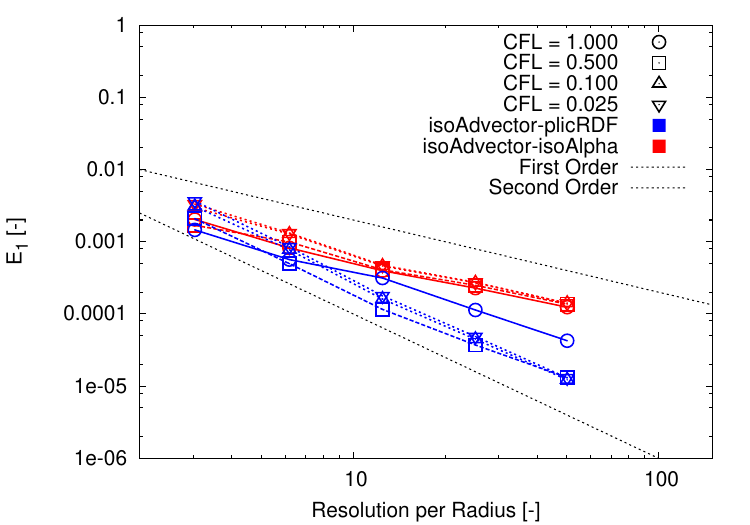}}
	\caption{$E_1$ error for circle advected in constant flow.}\label{fig:RecontriIterandHex} 
\end{figure}

Our first test case is a circular interface of radius 0.2 initially centred at $(0.25,0.25)$ and moving with a constant, uniform velocity field without changing its shape. The advection test is performed on triangle, square and polygonal prism meshes. The velocity field in the unit square domain is set to $(1.0 , 1.0)$. This simple test case was introduced by Rider and Kothe \cite{Rider.1998} and adopted by Harvie and Fletcher \cite{Harvie.2000} for structured meshes. Harvie and Fletcher \cite{Harvie.2000} showed that for fixed mesh resolution and simulation time, there is a range of time step sizes where the error will increase with decreasing time step size. This is due to low order errors the reconstruction step and the fact that the number of required reconstructions will increase inverse proportionally to the time steps size. At small enough time step the error will reach a constant (mesh resolution dependent) level. Figure \ref{fig:E1overCFL} shows the shape error for multiple CFL numbers on hexahedral meshes as obtained with EMPFA-SIR \cite{Lopez.2004}, UFVFC-Swartz \cite{Maric.2018}, and with isoAdvector using plicRDF for reconstruction. The two former algorithms both show second order convergence independently of the CFL number with EMPFA-SIR achieving the lowest absolute errors. IsoAdvector-plicRDF shows second order convergence for low CFL numbers with an absolute error lower than EMPFA-SIR \cite{Lopez.2004}. The convergence with IsoAdvector-plicRDF is, however, only first order for higher CFL numbers. A possible explanation could be that in the current isoAdvector implementation the isoAdvector advection treatment is only applied to downwind faces of interface cells. The interface may move into cells during a time step that share a vertex but no faces with an interface cell. In a more complete treatment such cells should also be treated using the isoAdvector advection step. This will be incorporated in future work and is expected to increase accuracy for larger time steps.

In Fig.~\ref{fig:RecontriIterandHex} we present the $E_{1}$ errors at the time $t = 0.5$ s as functions of the mesh resolution for the different mesh types and performed with four different CFL numbers. Fig.~\ref{fig:Recon-Circle-triIsoVsRDF} shows the results for the triangular prism meshes. It is observed that \verb|iso-Alpha| falls of with a convergence rate between 1 and 2 for the coarsest meshes but that the rate drops to first order at finer meshes. The \verb|iso-Alpha| errors are independent of the time step size but are significantly larger than the \verb|plic-RDF| errors on all resolutions. Both the \verb|plic-RDF| results with CFL = 0.1 and CFL = 0.025 exhibit second order convergence in the full range of resolution. With CLF = 0.5 and CFL=1.0 \verb|plic-RDF| falls off to first order at high resolutions but still has lower absolute error than \verb|iso-Alpha|. Fig.~\ref{fig:interfaceAdvectCricleTri3} shows a close--up of the interface at $t = 0.5$ s for the simulation with 10 cells per radius. It is evident that the interface segments of \verb|plic-RDF| (blue) are significantly improved  compared to \verb|iso-Alpha| (red).

Fig.~\ref{fig:Recon-Circle-hexIsoVsRDF} shows the $E_{1}$ errors as functions of mesh resolution for the square prism meshes. In contrast to the triangle prism results, \verb|iso-Alpha| improves with reduced time step size, especially for high resolutions. At lower resolutions \verb|iso-Alpha|  achieves second order convergence but at higher resolutions the convergence falls off, especially for large time steps. In contrast, \verb|plic-RDF| achieves second order convergence on the full resolution range with both CFL = 0.025 and 0.1. For CFL = 0.5 and CFL=1.0, the convergence rate drops to first order.

Fig.~\ref{fig:Recon-Circle-polyIsoVsRDF} shows the behaviour on polygonal prism meshes. The behaviour is similar to the triangular prism results with the \verb|iso-Alpha| giving larger absolute errors and dropping off to first order on finer meshes. Again, \verb|plic-RDF| retains second order convergence on the full resolution range with CFL = 0.025, 0.1 and 0.5. For CFL = 1.0 the convergence rate of \verb|plic-RDF| is roughly second order accurate but the absolute error is significantly higher.

In summary, \verb|plic-RDF| in combination with the isoAdvector advection scheme from \cite{Roenby.2016} consistently lowers the absolute errors and gives second order convergence rate for small enough time step size.

Fig.~\ref{fig:ResCFL01} depicts the time-averaged initial residual, $res$, from Eqn.~\eqref{eq:absRes},  as function of the mesh resolution at a CFL number of 0.1. Squares, triangles and circles represent, respectively, square, triangular and polygonal prism meshes. The estimation of the normal from previous time step drastically reduces the initial residual and has a convergence rate of second order for all mesh types. Guessing the initial normal with the gradient of the volume fraction (Youngs method) results in a constant error. The same behavior is observed at a CFL number of 0.5 as depicted in Fig.~\ref{fig:ResCFL05}. The new method converges with second order for polyhedral meshes. The convergence drops to rate below first order for triangular prism and square prism meshes. However, the method still improves the initial residual on the finest resolution by one order of magnitude for the square prism meshes and two orders of magnitude for the triangular prism meshes. The drop-off in accuracy is probably caused by the advection step where the same behaviour could be observed. The new estimation of the interface normals results in a reduced number reconstruction steps where on average about 2 to 2.5 interface reconstructions are needed to reach the specified tolerance. With the exception of the hexahedral meshes at a CFL number of 0.5 where 4 reconstruction steps are required. For comparison, using the Young method as initial guess, typically 4 reconstruction steps are required for all mesh types and CFL numbers.

\begin{figure}[ht]
	\subcaptionbox{CFL number = 0.1 \label{fig:ResCFL01}} 
	{\includegraphics[width=0.5\linewidth]{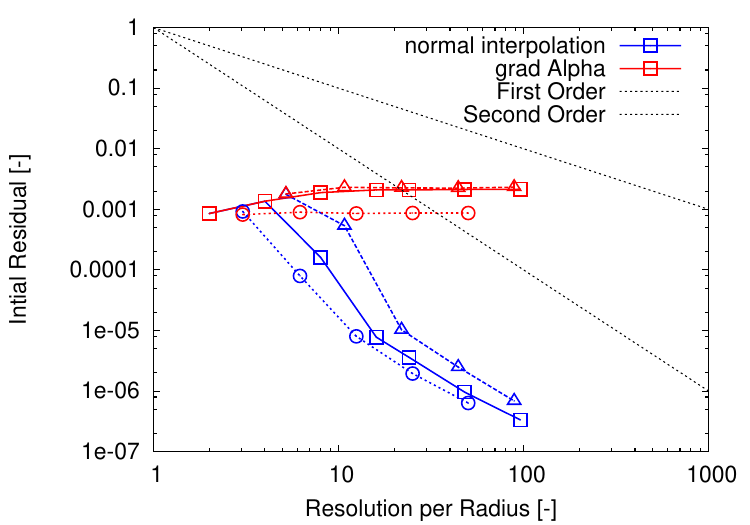}}
	\subcaptionbox{CFL number = 0.5 \label{fig:ResCFL05}} 
	{\includegraphics[width=0.5\linewidth]{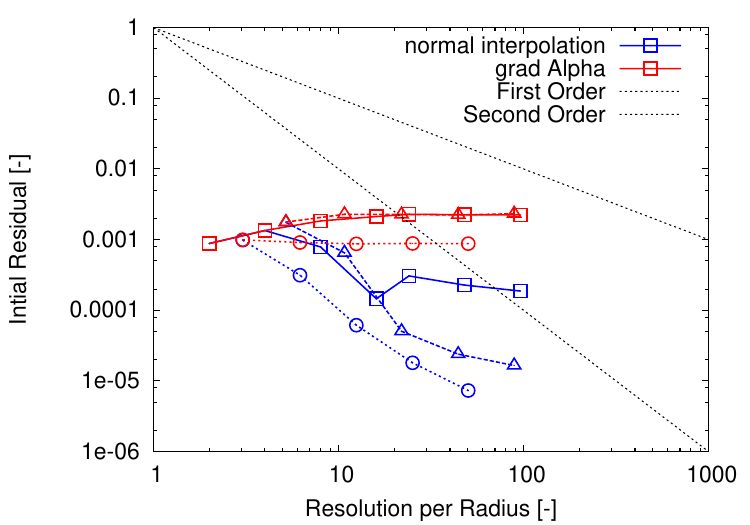}}
    \caption{Comparison of time-averaged initial residual of the gradient of volume fraction field with the interpolated normals from the previous time step.}
\end{figure}

\subsubsection{Rotation Test}\label{Rotation Test}

A frequently used benchmark is a sphere in a solid body rotation flow. The sphere with a radius of 0.15 is positioned at (0.5,0.75,0.5) in a unit cube, $[0,1]\times[0,1]\times[0,1]$, and advected with velocity field:

\begin{equation}
	\mathbf u(x,y,z) = (y - 0.5, -(x - 0.5), 0.0).
\end{equation}

The simulation runs for 2$\pi$ seconds which corresponds to one full rotation. In this test case, the flow speed is highest far from the origin. Therefore, the largest CLF numbers are attained at the corners of the cubical domain. For the hexahedral mesh we use a CFL limit of 1 as was done in {\cite{Jofre.2014}}. This gives rise to CFL$_{surf}$ numbers around 0.5  because the interface is closer to the origin. For our unstructured meshes we have very special cells in the corners of the domain so a CFL limit of 1 leads to a CFL$_{surf}$ around 0.1 and therefore non-comparable timestep sizes. Therefore, in the unstructured cases we have used a CFL$_{surf}$ number of 0.5 for the determination of the time step size. Table~\ref{tab:rotSphere} shows that the method conserves the volume and is numerically bounded. A direct comparison of the shape error, $E_1$, shows that the method of Jofre et al. \cite{Jofre.2014} is roughly 50 \% more accurate at the same resolutions. But a similar order of convergence is achieved. The reconstruction and execution time per iteration roughly quadruples with mesh refinement which is expected as the number of interface cells quadruples as well. On the highest resolution one more iteration for the reconstruction is required, which explains the higher reconstruction time.

\begin{table}[htb]
\centering
\caption{Rotation test with CFL = 1.0 or $\text{CFL}_{surf}$ = 0.5: Errors and timing (4 cores) for various mesh types and resolutions.}
\label{tab:rotSphere}
\begin{tabular}{l c c c c c c c}
\hline
\hline
Cartesian meshes & N & $E_{Vol}$ & $E_{bound}$ & $ E_{1} $ & $\mathcal O(E_{1})$ & $T_e$ & $T_r$\\
\hline
\hline
isoAdvector-plicRDF & 32 & 3.469e-18 &  5.393e-21 & 7.50e-04 & - & 0.034 & 0.027\\
Jofre et al. \cite{Jofre.2014} & 32  & - & - & 5.47e-04 & - & - & -\\
\hline
isoAdvector-plicRDF & 64 & 2.706e-16 &  7.412e-22   & 1.86e-04 & 2.03 & 0.10 & 0.08 \\
Jofre et al. \cite{Jofre.2014} & 64  & - & - & 1.29e-04 & 2.08 & - & -\\ 
\hline
isoAdvector-plicRDF & 128 & 3.565e-15 &  8.934e-23 & 4.77e-05 & 1.96 & 0.44  & 0.31\\
Jofre et al. \cite{Jofre.2014} & 128 & - & - & 3.46e-05 & 1.90 & - & -\\
\hline
isoAdvector-plicRDF & 256 & 2.504e-14 &  5.414e-16 & 1.41e-05 & 1.75 & 2.60  & 1.69\\
Jofre et al. \cite{Jofre.2014} & 256 & - & - & - & - & - & -\\
\hline
\hline
Tetrahedral meshes & N & $E_{Vol}$ & $E_{bound}$ & $ E_{1} $ & $\mathcal O(E_{1})$ & $T_e$ & $T_r$\\
\hline
\hline
isoAdvector-plicRDF & 32 & 8.327e-17 & 1.239e-19 & 1.99e-03 & - & 0.046 & 0.039\\
Jofre et al. \cite{Jofre.2014} & 32  & - & - & 6.30e-04 & - & - & -\\
\hline
isoAdvector-plicRDF & 64 & 1.676e-15 & 4.425e-20 & 4.45e-04 & 2.16 & 0.18 & 0.16\\
Jofre et al. \cite{Jofre.2014} & 64  & - & - & 2.31e-04 & 1.45 & - & -\\ 
\hline
isoAdvector-plicRDF & 128 & 1.020e-14 & 2.698e-21 & 1.03e-04 & 2.10 & 0.57  & 0.44\\
Jofre et al. \cite{Jofre.2014} & 128 & - & - & 6.79e-05 & 1.76 & - & -\\
\hline
isoAdvector-plicRDF & 256 & 4.603e-14 & 1.175e-21 & 2.783e-05 & 1.82 & 2.88 & 2.02\\
Jofre et al. \cite{Jofre.2014} & 256 & - & - & - & - & - & -\\
\hline
\hline
Polyhedral meshes & N & $E_{Vol}$ & $E_{bound}$ & $ E_{1} $ & $\mathcal O(E_{1})$ & $T_e$ & $T_r$\\
\hline
\hline
isoAdvector-plicRDF & 32 & 4.367e-10 & 1.884e-12 & 1.14e-03 & - & 0.078 & 0.066\\
\hline
isoAdvector-plicRDF & 64 & 9.749e-11 & 1.106e-13 & 2.09e-04 & 2.45 & 0.287 & 0.232\\
\hline
isoAdvector-plicRDF & 128 & 1.288e-09 & 7.563e-13 & 5.51e-05 & 1.93 & 1.45 & 1.07\\
\hline
isoAdvector-plicRDF & 256 & 4.522e-09 & 2.208e-13 & 1.566e-05 & 1.81 & 6.09 & 3.81\\
\hline
\hline
\end{tabular}
\end{table}

The error norms and timings for unstructured tetrahedral are depicted in the middle part of Table~\ref{tab:rotSphere}. As on the Cartesian meshes, good volume conservation and boundedness properties are obtained. Near second order convergence is observed for all resolutions and slightly higher convergence rates compared to Jofre et al. \cite{Jofre.2014} can be achieved. However, Jofre et al. \cite{Jofre.2014} achieves lower shape errors on coarse meshes, but the gap narrows with increasing mesh resolution. The timing data are as expected and quadruple with every mesh refinement. The results for the polyhedral meshes are given in the lower part of Table~\ref{tab:rotSphere}. Volume conservation is still good although errors are bigger. The shape errors are similar to the Cartesian mesh errors and are approximately half of those observed on the tetrahedral meshes, where computational times are also lower.

\subsubsection{Disc in reversed spiral flow}\label{SpirallingDisc}

We now move on to another standard advection test case, namely a circular disc of radius 0.15 initially centred at $(0.5,0.75)$ in the domain $[0,1]\times [0,1]$ and advected with the velocity field
\begin{equation}
	\mathbf u(x,y,t) = \cos(2\pi t/16)\left( -\sin^2(\pi x) \sin(2 \pi y), \sin(2 \pi x) \sin^2 (\pi y) \right).
\end{equation}
This flow causes a stretching and spiraling of the disc to a maximum level reached at time $t = 4$ s as depicted in Fig.~\ref{fig:vortexShearedDisc}. From time $t = 4$ s to 8 s, the flow is running backwards causing the fluid to end up at its initial configuration. The simulation is repeated for square, triangular and polygonal prism meshes of different resolutions with the linear cell size halved at each refinement. 

\begin{figure}
	\subcaptionbox{square prism mesh N=64} 
	{\includegraphics[width=0.32\columnwidth, bb=192 90 892 742, clip=true]{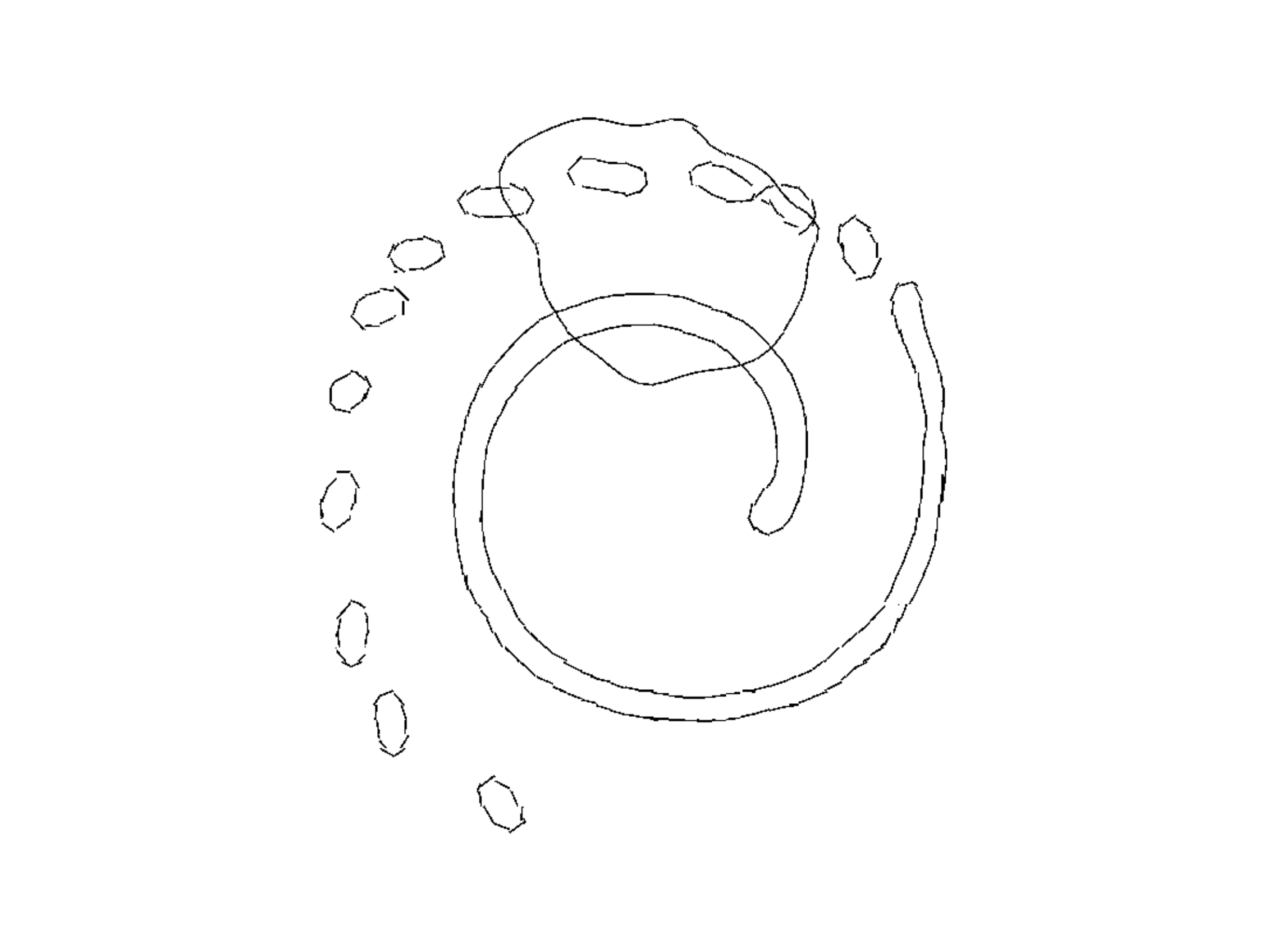}}
	\subcaptionbox{square prism mesh N=128} 
	{\includegraphics[width=0.32\columnwidth, bb=192 90 892 742, clip=true]{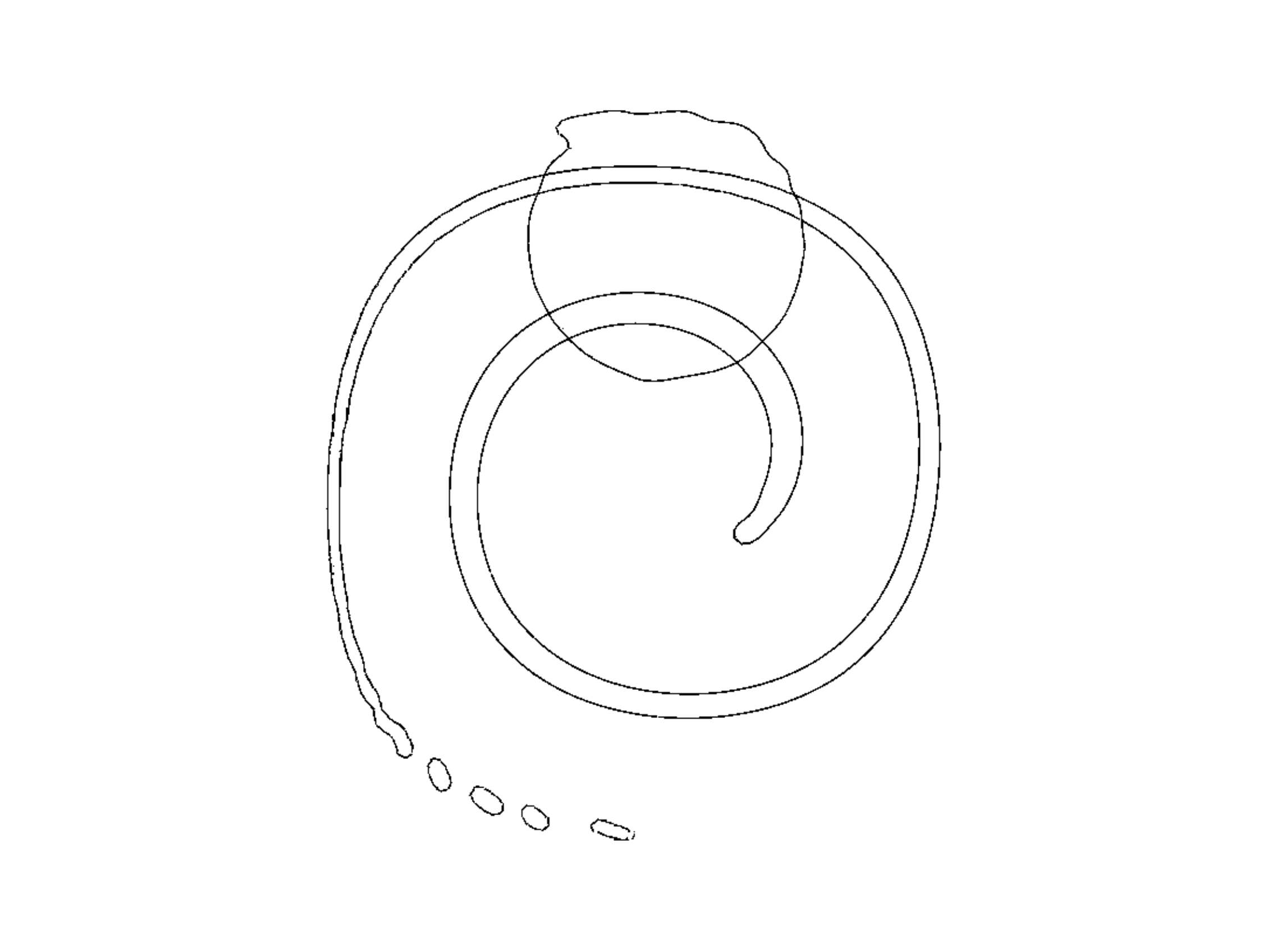}} 
	\subcaptionbox{square prism mesh N=256} 
	{\includegraphics[width=0.32\columnwidth, bb=192 90 892 742, clip=true]{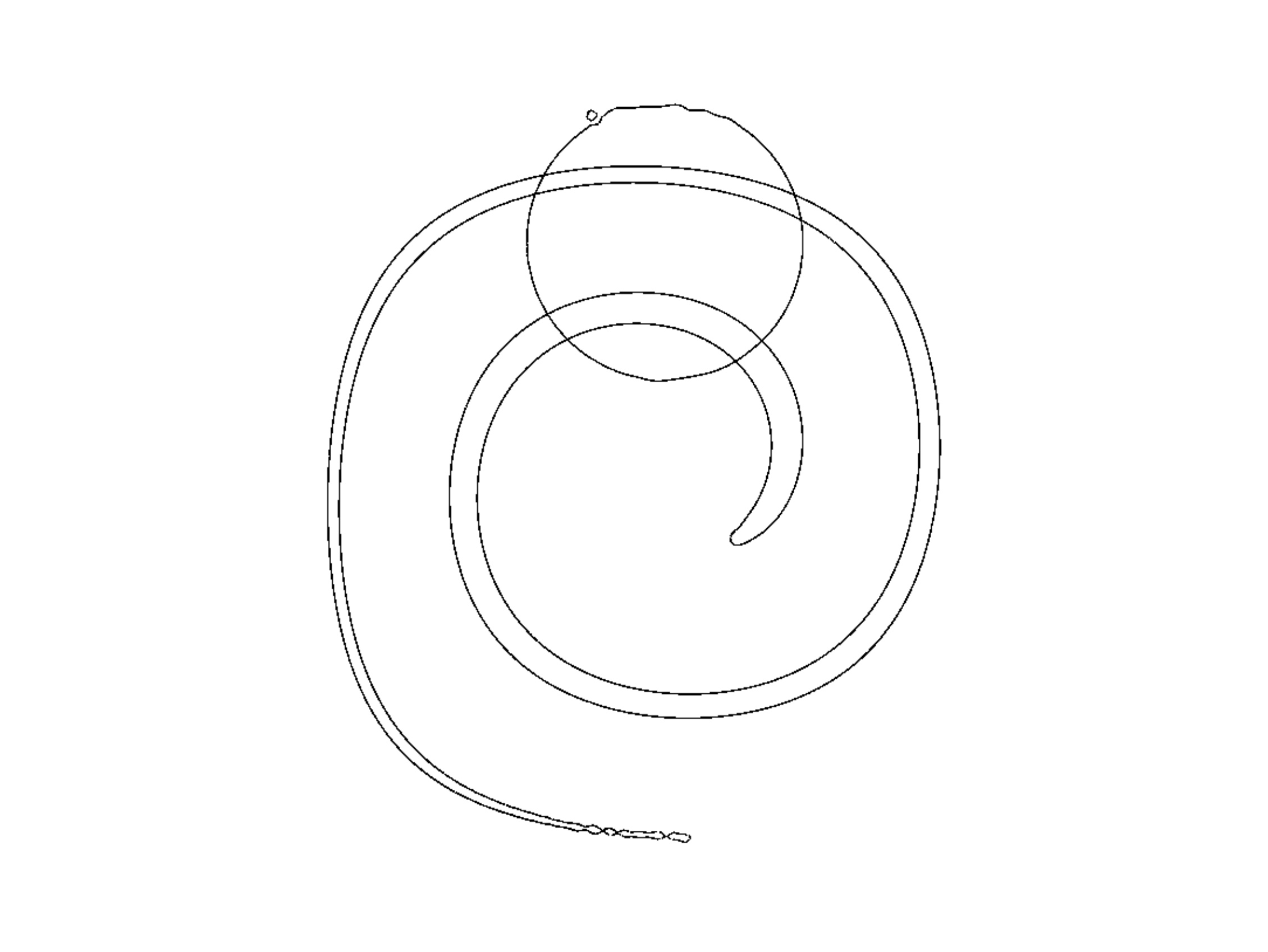}}
	\subcaptionbox{Triangular prism mesh N=64} 
	{\includegraphics[width=0.32\columnwidth, bb=192 90 892 742, clip=true]{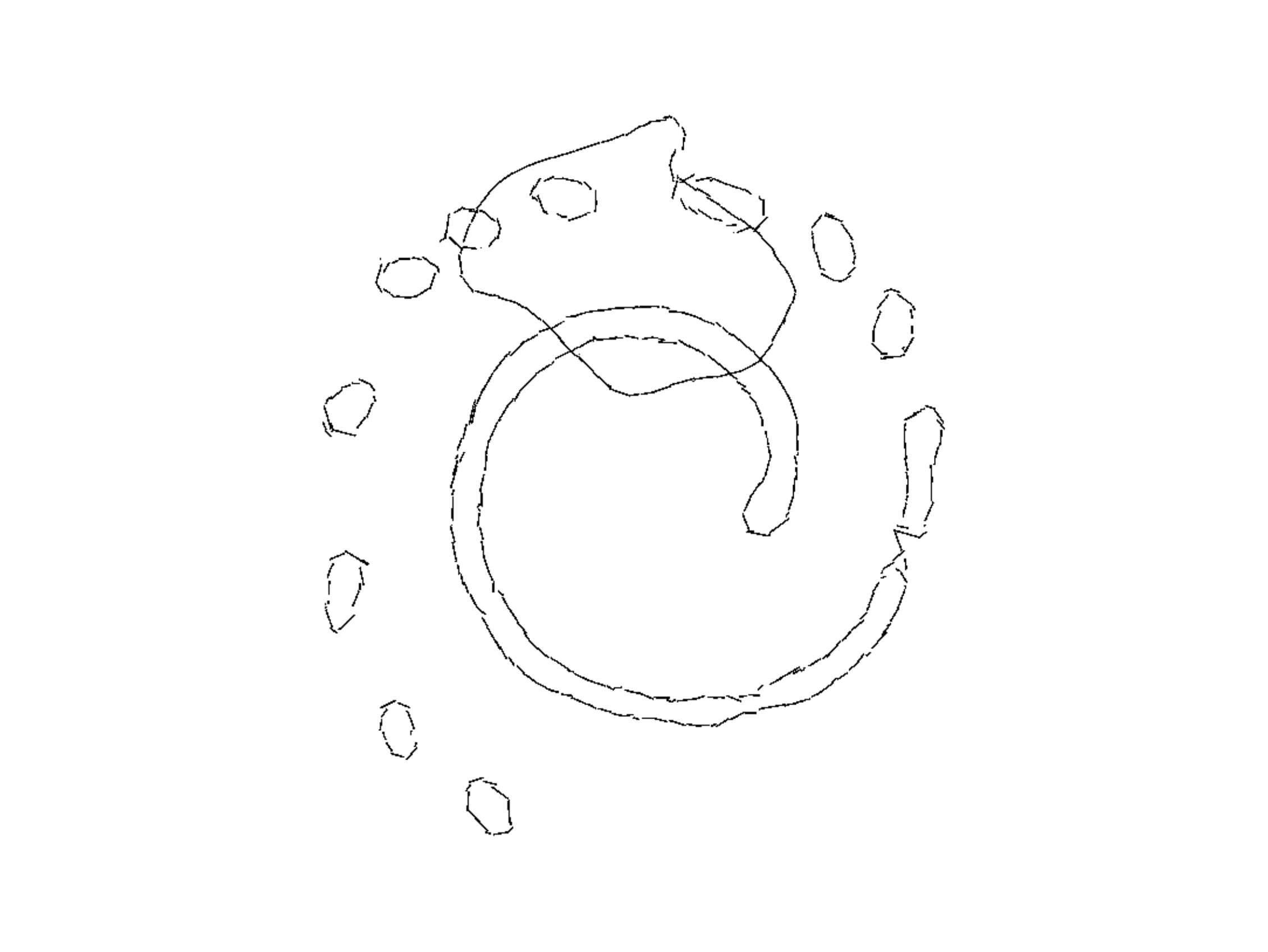}}
	\subcaptionbox{Triangular prism mesh N=128} 
	{\includegraphics[width=0.32\columnwidth, bb=192 90 892 742, clip=true]{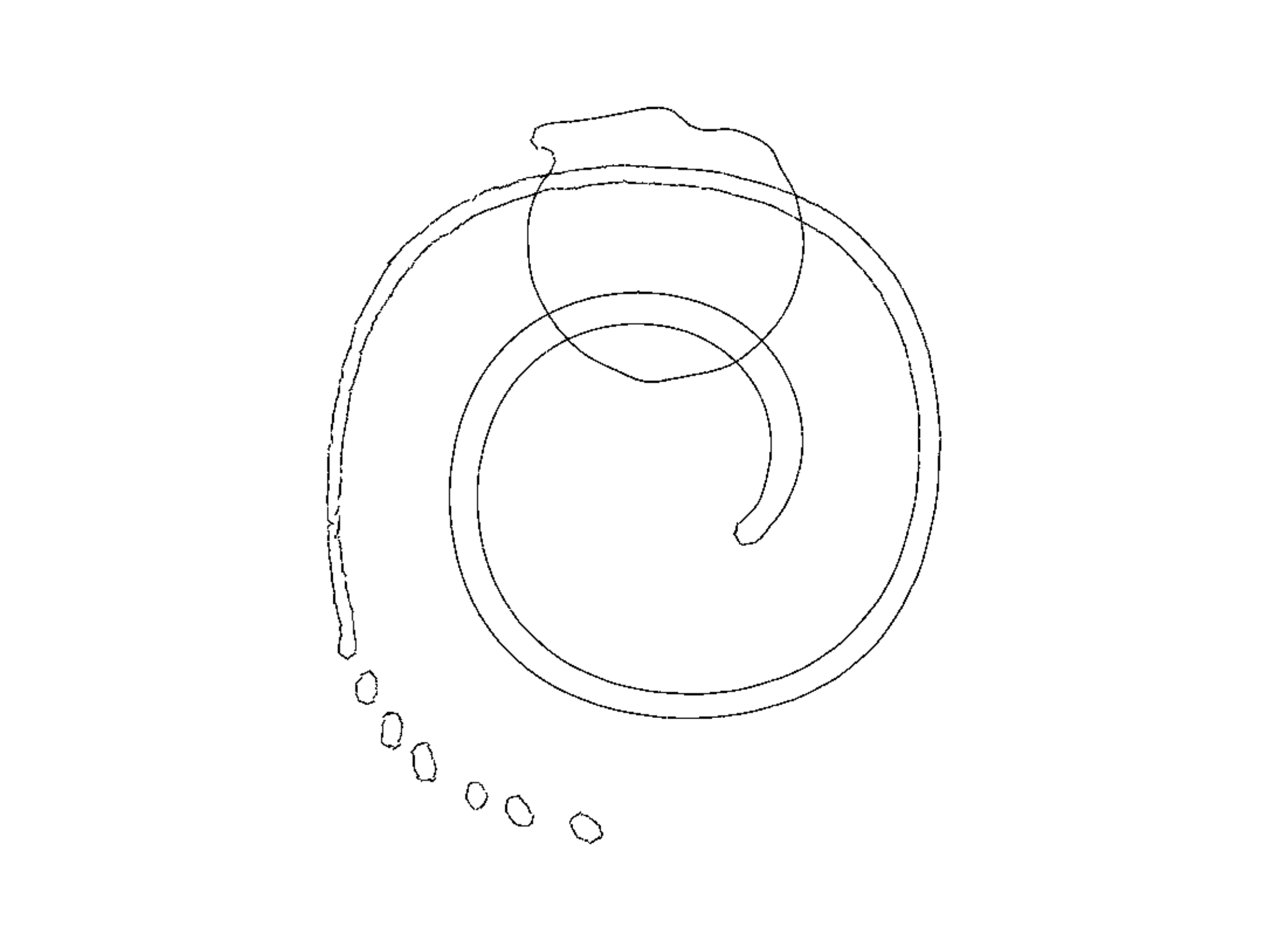}} 
	\subcaptionbox{Triangular prism mesh N=256} 
	{\includegraphics[width=0.32\columnwidth, bb=192 90 892 742, clip=true]{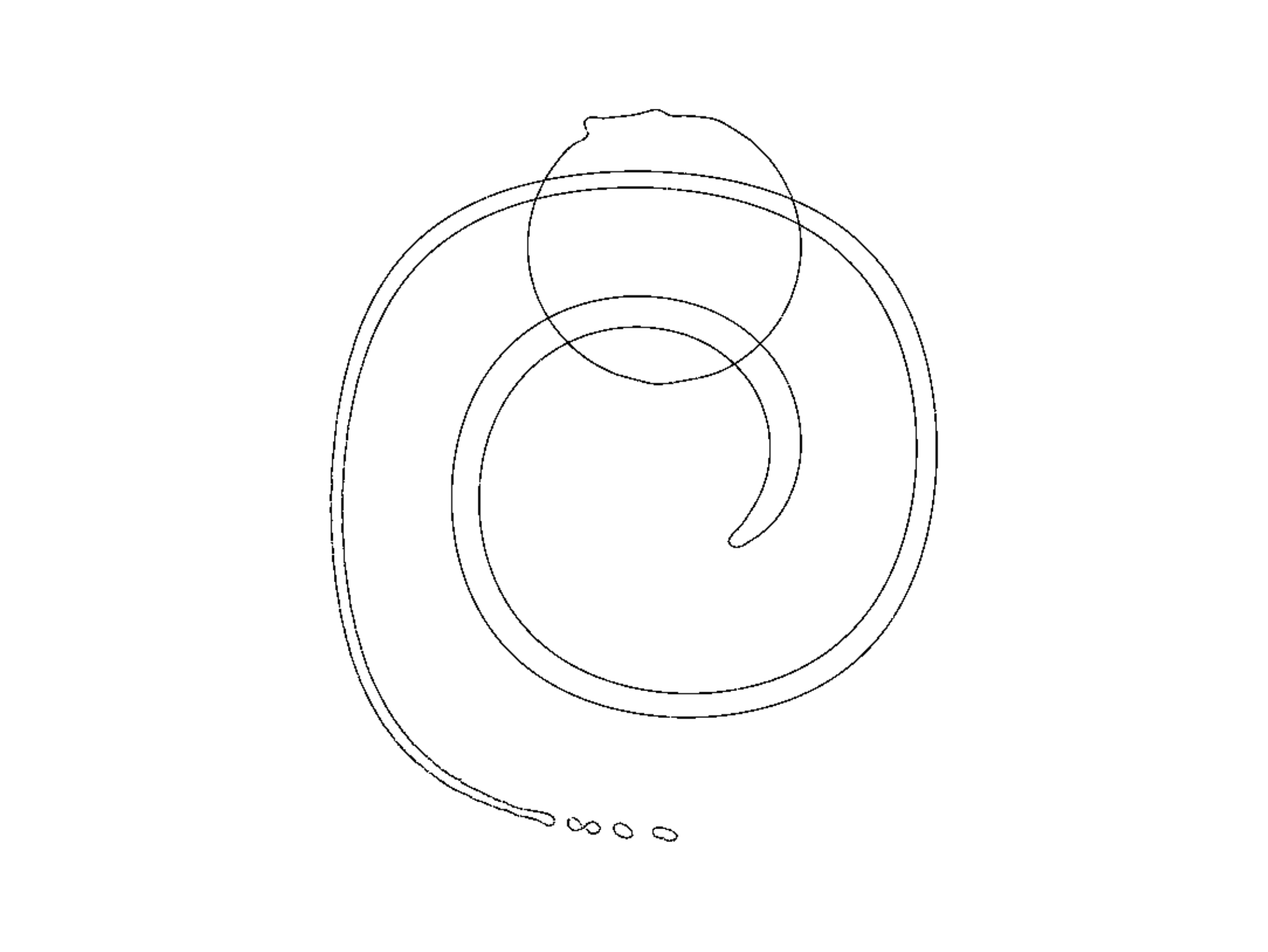}}	
	\subcaptionbox{polygonal prism mesh N=64} 
	{\includegraphics[width=0.32\columnwidth, bb=192 90 892 742, clip=true]{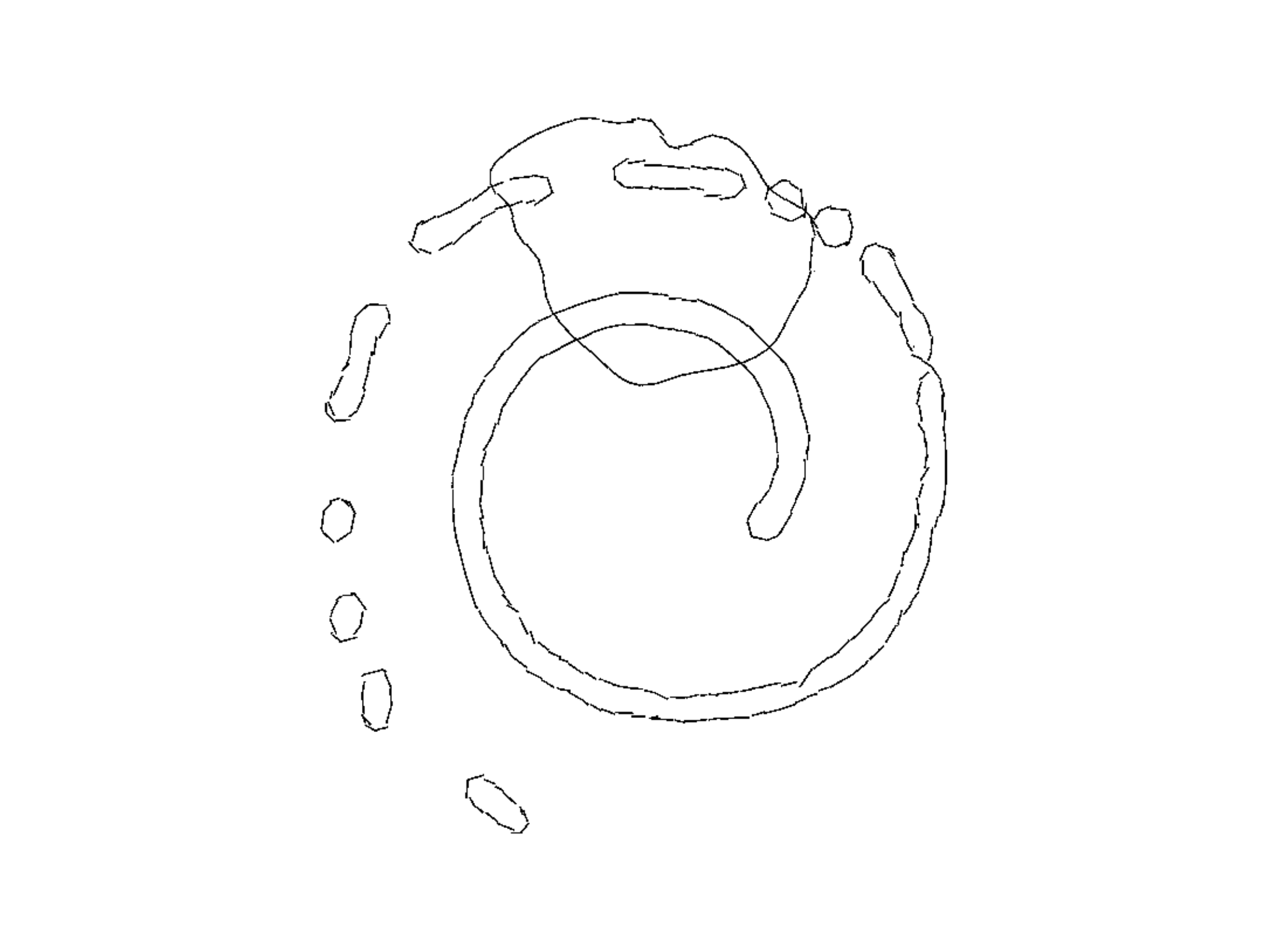}}
	\subcaptionbox{polygonal prism mesh N=128} 
	{\includegraphics[width=0.32\columnwidth, bb=192 90 892 742, clip=true]{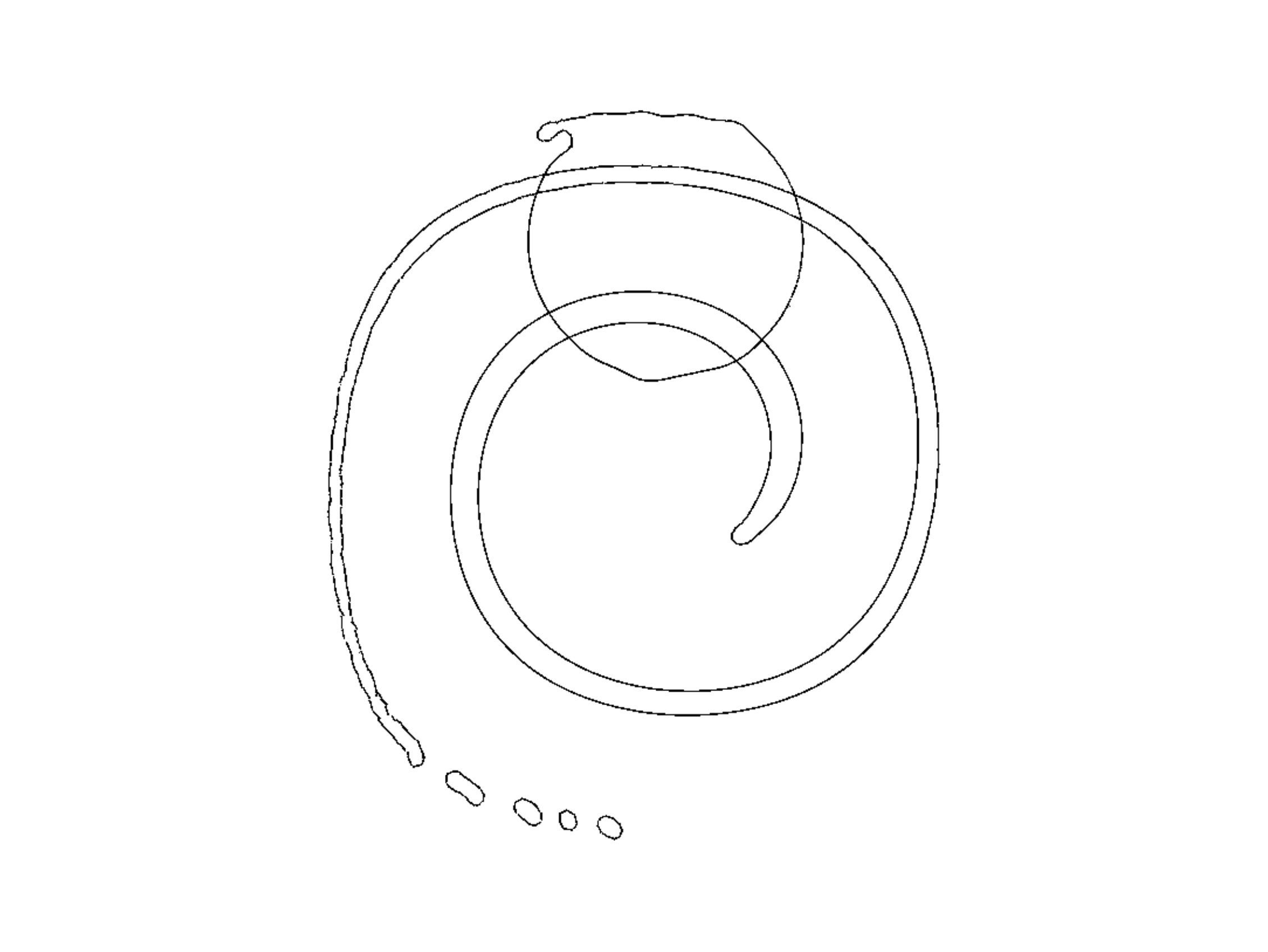}}
	\subcaptionbox{polygonal prism mesh N=256} 
	{\includegraphics[width=0.32\columnwidth, bb=192 90 892 742, clip=true]{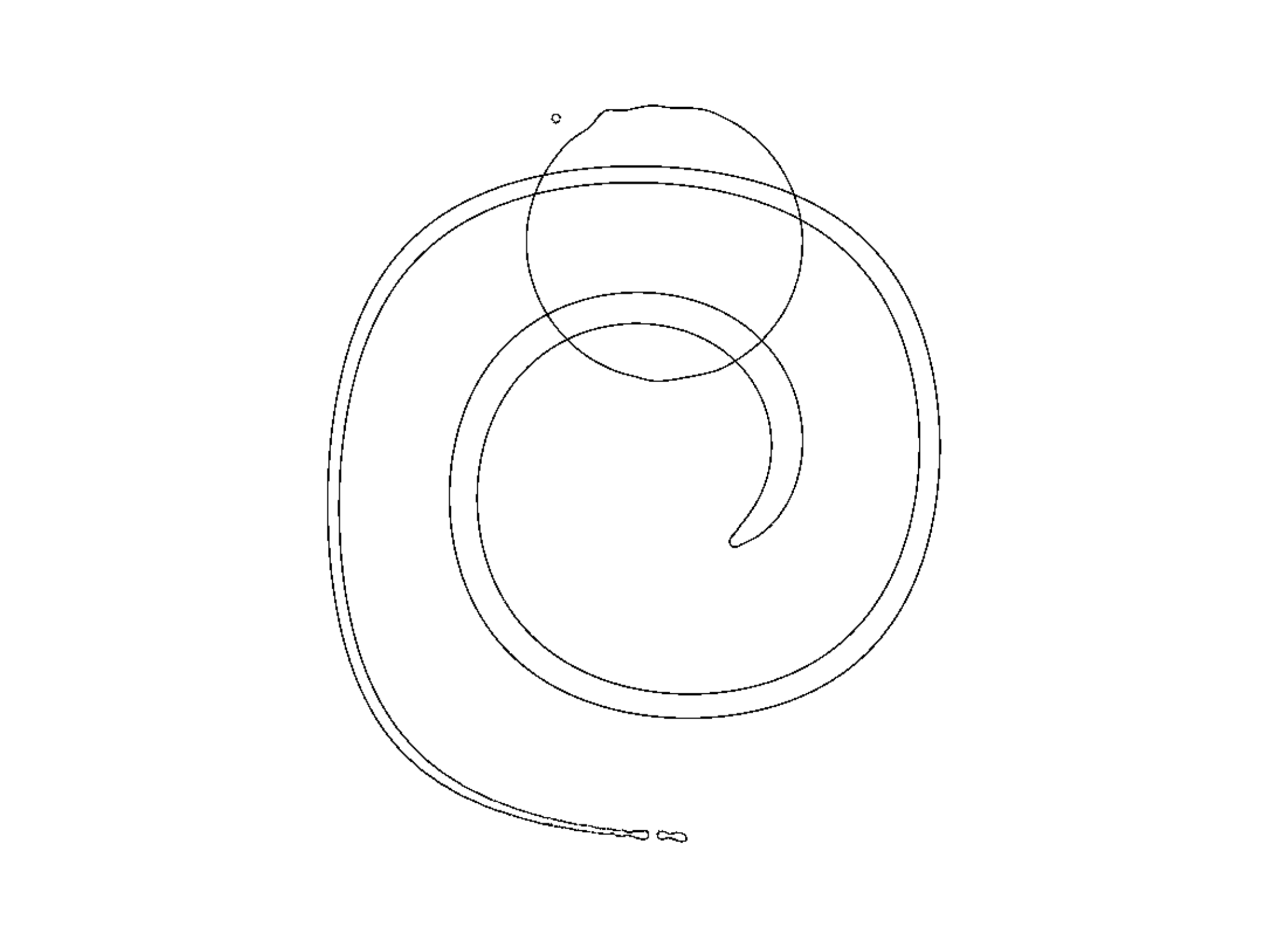}}
	\cprotect\caption{Interfaces at $t=8$ and $t=4$ second for the disc in reversed vortex flow test case with different resolutions and mesh types.}
	\label{fig:vortexShearedDisc}
\end{figure}

Table~\ref{tab:spirallingDiscCFL=0.5} shows errors and execution times for CFL = 0.5 on Cartesian meshes. The errors are compared to the results obtained by Owkes and Desjardins {\cite{Owkes.2014}} who also used CFL = 0.5. The execution times of Owkes and Desjardins {\cite{Owkes.2014}} are not shown here because Maric et al. {\cite{Maric.2018}} pointed out inconsistencies in the values reported by Owkes and Desjardins {\cite{Owkes.2014}}. The table shows that both methods perform well in terms of volume conservation and bounding. The isoAdvector-plicRDF $E_1$ errors are roughly 50 \% higher than those of Owkes and Desjardins {\cite{Owkes.2014}}. 

Although the isoAdvector accuracy is known to get significantly worse as the CFL number approaches 1, we show results for CFL = 1 in Table~\ref{tab:spirallingDiscCFL=1}. This is done for comparison with the results reported for this case by Maric et al. {\cite{Maric.2018}} and Ivey and Moin {\cite{Ivey.2017}} who both used CFL = 1. Maric et al. {\cite{Maric.2018}} also listed calculation times which are also reproduced in Table~\ref{tab:spirallingDiscCFL=1} for comparison. All methods perform well in terms of volume conservation and bounding. As for $E_1$, comparison with Ivey and Moin {\cite{Ivey.2017}} shows that we obtain 23\% and 15\% lower $E_1$ errors on the coarse and intermediate resolutions, respectively, but twice as large errors on the fine mesh. Our $E_1$ error is 50\% larger than those of Maric et al. {\cite{Maric.2018}} for the coarse and intermediate resolutions, but more than 3 times larger on the fine mesh. Comparing the time spent per time step, we see that our method is 10 times faster than the UFVFC-Swartz method of Maric et al. {\cite{Maric.2018}}. As can be seen by comparing the $\mathcal O(E_1)$ columns of Table~\ref{tab:spirallingDiscCFL=0.5} and Table~\ref{tab:spirallingDiscCFL=1}, the convergence of isoAdvector-plicRDF can be restored by halving the CFL to 0.5, which should still give a significant speed up compared to Maric et al. {\cite{Maric.2018}}.

The errors for unstructured triangular prism meshes are shown in Table~\ref{tab:spirallingDiscUnstruct} for the $\text{CFL}_{surf}$ number of 0.5. They show a similar behavior with good boundedness and volume conservation, as well as second order convergence. For the coarse meshes the absolute shape error is roughly twice as large as for the Cartesian meshes. For the finest meshes the errors are similar to those of the Cartesian meshes. On polyhedral prism meshes, the data shown in  Table~\ref{tab:spirallingDiscUnstruct} shows similar errors as for Cartesian meshes, albeit at a higher computational cost on the finer meshes. 

\begin{table}
\centering
\caption{Disc in reversed vortex flow: Error and time (on 1 core) for Cartesian meshes with CFL = 0.5.}
\label{tab:spirallingDiscCFL=0.5}
\begin{tabular}{l c c c c c c } 		
\hline
\hline
Cartesian meshes & N & $E_{Vol}$ & $E_{bound}$ & $E_1$ & $\mathcal O(E_{1})$ & $T_e$\\
\hline
\hline
isoAdvector-plicRDF & 64 & 1.665e-16 & 1.514e-20 & 1.26e-02 & - & 0.010 \\
Owkes and Desjardins \cite{Owkes.2014} & 64 & 9.755e-15 & 6.51e-17 & 7.75e-03 & - & -\\
\hline
isoAdvector-plicRDF & 128 & 9.714e-17 & 4.689e-13 & 2.61e-03 & 2.27 & 0.023\\
Owkes and Desjardins \cite{Owkes.2014} & 128 & 1.290e-14 & 6.328e-17 & 1.87e-03 & 2.04 & -\\
\hline
isoAdvector-plicRDF & 256 & 1.207e-15 & 5.266e-12 & 5.71e-04 & 2.19 & 0.05\\
Owkes and Desjardins \cite{Owkes.2014} & 256 & 1.736e-14 & 9.33e-17 & 4.04e-04 & 2.21 & -\\
\hline
isoAdvector-plicRDF & 512 & 9.590e-15 &  4.142e-14 &  1.04e-04 &  2.44 &  0.12\\
Owkes and Desjardins \cite{Owkes.2014} & 512 &  1.736e-14 & 9.325e-17 & 8.32e-05 & 2.28 & -\\
\hline
isoAdvector-plicRDF & 1024 & 7.924e-14 &  3.686e-13 &  3.50e-05 &  1.57 &  0.35\\
Owkes and Desjardins \cite{Owkes.2014} & 1024 & 1.678e-14 & 1.043e-16 & 2.35e-05 & 1.82 & -\\
\hline
\hline
\end{tabular}
\end{table}

\begin{table}
\centering
\caption{Disc in reversed vortex flow: Error and time (on 1 core) for Cartesian meshes with CFL = 1.}
\label{tab:spirallingDiscCFL=1}
\begin{tabular}{l c c c c c c } 		
\hline
\hline
Cartesian meshes & N & $E_{Vol}$ & $E_{bound}$ & $E_1$ & $\mathcal O(E_{1})$ & $T_e$\\
\hline
\hline
isoAdvector-plicRDF & 64 &  2.637e-16 &  4.345e-12 &  8.700e-03 &    - &  0.011 \\
UFVFC-Swartz  \cite{Maric.2018} & 64 & 5.89e-15 & 0.0 & 5.74e-03 & - & 0.12 \\
Ivey and Moin  NIFPA-1 \cite{Ivey.2017} & 65 & - & - & 1.13e-02 & - & - \\
\hline
isoAdvector-plicRDF & 128 &  5.967e-16 &  8.635e-15 &  2.266e-03 &   1.94 &  0.023 \\
UFVFC-Swartz  \cite{Maric.2018} & 128 & 1.56e-14 & 0.0 & 1.45e-03 & 1.98 & 0.27\\
Ivey and Moin  NIFPA-1 \cite{Ivey.2017} & 129 & - & - & 2.67e-03 & 2.08 & -\\
\hline
isoAdvector-plicRDF & 256 &  1.360e-15 &  1.776e-12 &  1.614e-03 &   0.49 &  0.06  \\
UFVFC-Swartz \cite{Maric.2018} & 256 & 6.16e-14 & 0.0 & 3.77e-04 & 1.94 &  0.58\\
Ivey and Moin  NIFPA-1 \cite{Ivey.2017} & 257 & - & - & 5.36e-04 & 2.24  & -\\
\hline
isoAdvector-plicRDF & 512 &  7.577e-15 &  9.177e-11 &  1.929e-03 &  -0.26 &  0.17  \\
\hline
isoAdvector-plicRDF & 1024 &  2.229e-14 &  2.931e-11 &  2.753e-03 &  -0.51 &  0.61  \\
\hline
\hline
\end{tabular}
\end{table}

\begin{table}
\centering
\caption{Disc in reversed vortex flow: Error and time (on 1 core) for unstructured meshes with $\text{CFL}_{surf}$ = 0.5.}
\label{tab:spirallingDiscUnstruct}
\begin{tabular}{l c c c c c c } 		
\hline
\hline
Triangular meshes & N & $E_{Vol}$ & $E_{bound}$ & $E_1$ & $\mathcal O(E_{1})$ & $T_e$\\
\hline
\hline
isoAdvector-plicRDF & 64 & 9.714e-17 & 2.727e-20 & 2.208e-02 & - &  0.012\\

isoAdvector-plicRDF & 128 & 3.469e-16 & 8.106e-21 & 3.582e-03 & 2.62 & 0.024\\

isoAdvector-plicRDF & 256 & 3.594e-15 & 1.163e-21 & 7.514e-04 & 2.25 & 0.058\\

isoAdvector-plicRDF & 512 & 5.854e-14 & 5.323e-22 & 1.309e-04 & 2.52 & 0.155\\

isoAdvector-plicRDF & 1024 & 1.132e-13 & 1.857e-18 & 3.280e-05 & 1.99 & 0.45\\
\hline
\hline
Polygonal meshes & N & $E_{Vol}$ & $E_{bound}$ & $E_1$ & $\mathcal O(E_{1})$ & $T_e$\\
\hline
\hline
isoAdvector-plicRDF & 64 & 2.498e-16 & 2.358e-20 & 1.292e-02 & - &  0.014\\

isoAdvector-plicRDF & 128 & 6.800e-16 & 8.169e-21 & 2.487e-03 & 2.37 & 0.028\\

isoAdvector-plicRDF & 256 & 4.399e-15 & 2.488e-21 & 4.186e-04 & 2.57 & 0.068\\

isoAdvector-plicRDF & 512 & 5.168e-14 & 6.084e-22 & 9.292e-05 & 2.17 & 0.206\\

isoAdvector-plicRDF & 1024 & 1.884e-13 & 1.657e-22 & 2.661e-05 & 1.80 &  0.71\\
\hline
\hline
\end{tabular}
\end{table}

We note that with the resolutions used here and elsewhere in literature, the interface is not well-resolved at the time of maximum deformation of the circle. This is illustrated in Fig.~\ref{fig:vortexShearedDisc}, where pinched off ``droplets'' are clearly visible at the tail of the spiral. In the exact solution, and in simulations where the thin tail is well-resolved, this does not happen. Inspection of the simulations reveals that the fluid particles in the top part of the circular disc that is most distorted in Fig.~\ref{fig:vortexShearedDisc} are indeed the ones forming the pinching tail at time $t = 4$ s. If we compare the different mesh types at the same resolution, most droplets are formed on the triangular prism meshes which also have the highest shape error.

\subsubsection{3D Shear Flow}\label{Spiral Test}

In this deformation test case a sphere of radius 0.15 is positioned at (0.5,0.75,0.5) in a box of dimensions $[0,1] \times [0,1] \times [0,2]$ and transported in the velocity field:

\begin{equation}
\mathbf u(x,y,z,t) = \cos(\pi t/T)
\left(
\begin{matrix}
\sin^2{(\pi x)} \sin{(2 \pi y)}\\
-\sin{(2 \pi x)} \sin^2{(\pi y)}\\
\left(1 - 2r\right)^2
\end{matrix}
\right),
\end{equation}
where $r = \sqrt{(x-0.5)^2 + (y-0.5)^2}$ and $T = 3$. The simulation is run with CFL number 0.5 and $\text{CFL}_{surf}$ of 0.5 for unstructured meshes until $t = 3$ at which point all fluid particles have returned to their initial position due to the reversal of the flow at $t = 1.5$. The proposed method achieves near perfect mass conservation and boundedness as shown in Table~\ref{tab:shearFlow}. The comparison with Jofre et al. \cite{Jofre.2014} in Table~\ref{tab:shearFlow} shows very similar shape errors both on Cartesian and tetrahedral meshes. As in the previous test case the UFVFC-Swartz algorithm \cite{Maric.2018} is the most accurate, but also roughly one order of magnitude slower. The reduction in computational cost is mainly due to the advection algorithm, but compared to the modified Swartz algorithm from \cite{Maric.2018} our reconstruction method is 3-5 times faster. It must be noted that our CPU had a 35\% higher base clock but also uses slower memory (see beginning of Section \ref{sec:NumTests}).

The middle and lower parts of Table~\ref{tab:shearFlow} show the results for the tetrahedral and polyhedral meshes. The execution and reconstruction times for the tetrahedral meshes are higher than for the Cartesian meshes and the shape errors are also almost twice as high. On polyhedral meshes we observed an accuracy similar to the results achieved on Cartesian meshes. The reconstructed interface for $t=1.5$ and $t=3$ seconds are shown in Fig. {\ref{fig:shearFlow3D}}.
\begin{table}
\centering
\caption{3D shear flow with CFL = 0.5 or $\text{CFL}_{surf}$ = 0.5: Errors and timing (4 cores) on various mesh types and resolutions.}
\label{tab:shearFlow}
\begin{tabular}{l c c c c c c c}                       
\hline
\hline
Cartesian meshes & N & $E_{Vol}$ & $E_{bound}$ & $ E_{1} $ & $\mathcal O(E_{1}) $ & $T_e$ & $ T_r$\\
\hline
\hline
isoAdvector-plicRDF & 32 & 2.654e-16 & 1.011e-19 & 4.06e-03 & - & 0.040 & 0.031\\
UFVFC-Swartz & 32  & 2.33e-15 & 0.0 & 1.97e-03 & - & 0.57 & 0.11\\
Jofre et al. \cite{Jofre.2014} & 32  & - & - & 4.08e-03 & - & - & -\\
\hline
isoAdvector-plicRDF & 64 & 3.313e-16 & 1.846e-13 & 1.079e-03 & 1.91 & 0.139 & 0.11 \\
UFVFC-Swartz \cite{Maric.2018} & 64  & 4.91e-15 & 0.0 & 4.25e-04 & 2.21 & 2.31 & 0.41\\
Jofre et al. \cite{Jofre.2014} & 64  & - & - & 1.46e-03 & 1.48 & - & -\\
\hline
isoAdvector-plicRDF & 128 & 2.897e-16 & 5.881e-15 & 2.44e-04 & 2.14 & 0.65 & 0.44\\
UFVFC-Swartz \cite{Maric.2018} & 128 & 1.22e-14 & 0.0 & 1.24e-04 & 1.77 & 12.00 & 2.38\\
Jofre et al. \cite{Jofre.2014} & 128 & - & - & 3.53e-04 & 2.05 & - & -\\
\hline
\hline
Tetrahedral meshes & N & $E_{Vol}$ & $E_{bound}$ & $ E_{1} $ & $\mathcal O(E_{1}) $ & $T_e$ & $ T_r$\\
\hline
\hline
isoAdvector-plicRDF & 32 &1.214e-16 &  6.524e-18 &  8.43e-03 &   - &  0.057 &  0.049\\
Jofre et al. \cite{Jofre.2014}& 32  & - & - & 5.97e-03 & - & - & -\\
\hline
isoAdvector-plicRDF & 64 & 6.939e-16 & 6.326e-15 & 2.88e-03 & 1.55 & 0.209 & 0.182\\
Jofre et al. \cite{Jofre.2014} & 64  & - & - & 1.64e-03 & 1.87 & - & -\\ 
\hline
isoAdvector-plicRDF & 128 & 6.477e-15 & 1.490e-17 & 5.50e-04 & 2.39 & 1.142 & 0.869\\
Jofre et al. \cite{Jofre.2014} & 128 & - & - & 5.37e-04 & 1.61 & - & -\\
\hline
\hline
Polyhedral meshes & N & $E_{Vol}$ & $E_{bound}$ & $ E_{1} $ & $\mathcal O(E_{1}) $ & $T_e$ & $ T_r$\\
\hline
\hline
isoAdvector-plicRDF & 32 & 6.245e-17 & 9.715e-18 & 5.99e-03 & - & 0.058 & 0.047 \\
\hline
isoAdvector-plicRDF & 64 & 6.852e-16 & 1.100e-19 & 1.17e-03 & 2.35 & 0.269 & 0.198 \\
\hline
isoAdvector-plicRDF & 128 & 6.025e-15 & 7.700e-21 & 2.59e-04 & 2.18 & 1.40 & 0.831 \\
\hline
\hline
\end{tabular}
\end{table}

\begin{figure}[htp]
	\subcaptionbox{Hexahedral mesh N=64} 
	{\includegraphics[width=0.33\columnwidth, bb=242 150 842 742, clip=true]{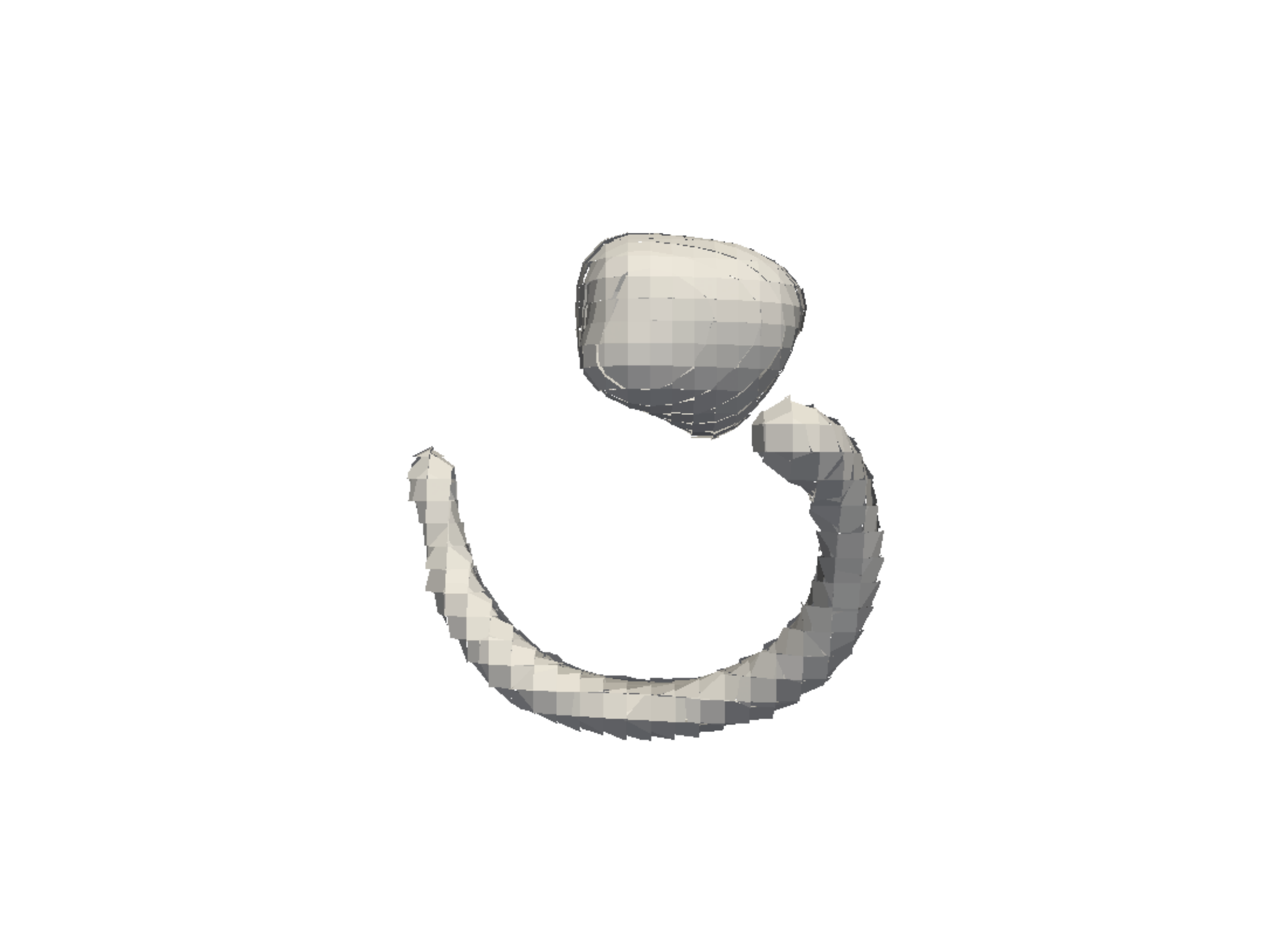}}
	\subcaptionbox{Hexahedral mesh N=128} 
	{\includegraphics[width=0.33\columnwidth, bb=291 150 842 792, clip=true]{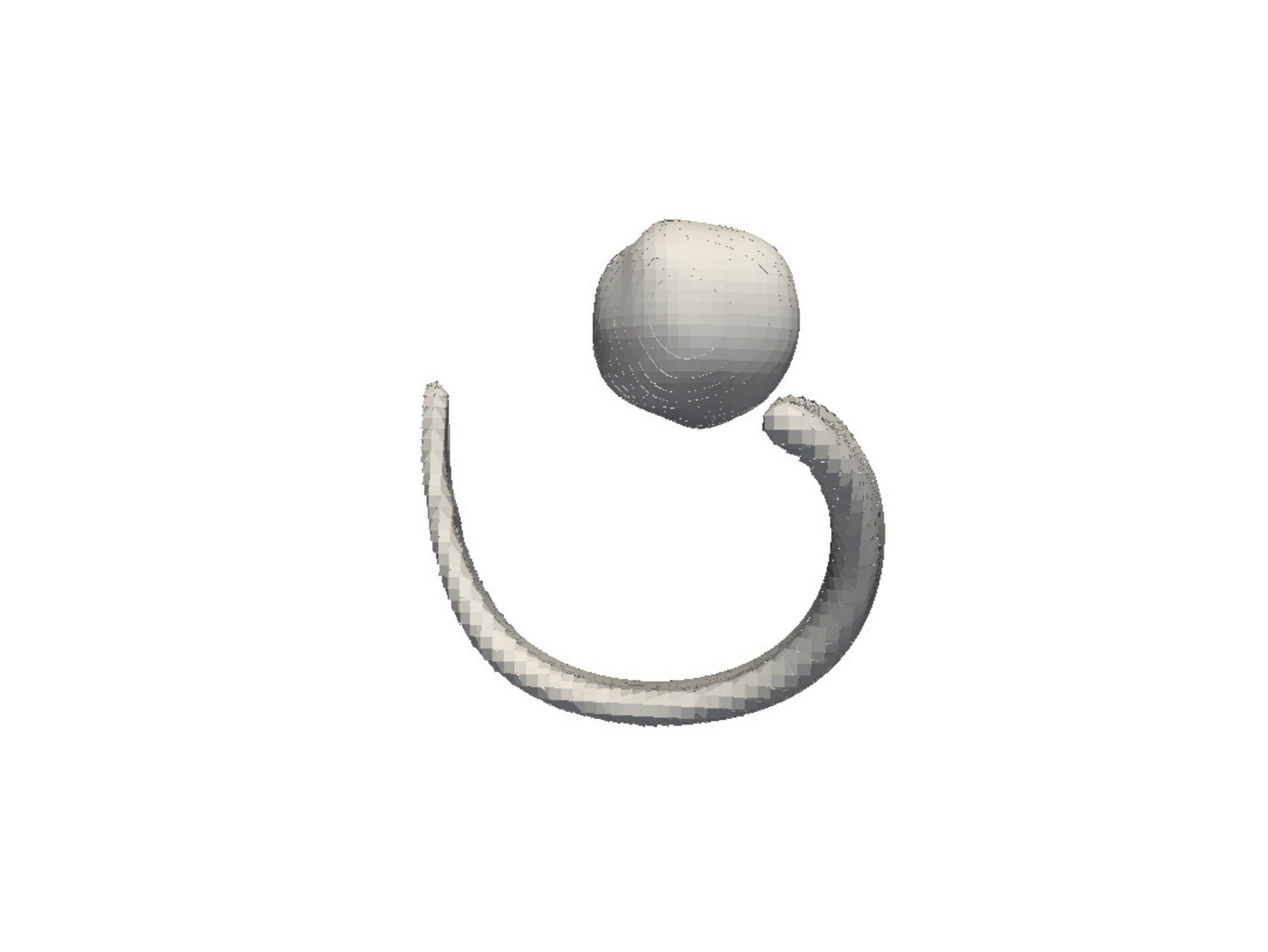}} 
	\subcaptionbox{Hexahedral mesh N=256} 
	{\includegraphics[width=0.33\columnwidth, bb=291 150 842 792, clip=true]{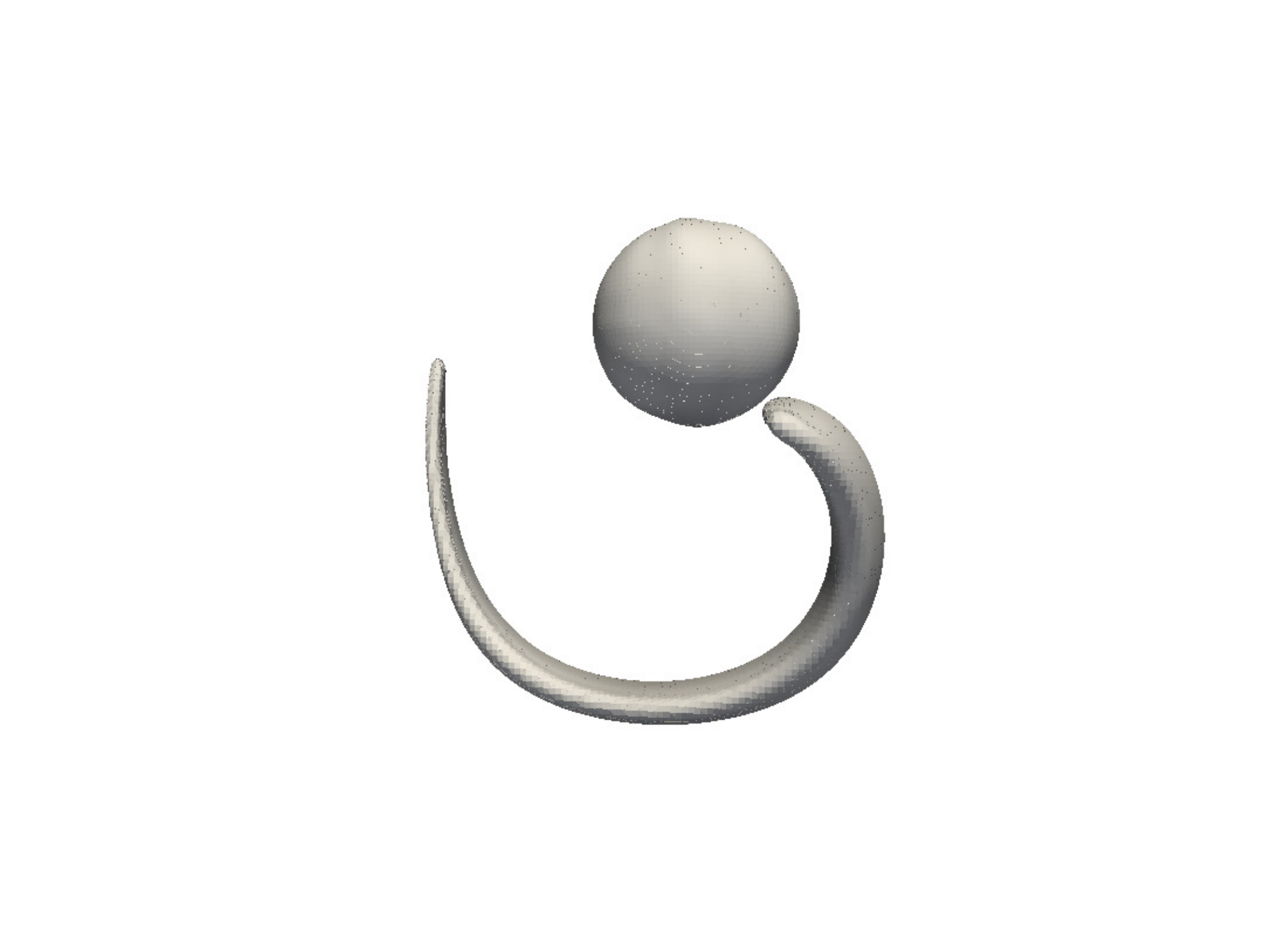}}
	
	\subcaptionbox{Tetrahedral mesh N=64} 
	{\includegraphics[width=0.33\columnwidth, bb=291 150 842 792, clip=true]{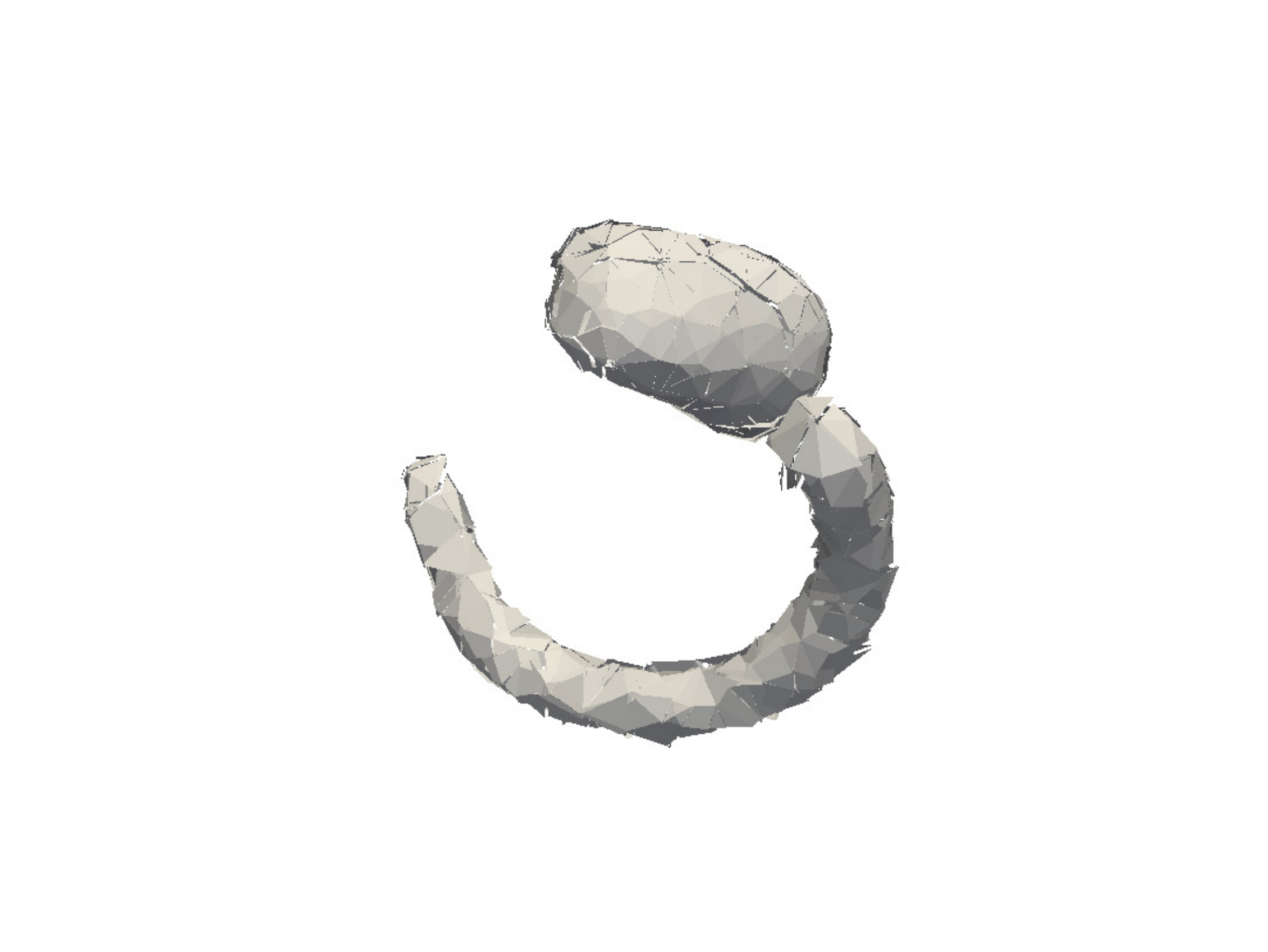}}
	\subcaptionbox{Tetrahedral mesh N=128} 
	{\includegraphics[width=0.33\columnwidth, bb=291 150 842 792, clip=true]{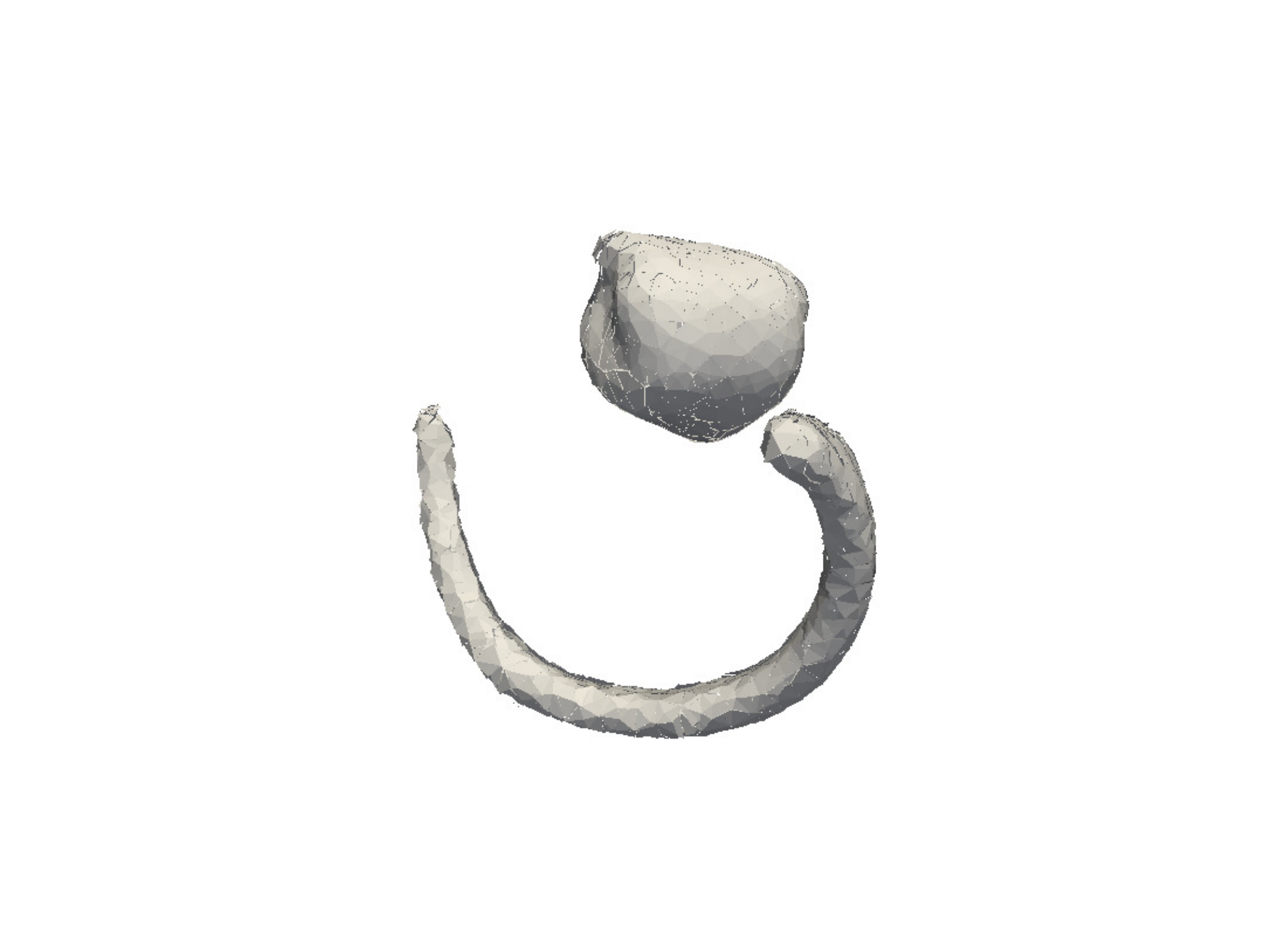}} 
	\subcaptionbox{Tetrahedral mesh N=256} 
	{\includegraphics[width=0.33\columnwidth, bb=291 150 842 792, clip=true]{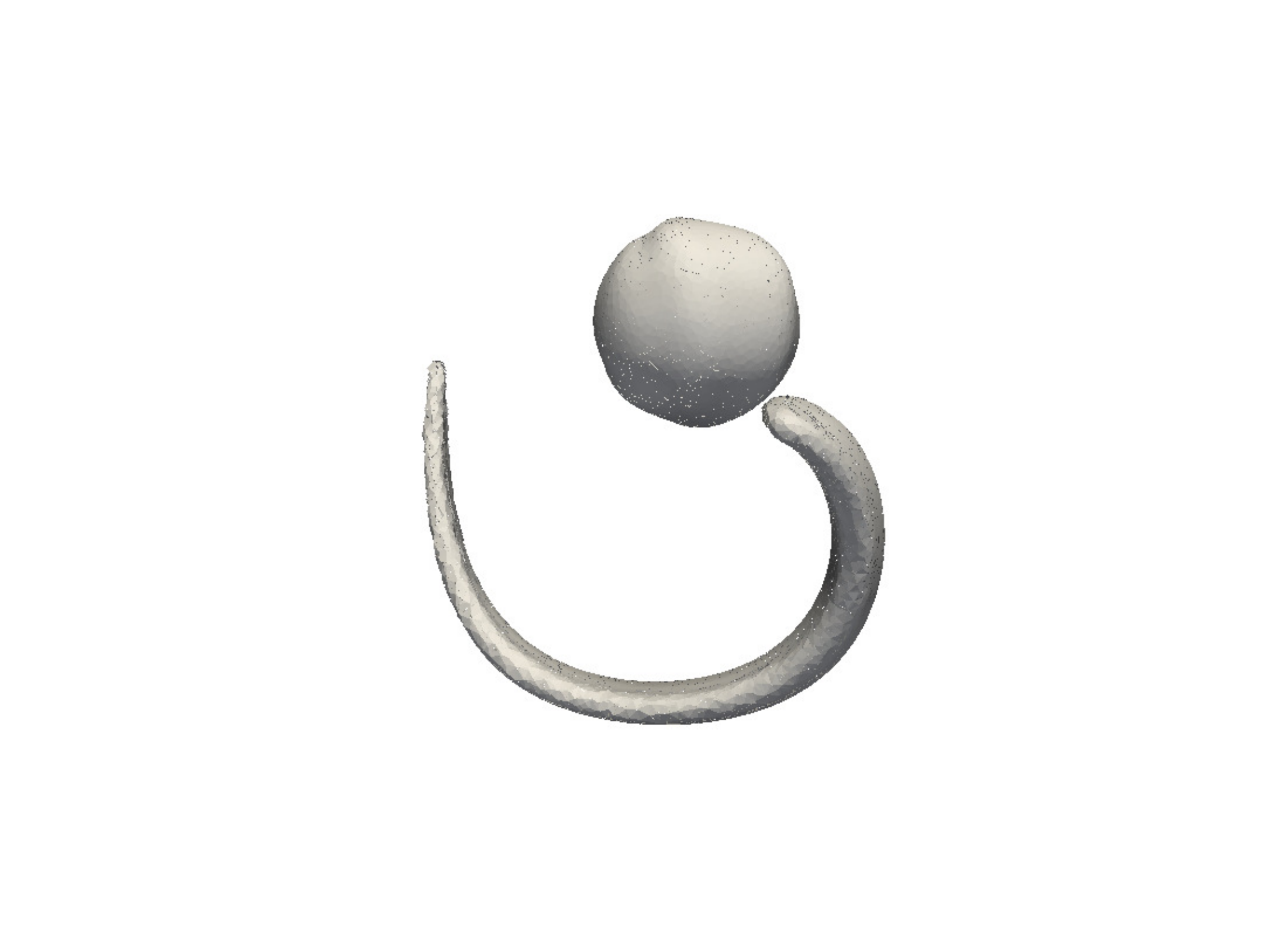}}
	
	\subcaptionbox{Polyhedral mesh N=64} 
	{\includegraphics[width=0.33\columnwidth, bb=291 150 842 792, clip=true]{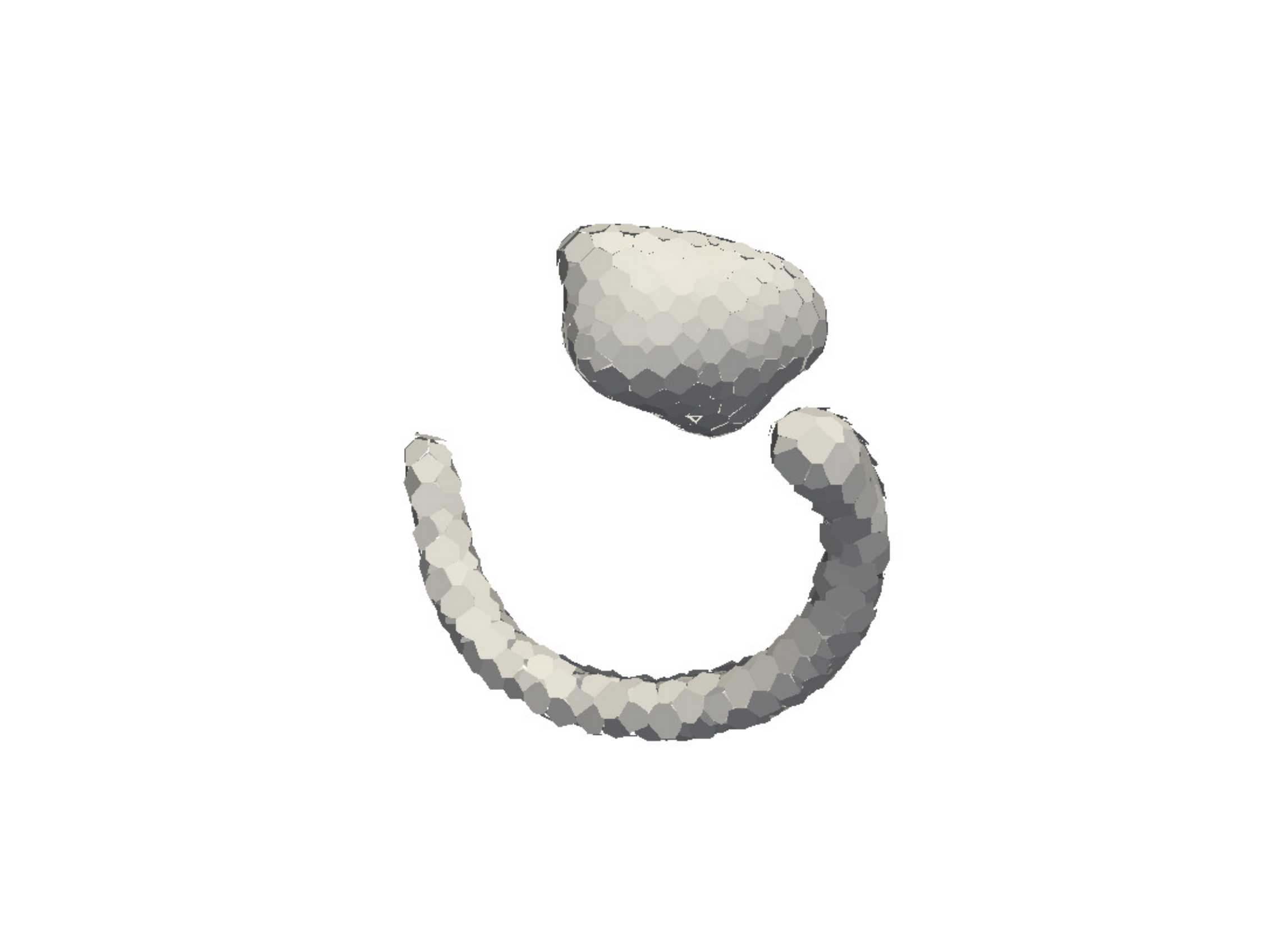}}
	\subcaptionbox{Polyhedral mesh N=128} 
	{\includegraphics[width=0.33\columnwidth, bb=291 150 842 792, clip=true]{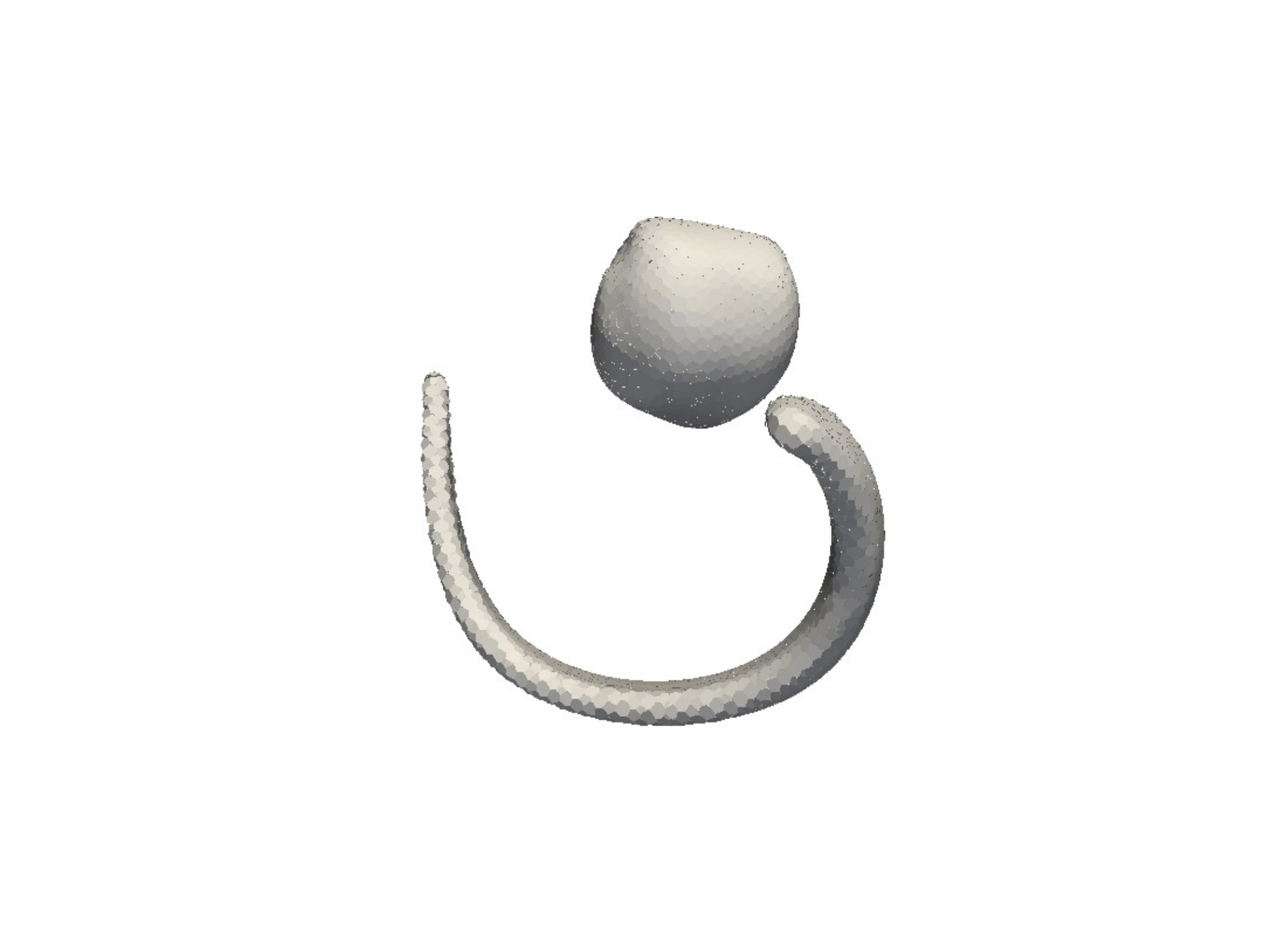}} 
	\subcaptionbox{Polyhedral mesh N=256} 
	{\includegraphics[width=0.33\columnwidth, bb=291 150 842 792, clip=true]{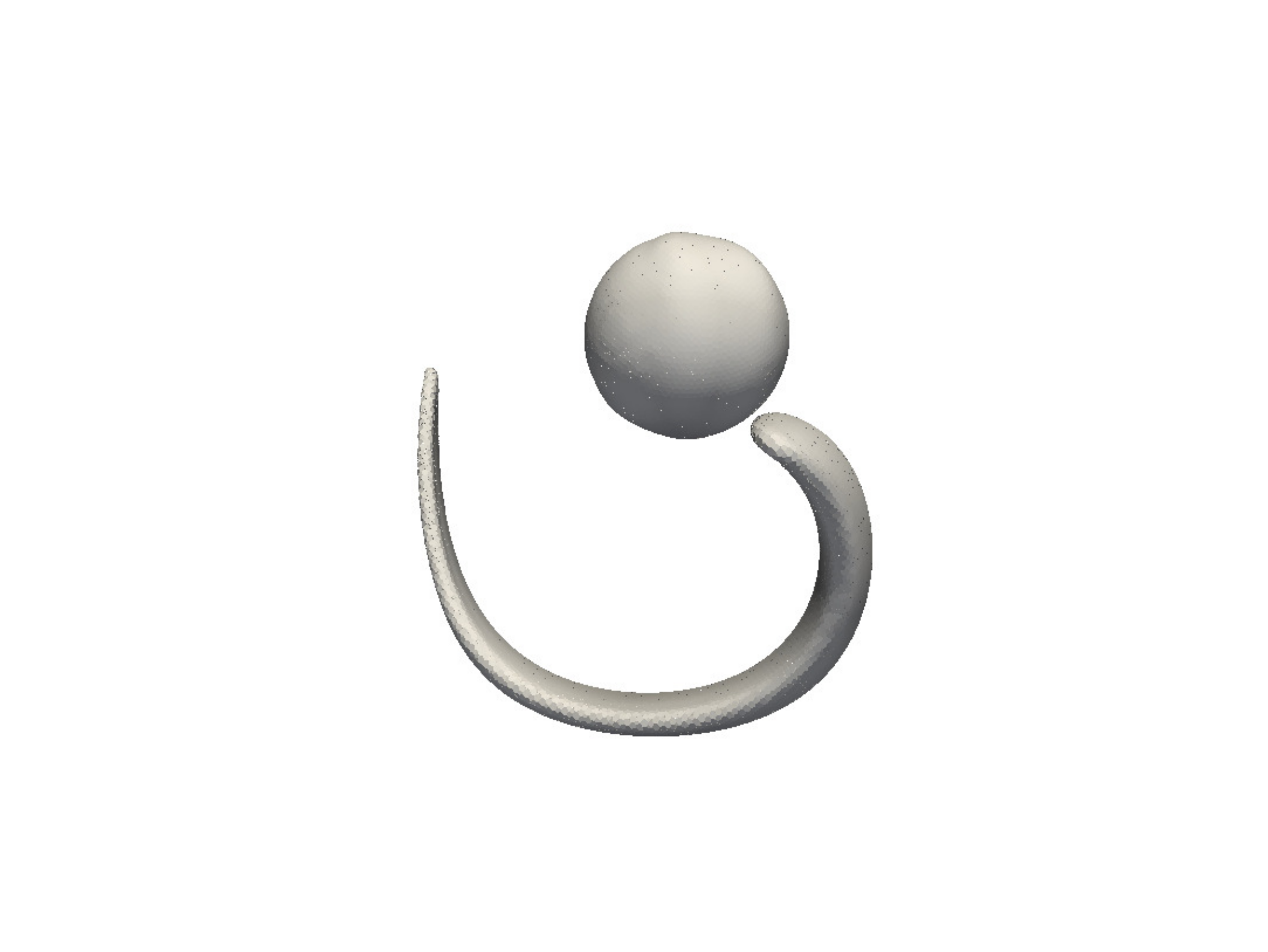}}
	\cprotect\caption{Interfaces at $t=1.5$ and $t=3$ second for the 3D shear flow test case with different resolutions and mesh types.}
	\label{fig:shearFlow3D}
\end{figure}

\subsubsection{3D deformation flow}\label{Leveque Test}

Leveque \cite{Leveque.1996} introduced another popular benchmark test case which was adopted by several authors \cite{Owkes.2014} \cite{Jofre.2014} \cite{Maric.2018} \cite{Lopez.2008}: A sphere of radius 0.15 centred at (0.35,0.35,0.35) in a unit cube domain is deformed with the velocity field:

\begin{equation}
	\mathbf u(x,y,z,t) = \cos(\pi t/T)
	\left(
    	\begin{matrix}
            2\sin^2{(\pi x)} \sin{(2 \pi y)} \sin{(2 \pi z)} \\
            -\sin{(\pi x)} \sin^2{(\pi y)} \sin{(2 \pi z)} \\
            -\sin^2{(\pi x)} \sin{(2 \pi y)} \sin^2{(\pi z)}
        \end{matrix}
    \right).
\end{equation}

\begin{figure}[htp]
	\subcaptionbox{Hexahedral mesh N=64} 
	{\includegraphics[width=0.30\columnwidth , bb=212 60 912 742, clip=true]{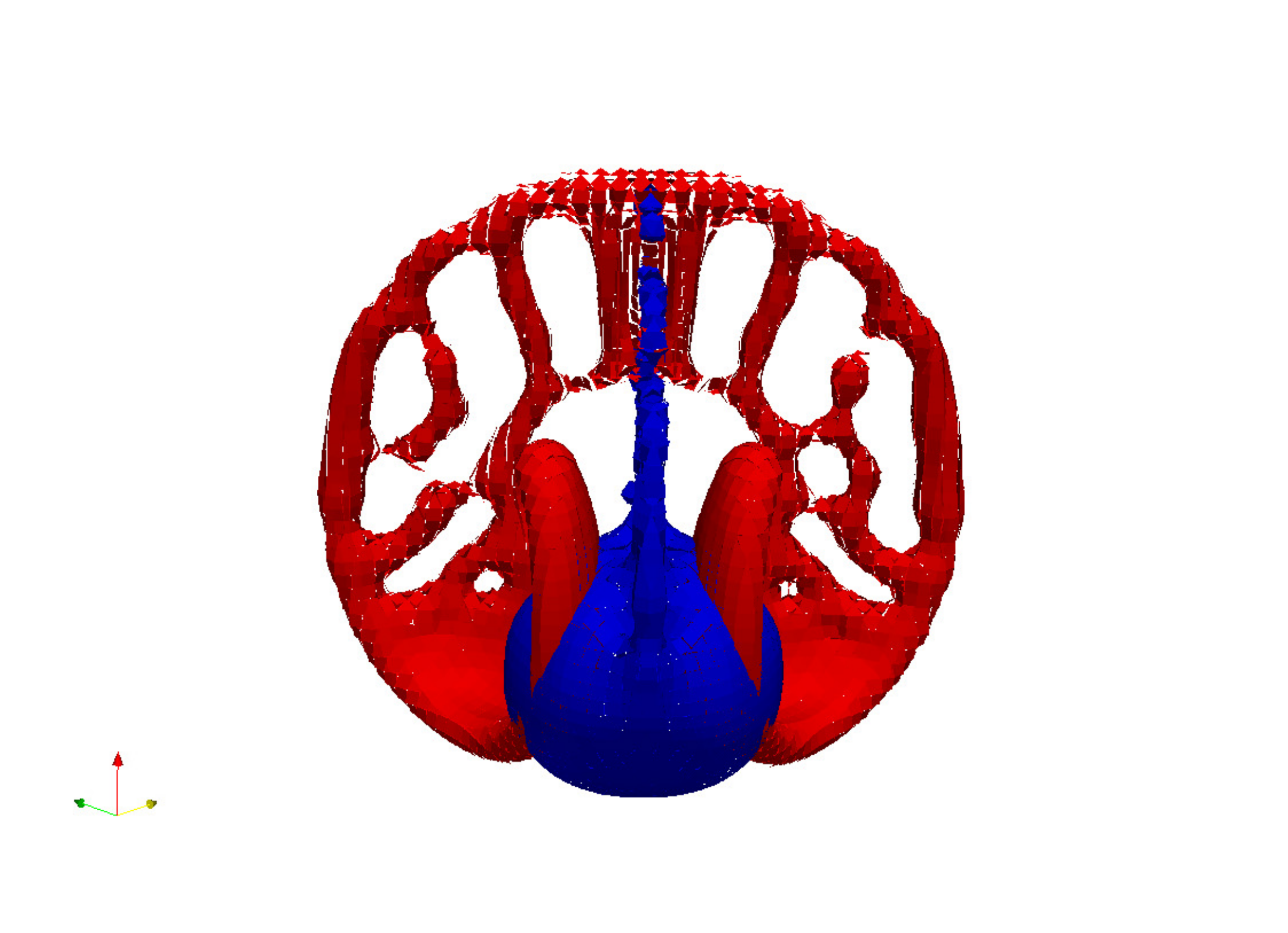}}
	\subcaptionbox{Hexahedralmesh N=128} 
	{\includegraphics[width=0.30\columnwidth , bb=272 60 912 700, clip=true]{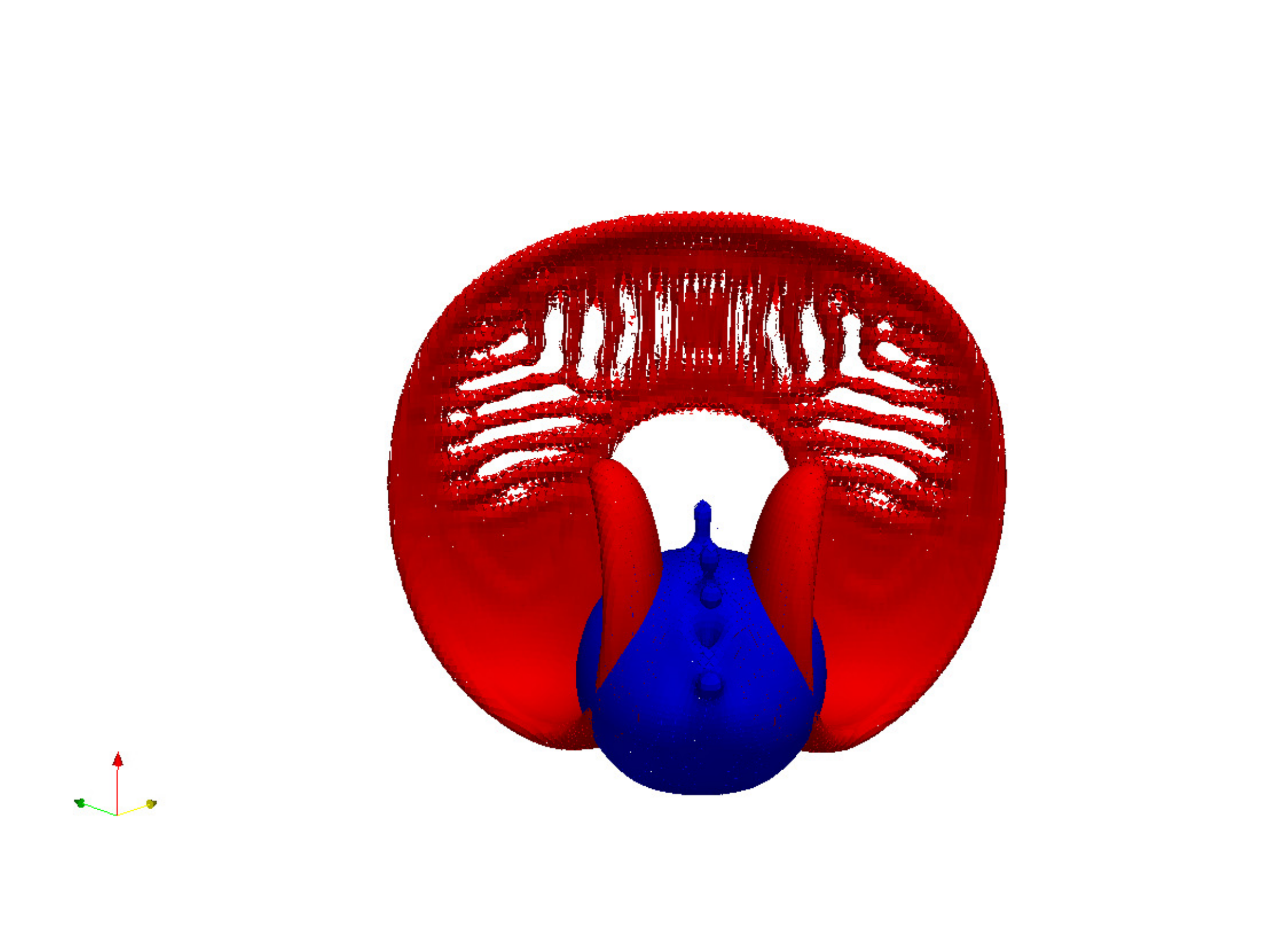}} 
	\subcaptionbox{Hexahedral mesh N=256} 
	{\includegraphics[width=0.30\columnwidth , bb=212 60 912 742, clip=true]{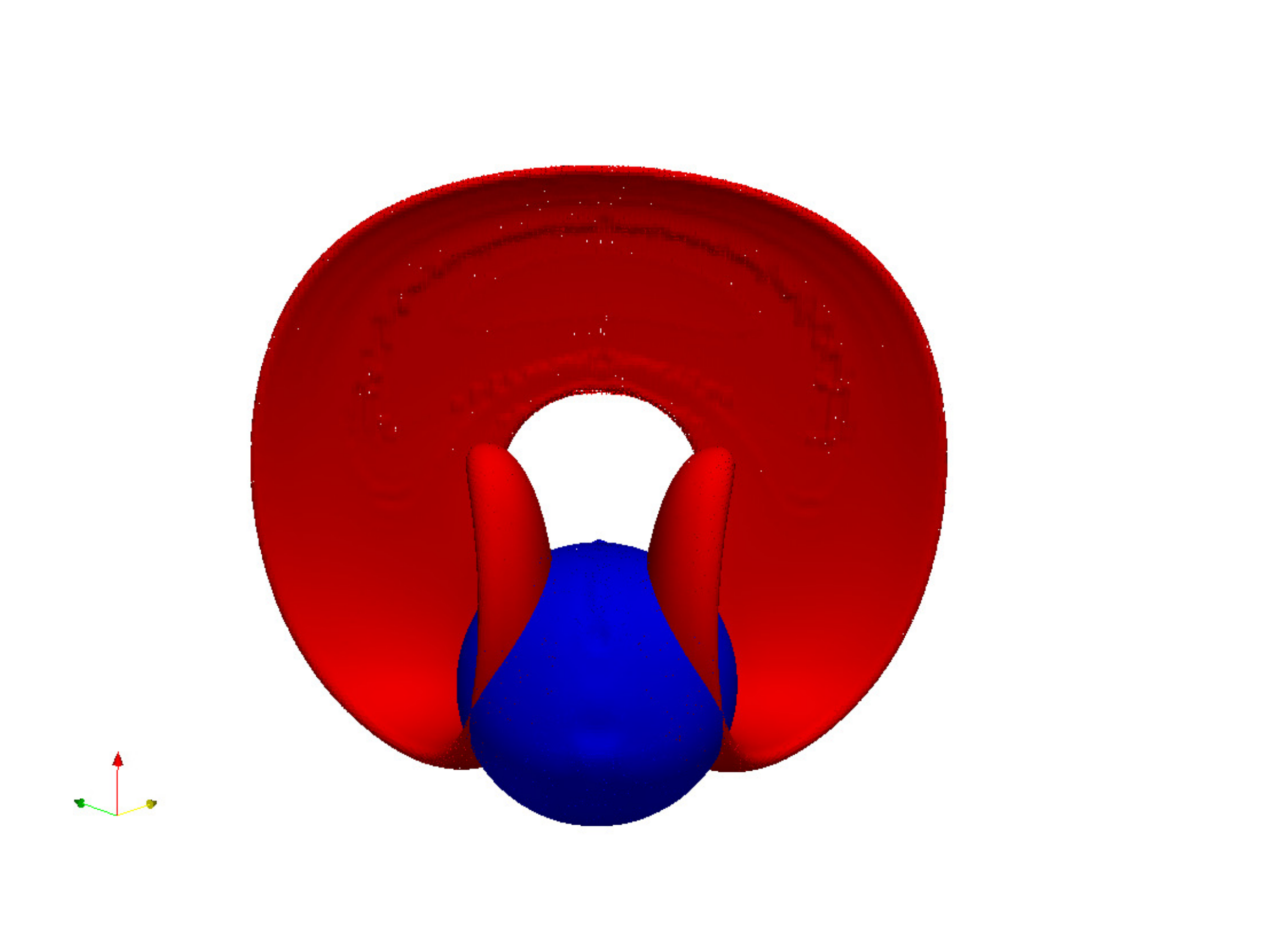}}
	
	\subcaptionbox{Tetrahedral mesh N=64} 
	{\includegraphics[width=0.30\columnwidth , bb=212 60 912 742, clip=true]{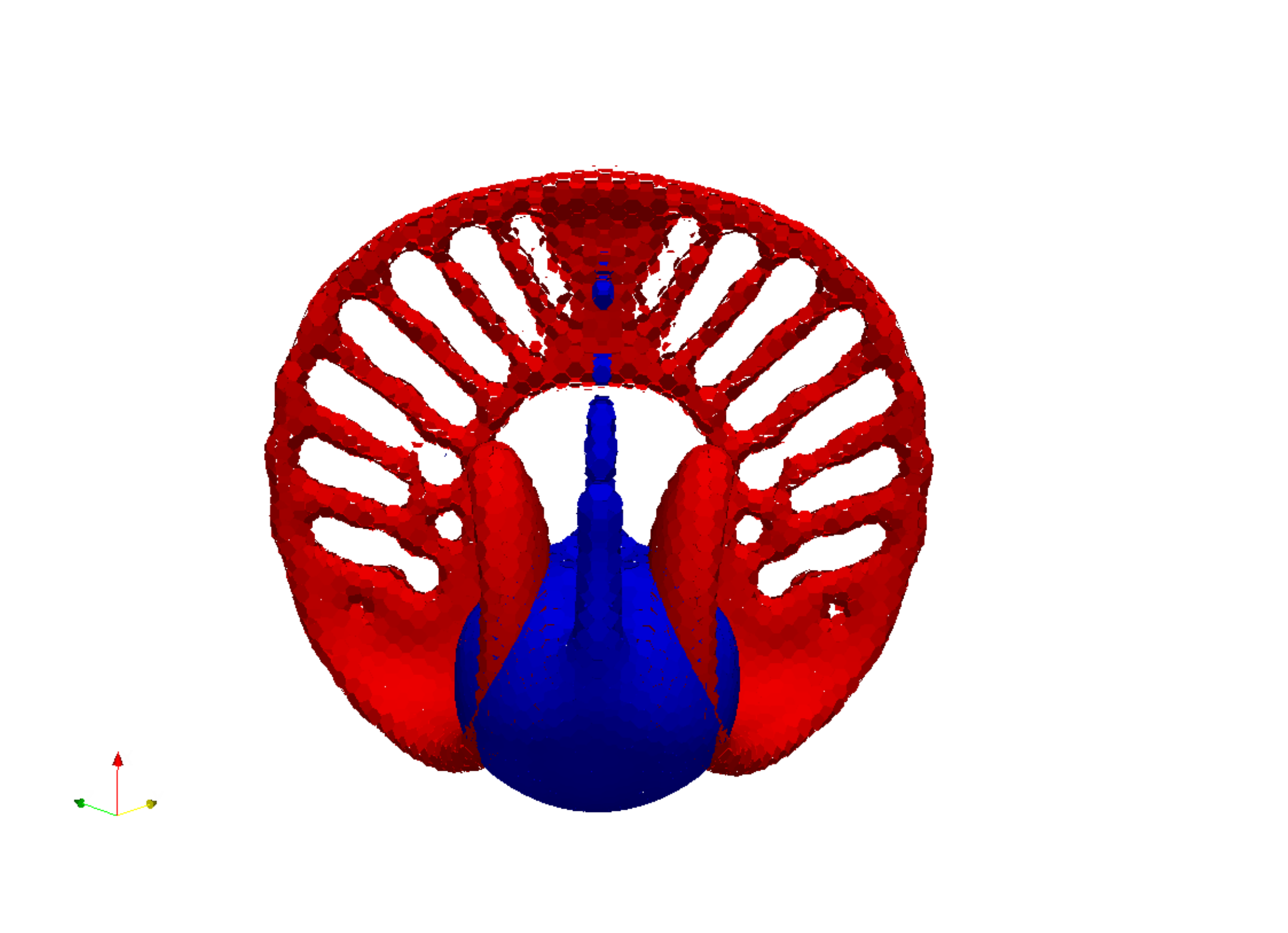}}
	\subcaptionbox{Tetrahedral mesh N=128} 
	{\includegraphics[width=0.30\columnwidth , bb=212 60 912 742, clip=true]{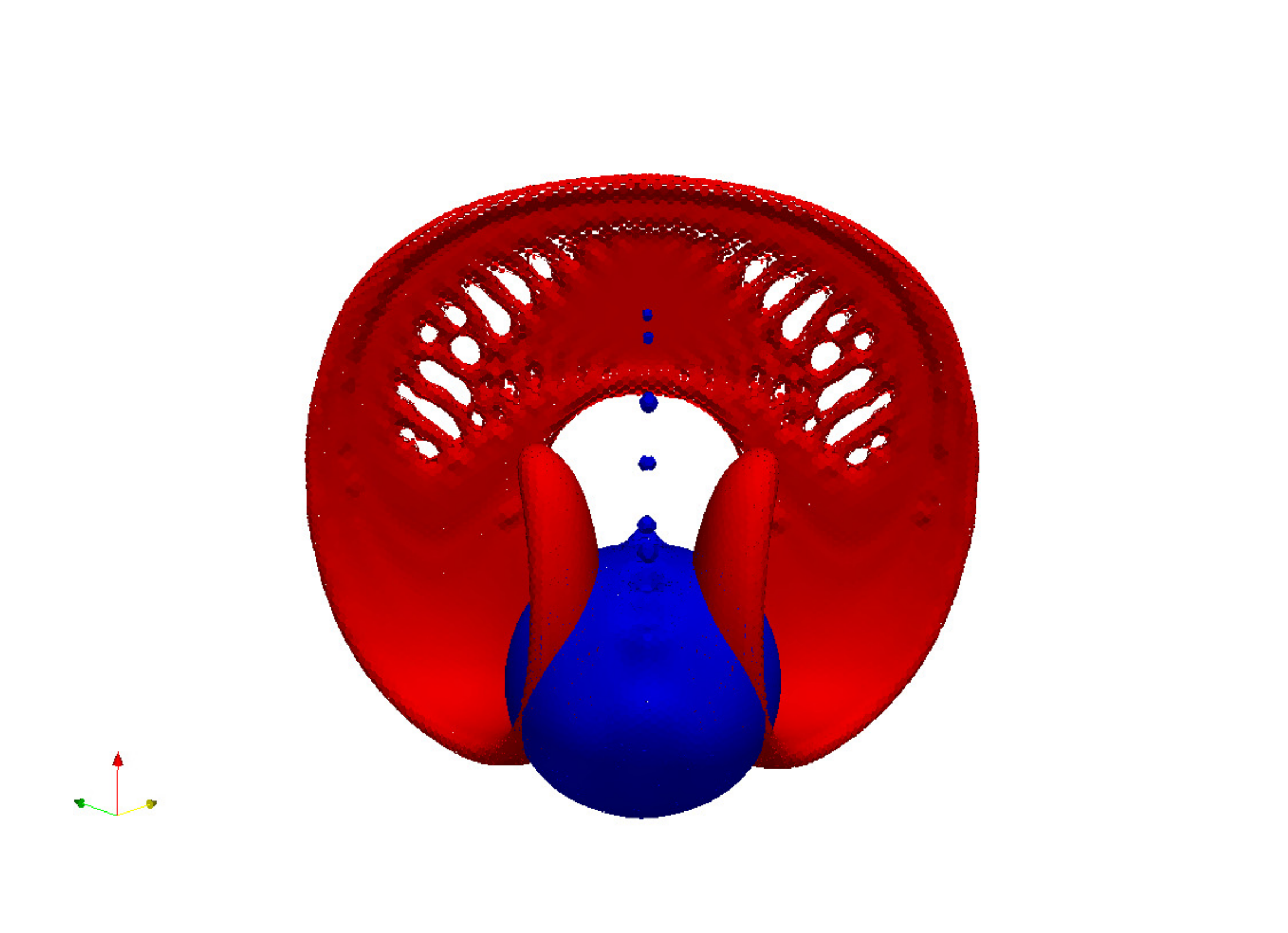}} 
	\subcaptionbox{Tetrahedral mesh N=256} 
	{\includegraphics[width=0.30\columnwidth , bb=212 60 912 742, clip=true]{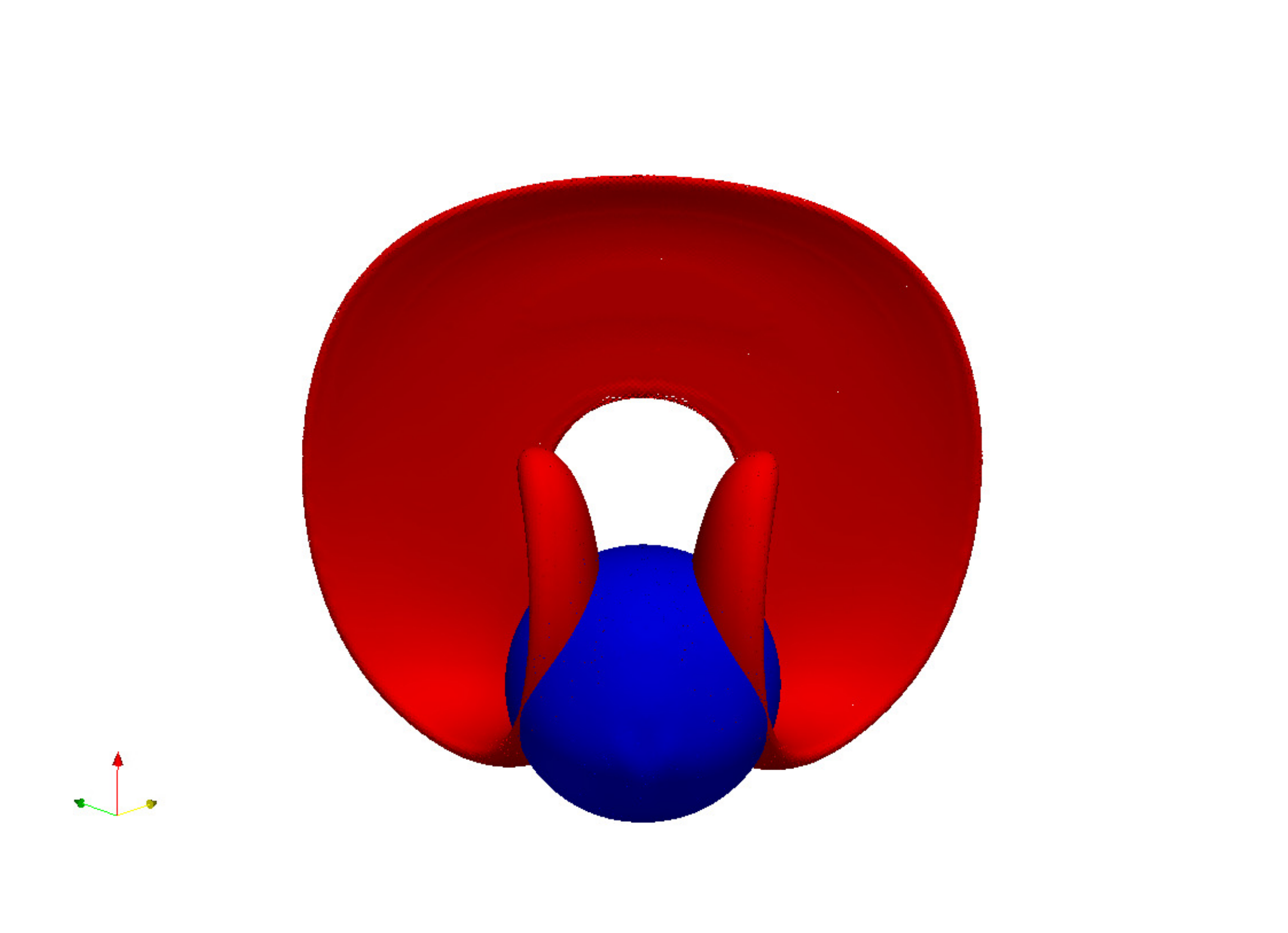}}
	
	\subcaptionbox{Polyhedral mesh N=64} 
	{\includegraphics[width=0.30\columnwidth , bb=212 60 912 742, clip=true]{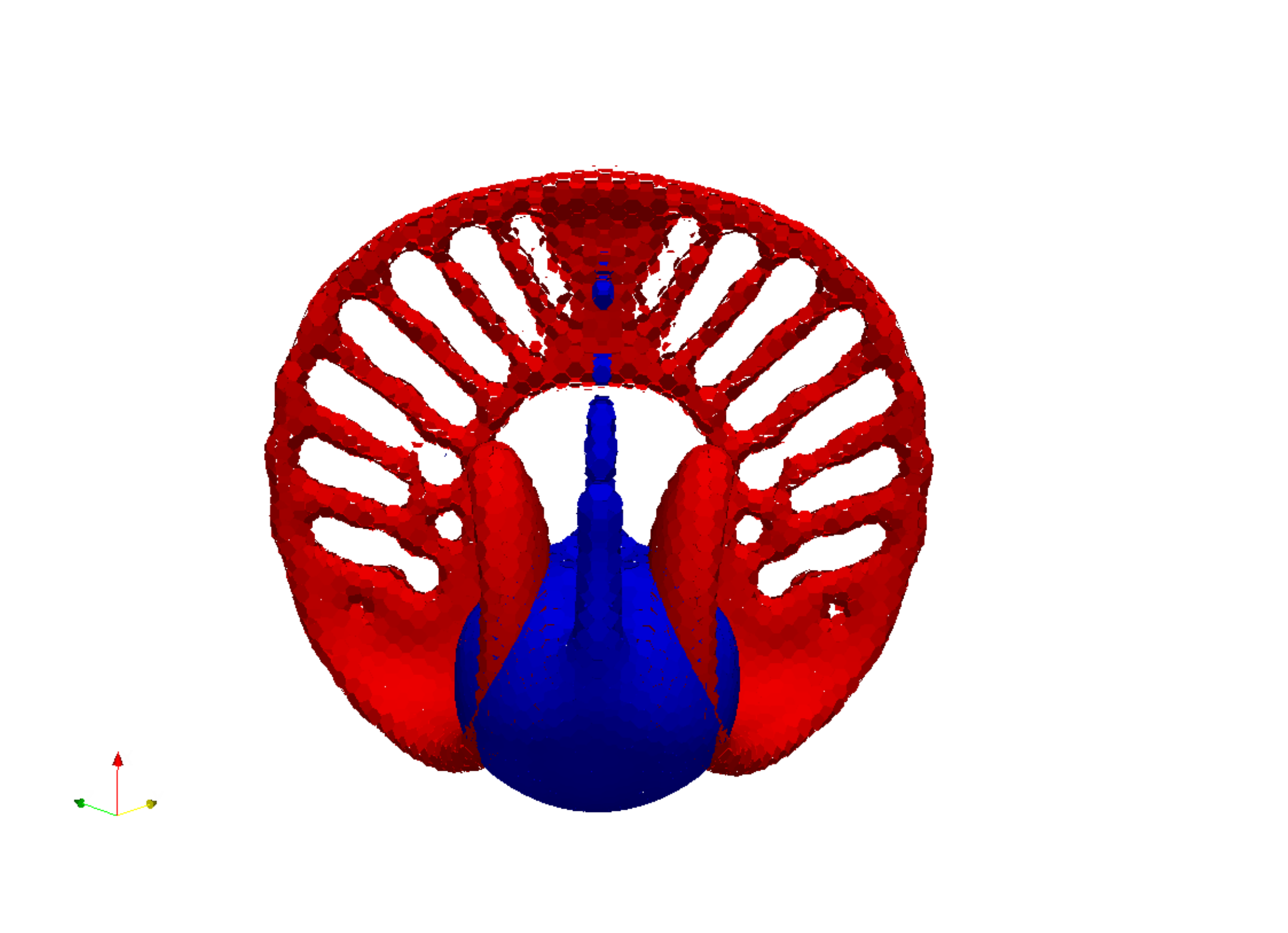}}
	\subcaptionbox{Polyhedral mesh N=128} 
	{\includegraphics[width=0.30\columnwidth , bb=212 60 912 742, clip=true]{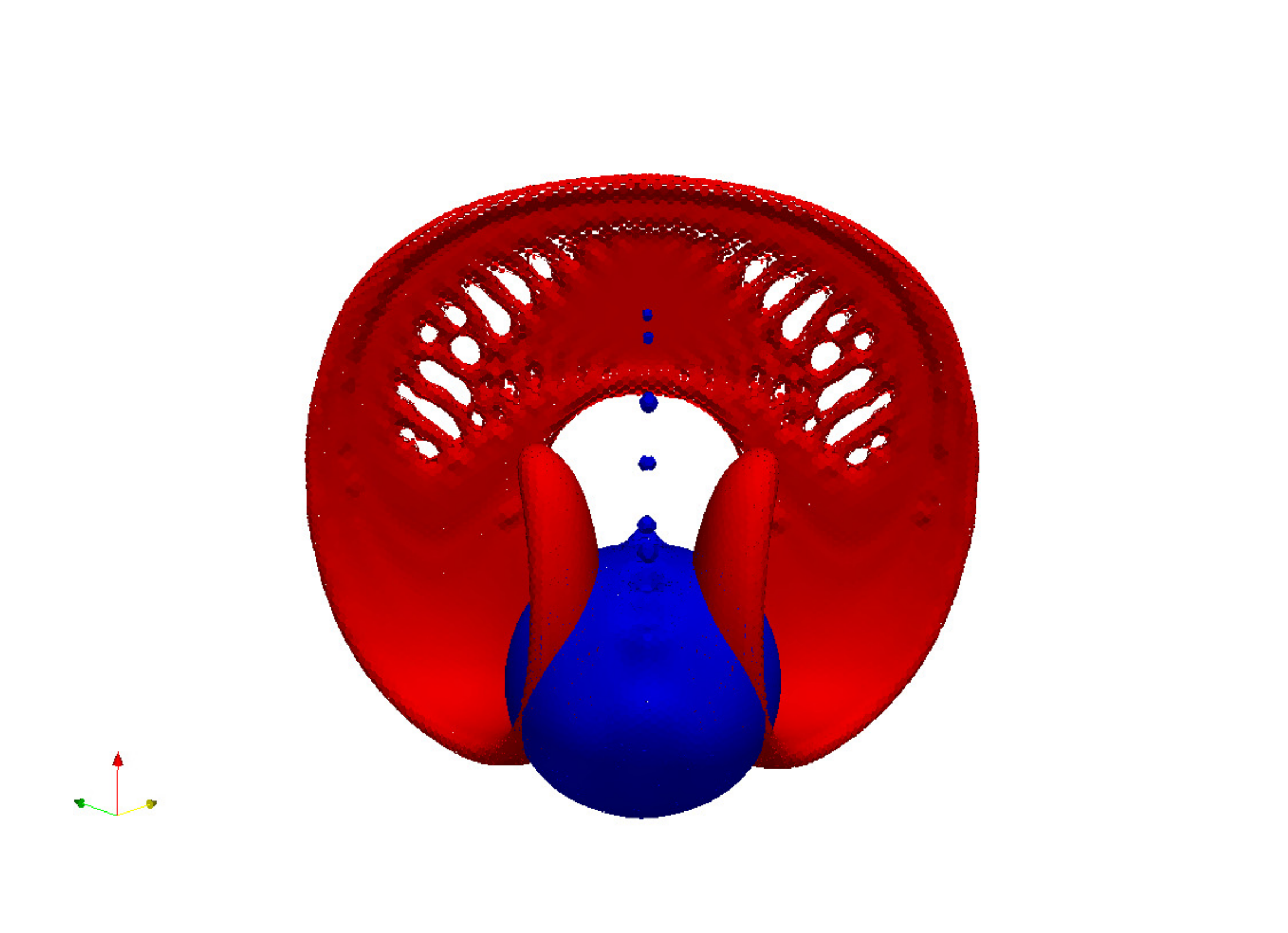}} 
	\subcaptionbox{Polyhedral mesh N=256} 
	{\includegraphics[width=0.30\columnwidth , bb=212 60 912 742, clip=true]{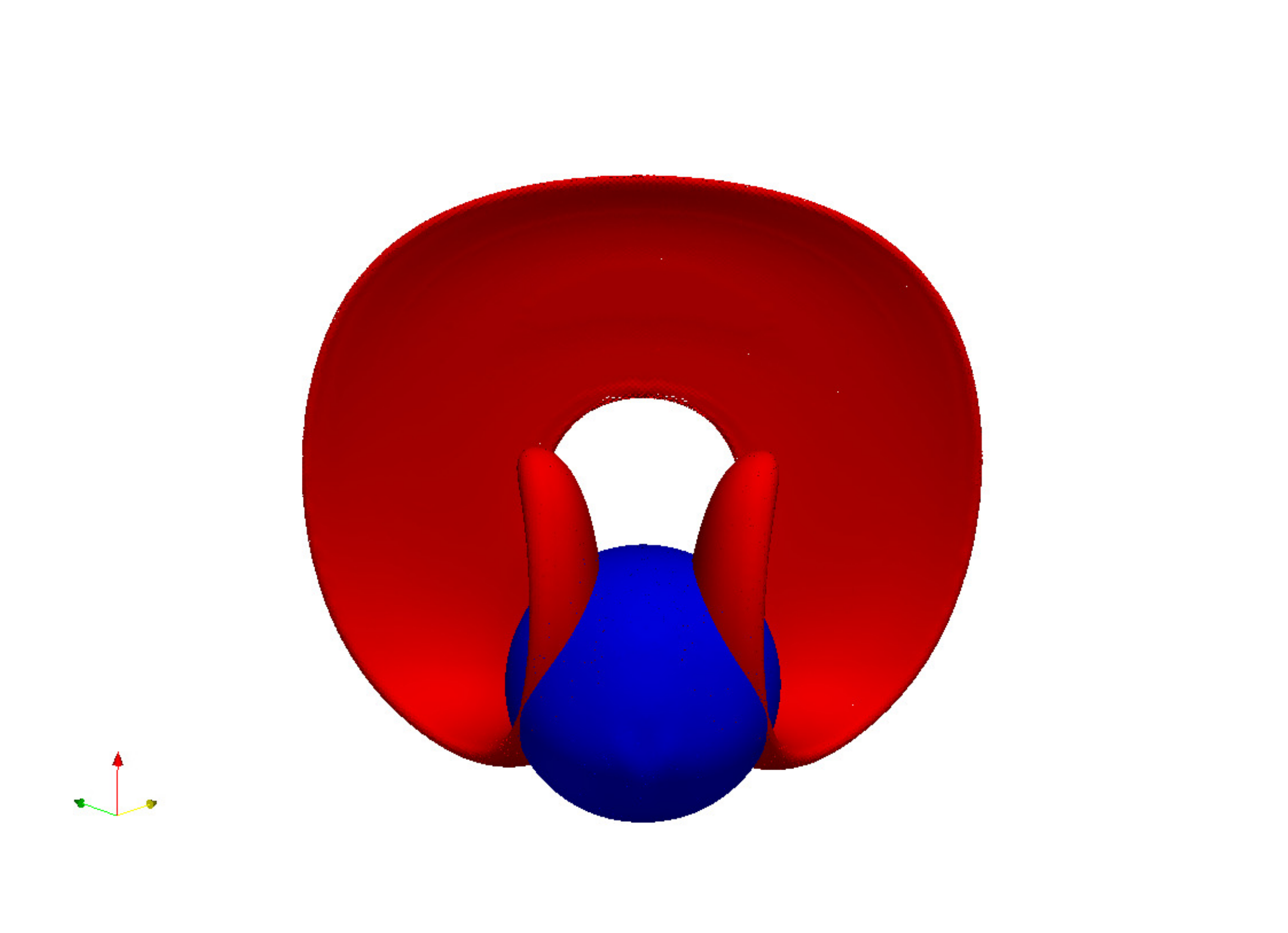}}
	\cprotect\caption{Interfaces at $t=1.5$ (red) and $t=3$ (blue) second for the 3D deformation flow test case with different resolutions and mesh types.}
	\label{fig:Leveque}	
\end{figure}

\begin{table}[hp]
\centering
\caption{3D deformation flow with CFL = 0.5 or $\text{CFL}_{surf}$ = 0.5: Errors and timings (4 cores) on various mesh types and resolutions.}
\label{tab:deformation}
\begin{tabular}{l c c c c c c c} 
\hline
\hline
Cartesian meshes & N & $E_{Vol}$ & $E_{bound}$ & $E_{1}$ & $\mathcal O(E_{1}) $ &  $T_e$ & $T_r$\\
\hline
\hline
isoAdvector-plicRDF & 32  & 1.076e-16 &  2.115e-14 &  8.36e-03 &  - &  0.058 &  0.049\\
UFVFC-Swartz \cite{Maric.2018} & 32  & 2.46e-15 & 0.0 & 5.86e-03 & -  & 0.69 & 0.14\\
Owkes and Desjardins \cite{Owkes.2014} & 32  & 2.79e-15 & 2.341e-17 & 6.97e-03 & - & 0.78 & -\\
Jofre et al. \cite{Jofre.2014} & 32  & - & - & 6.92e-03 & - & - & -\\
\hline
isoAdvector-plicRDF & 64  & 9.749e-16 &  6.406e-20 &  3.25e-03 &  1.36 &  0.15&  0.13\\
UFVFC-Swartz \cite{Maric.2018} & 64  & 6.01e-15 & 0.0 & 1.56e-03 & 2.34 & 1.91 & 0.51\\
Owkes and Desjardins \cite{Owkes.2014} & 64 & 1.675e-14 & 2.752e-17 & 2.09e-03 & 1.73 & 2.85 & -\\
Jofre et al. \cite{Jofre.2014} & 64  & - & - & 2.43e-03 & 1.51 & - & -\\
\hline
isoAdvector-plicRDF & 128 & 3.705e-15 & 1.11e-15 & 6.574e-04 & 2.31 & 0.52 & 0.39\\
UFVFC-Swartz \cite{Maric.2018} & 128 & 1.56e-14 & 0.0 & 3.08e-04 & 2.81 & 12.00 & 2.37\\
Owkes and Desjardins \cite{Owkes.2014} & 128 & 1.675e-14 & 2.752e-17 & 5.62e-04 & 1.89 & 12.2 & -\\
Jofre et al. \cite{Jofre.2014} & 128 & - & - & 6.37e-04 & 1.93 & - & -\\
\hline 
isoAdvector-plicRDF & 256 & 2.030e-14 &  2.366e-16 &  9.54e-05 &  2.78 &  2.63 &  1.67\\
Owkes and Desjardins \cite{Owkes.2014} & 256 & 3.870e-14 & 4.690e-17 & 1.01e-04 & 2.47 &  45.5 & -\\
\hline
\hline
Tetrahedral meshes & N & $E_{Vol}$ & $E_{bound}$ & $ E_{1} $ & $\mathcal O(E_{1})$ & $T_e$ & $T_r$\\
\hline
\hline
isoAdvector-plicRDF & 32 & 1.648e-16 & 4.501e-18 & 1.31e-02 & - &  0.098 & 0.09\\
Jofre et al. \cite{Jofre.2014} & 32 & - & - & 1.02e-02 & - & - & - \\
\hline
isoAdvector-plicRDF & 64 & 9.558e-16 & 4.143e-19 & 6.34e-03 & 1.06 & 0.32 & 0.30\\
Jofre et al. \cite{Jofre.2014} & 64 & - & - & 3.54e-03 & 1.53 & - & - \\
\hline
isoAdvector-plicRDF & 128 & 7.752e-15 & 7.656e-21 & 1.31e-03 & 2.27 & 1.45 &  1.25\\
Jofre et al. \cite{Jofre.2014} & 128 & - & - & 7.20e-04 & 2.30 & - & - \\
\hline
isoAdvector-plicRDF & 256 & 7.206e-14 & 2.516e-20 & 1.93e-04 & 2.77 & 5.58 &  4.12\\
\hline
\hline
Polyhedral meshes & N & $E_{Vol}$ & $E_{bound}$ & $ E_{1} $ & $\mathcal O(E_{1})$ & $T_e$ & $T_r$\\
\hline
\hline
isoAdvector-plicRDF & 32 & 1.023e-16 & 2.768e-18 & 1.21e-02 & - & 0.072 & 0.061\\
\hline
isoAdvector-plicRDF & 64 & 9.524e-16 & 2.893e-15 & 3.32e-03 & 1.88 & 0.25 &  0.21\\
\hline
isoAdvector-plicRDF & 128 & 5.933e-16 & 2.323e-15 & 5.50e-04 & 2.59 & 0.89 & 0.61\\
\hline
isoAdvector-plicRDF & 256 & 1.581e-13 & 7.349e-22 & 8.55e-05 & 2.69 & 5.59 &  2.58\\
\hline
\hline
\end{tabular}
\end{table}

The top part of Table~\ref{tab:deformation} shows the error for Cartesian meshes with a CFL number of 0.5. UFVFC-Swartz \cite{Maric.2018} shows the lowest errors on all resolutions. The $\text{CFL}_{surf}$ was used for unstructured meshes. isoAdvector-plicRDF shows the highest errors on coarser meshes, but catches up with Jofre et al. \cite{Jofre.2014} and Owkes and Desjardins \cite{Owkes.2014} on the fine meshes. Our execution times are more than an order of magnitude lower than the ones reported in \cite{Maric.2018} and Owkes and Desjardins \cite{Owkes.2014}. The reconstruction alone is a factor of 3-5 faster than Maric et al. \cite{Maric.2018} with the largest factor for fine meshes.

On tetrahedral meshes (middle rows of Table~\ref{tab:deformation}) our method gives 30-50\% larger errors than Jofre et al. \cite{Jofre.2014}. Also our method gives larger errors than on the Cartesian meshes with similar resolution and calculation times are almost doubled. For the polyhedral meshes (lower part of Table~\ref{tab:deformation}) the errors are closer to those on Cartesian meshes but calculation times are still higher.A visualization of the results is shown in Fig.~{\ref{fig:Leveque}} confirming the similar shape errors at the same resolution for the different mesh types.

\section{Conclusion}\label{Sec:Conclusion}

We have presented an iterative residual based interface reconstruction procedure utilizing a reconstructed distance function (RDF) to estimate the local interface position and orientation from the raw volume fraction data. The new method is demonstrated to exhibit second order convergence with mesh refinement in the interface position and orientation on both 2D and 3D structured and unstructured meshes. The algorithm has been developed in two variants based on RDF isosurface reconstruction and on piecewise linear interface construction (PLIC), respectively. Especially on unstructured meshes and for the local interface orientation both methods have significantly improved convergence properties compared to the reconstruction method presented in \cite{Roenby.2016}, which was based on isosurfaces of the volume fraction data. A comparison with literature shows a similar level of accuracy as Height Function based methods and more accurate results than the LVIRA method. The implementation of the new reconstruction methods is straightforward and the increased computational cost caused by the introduction of iterations is limited to an acceptable level. To further reduce the computational costs the initial residuals are reduced by using interpolated interface normals from the previous time step.

The new reconstruction scheme is combined with the isoAdvector advection step \cite{Roenby.2016}, and a number of simple advection test cases are investigated. The overall conclusion is that second order convergence with mesh refinement is achieved on all mesh types if the CFL number is sufficiently low (0.2 or below). Work is in progress to improve accuracy and convergence with higher CFL numbers. On unstructured tetrahedral meshes depending on the test case and resolution our errors vary from similar in seize to twice as large as the those reported by Jofre et al. \cite{Jofre.2014}. On Cartesian meshes, we obtain errors between half to twice as large as those reported by Jofre et al. \cite{Jofre.2014}. The recent method by Maric et al. \cite{Maric.2018} yields significantly reduced errors. The main advantage of our method compared to Maric et al. \cite{Maric.2018} is the computational cost which is roughly an order of magnitude lower.

The new methods are implemented as an OpenFOAM extension, and released as open source together with this paper \cite{Scheufler.2018}. The framework is structured in a way allowing easy implementation and testing of new reconstruction and advection schemes and with a special emphasis on future extensions to adaptive mesh refinement. It is our hope that the released library will be used, tested and further developed by the CFD community, and eventually result in improved simulation quality in the broad field of applications involving interfacial and free surface flows.

\section*{Acknowledgement}

This work was supported by German Aerospace Center - DLR and by the GTS grant to DHI (where the initial part of JR's work on this article was done) from the Danish Agency for Science, Technology and Innovation.

\bibliographystyle{elsarticle-num}
\bibliography{citations}
	
\end{document}